\documentclass[aps,pra,twocolumn,nofootinbib,floatfix,superscriptaddress]{revtex4-1}
\usepackage{amsmath,mleftright,mathtools}
\usepackage{bm}
\usepackage{stix,microtype}
\usepackage{graphicx}
\usepackage{xspace}
\usepackage{units}
\usepackage[colorlinks,linkcolor=blue,citecolor=blue,urlcolor=blue]{hyperref}
\usepackage[dvipsnames]{xcolor}
\usepackage{cleveref}
\usepackage{array}   
\newcolumntype{L}{>{$}c<{$}} 
\newcolumntype{C}{>{$}c<{$}} 
\usepackage{dcolumn}
\usepackage{bbm}

\newcommand{\be}{\begin{equation}}
\newcommand{\ee}{\end{equation}}
\newcommand{\bea}{\begin{eqnarray}}
\newcommand{\eea}{\end{eqnarray}}

\def\a{\alpha}
\def\b{\beta}
\def\g{\gamma}
\def\G{\Gamma}
\def\d{\delta}
\def\D{\Delta}

\def\th{\theta}
\def\Th{\Theta}

\def\l{\lambda}
\def\L{\Lambda}
\def\m{\mu}
\def\n{\nu}

\def\p{\pi}
\def\P{\Pi}
\def\r{\rho}
\def\s{\sigma}
\def\S{\Sigma}
\def\t{\tau}
\def\f{\phi}
\def\vf{\varphi}
\def\F{\Phi}

\def\w{\omega}
\def\W{\Omega}
\def\q{\psi}
\def\Q{\Psi}

\def\bgf{\mbox{\boldmath $\phi$}}
\def\bgvf{\mbox{\boldmath $\varphi$}}

\def\ble{{\mathbf e}}

\def\blk{{\mathbf k}}

\def\bln{{\mathbf n}}

\def\blq{{\mathbf q}}
\def\blr{{\mathbf r}}

\def\blu{{\mathbf u}}

\def\blx{{\mathbf x}}

\def\bcallg{\mbox{\boldmath $g$}}

\def\bcalls{\mbox{\boldmath $s$}}

\def\blP{{\mathbf P}}

\def\blR{{\mathbf R}}

\def\blU{{\mathbf U}}

\def\callH{\mbox{$\mathcal{H}$}}

\def\callT{\mbox{$\mathcal{T}$}}
\def\callU{\mbox{$\mathcal{U}$}}

\def\callZ{\mbox{$\mathcal{Z}$}}

\def\ra{\rightarrow}

\def\de{\partial}
\def\iif{\infty}
\def\bra{\langle}
\def\ket{\rangle}
\def\grad{\mbox{\boldmath $\nabla$}}

\def\Tr{{\rm Tr}}

\def\iu{{\rm i}}
\def\1op{\hat{\mathbbm{1}}}
\def\nn{\nonumber}

\keywords{}
\begin{document}
\title{In- and out-of-equilibrium {\em ab initio} theory of electrons and phonons}
\author{Gianluca Stefanucci}
\affiliation{Dipartimento di Fisica, Universit{\`a} di Roma Tor Vergata, Via della Ricerca Scientifica 1,
00133 Rome, Italy}
\affiliation{INFN, Sezione di Roma Tor Vergata, Via della Ricerca Scientifica 1, 00133 Rome, Italy}
\author{Robert van Leeuwen}
\affiliation{Department of Physics, Nanoscience Center P.O.Box 35
FI-40014 University of Jyv\"{a}skyl\"{a}, Finland}
\author{Enrico Perfetto}
\affiliation{Dipartimento di Fisica, Universit{\`a} di Roma Tor Vergata, Via della Ricerca Scientifica 1,
00133 Rome, Italy}
\affiliation{INFN, Sezione di Roma Tor Vergata, Via della Ricerca Scientifica 1, 00133 Rome, Italy}
\date{\today}

\begin{abstract}  
In this work we lay down the {\em ab initio} many-body quantum theory of electrons 
and phonons in- and out-of-equilibrium at any temperature.
Our focus is on the harmonic approximation but the developed 
tools clearly indicate how to incorporate anharmonic effects. 
We begin by addressing a fundamental issue concerning the {\em ab 
initio} Hamiltonian in the harmonic approximation, 
which we show must be determined {\em self-consistently} to avoid 
inconsistencies.
After identifying the most suitable partitioning into a 
``noninteracting'' and an ``interacting'' part we embark on the 
Green's function diagrammatic analysis. We single out key 
diagrammatic structures to carry on the expansion in terms of 
dressed propagators and screened interaction. The final outcome is 
the finite-temperature nonequilibrium extension of the Hedin 
equations, featuring the appearance of the time-local Ehrenfest 
diagram in the electronic self-energy. 
The Hedin equations have limited practical utility for real-time 
simulations of systems driven out of equilibrium by external fields.
To overcome this limitation, we 
leverage the versatility of diagrammatic expansion to generate a 
closed system of integro-differential equations for the Green's 
functions and nuclear displacements.     
 These are the 
Kadanoff-Baym equations for electrons and phonons. 
Another advantage of the diagrammatic derivation is the ability to 
use conserving approximations, which ensure the satisfaction of the 
continuity equation and the conservation of total energy during time 
evolution.    
 As an example we show that the adiabatic 
Born-Oppenheimer approximation is not conserving whereas its 
dynamical extension is conserving only provided that the electrons 
are treated in the Fan-Migdal approximation with a dynamically 
screened electron-phonon coupling. We also derive the formal solution 
of the Kadanoff-Baym equations in the long time limit and at the 
steady state. The expansion of the phononic Green's function around the 
quasi-phonon energies points to a possible correlation-induced 
splitting of the phonon dispersion in materials 
with no time-reversal invariance.

\end{abstract}
\maketitle

\section{Introduction}

The concept of phonons as quasi-particles describing independent 
excitations of the nuclear lattice dates back to almost a century 
ago~\cite{bloch_uber_1929,Frenkelbook}. Nonetheless, 
the first rigorous theory of electrons and phonons 
saw the light of day in 1961~\cite{baym_field-theoretic_1961}. In a seminal paper 
Baym showed how to map the original electron-nuclear Hamiltonian onto 
a low-energy or equivalently electron-phonon ($e$-$ph$) Hamiltonian, and derived 
a set of equations for the electronic and phononic Green's 
functions (GF) $G$ and $D$. The Baym equation for  
the electronic GF  was rather implicit though. 
In the mid-sixties Hedin 
used the same technique as Baym, the so called source-field method or 
field-theoretic approach, to generate a more explicit set of 
equations for the electronic GF at clamped 
nuclei~\cite{hedin_new_1965}. The contributions by Baym and Hedin 
have been largely ignored by the 
electron-phonon community (including ourselves) in favor of 
semi-empirical Hamiltonians. Only recently the works of Baym, Hedin 
and a few 
others~\cite{keating_dielectric_1968,vanleeuwen_first-principles_2004,marini_many-body_2015} 
have been rigorously merged by Giustino in a unified many-body 
GF framework~\cite{feliciano_electron-phonon_2017}, which 
we keep naming the {\em Hedin-Baym equations} (instead of  ``Hedin 
equations'') after Giustino.

Despite these recent notable advances the {\em formal} theory of 
electrons and phonons is still not complete. 
We stress here the word ``formal'' as this paper does not 
address specific aspects or computational 
strategies related to the $e$-$ph$ interaction, 
for which we refer the reader to excellent 
textbooks~\cite{ziman_electrons_1960,grimvall_the-electron-phonon_1981,schrieffer_theory_1983,Mahan-book,BruusFlensberg:04,giustinobook,czycholl_solid_2023} and modern 
comprehensive reviews~\cite{feliciano_electron-phonon_2017},
rather it establishes a mathematically rigorous apparatus for the 
quantum treatment of
electrons and nuclei in the so called harmonic approximation.

Three pivotal 
issues are still waiting to be clarified and solved. The first issue has to do 
with the {\em ab initio} $e$-$ph$ Hamiltonian, often replaced 
by semi-empirical Hamiltonians or left unspecified as unnecessary for the 
implementation of approximate formulas for the phononic dispersions, 
life-times, etc.. The {\em ab initio} $e$-$ph$ Hamiltonian 
is, however, of paramount relevance. It is necessary for assessing  
the validity of semi-empirical approaches, for improving approximations based on 
perturbation theory, 
for making fair comparisons between different methods and between different 
approximations within the same method as well as for benchmarking the 
harmonic approximation
against other methods like, e.g., 
the surface hopping approach~\cite{tully_molecular_1990} 
or the exact-factorization scheme~\cite{abedi_exact_2010,requist_exact_2016}.
A plausible explanation for the scarce attention given to the  
{\em ab initio} $e$-$ph$ Hamiltonian is that the minimal sensible 
approximation is {\em nonperturbative}. In fact, the {\em ab initio} 
$e$-$ph$ Hamiltonian evaluated at zero $e$-$ph$ coupling does not even 
contain phonons.
A clean derivation of the $e$-$ph$ Hamiltonian can be found in the Baym's 
work~\cite{baym_field-theoretic_1961}.
However, Baym's original expression 
as well as equivalent expressions designed for having 
a good starting point for perturbative expansions~\cite{marini_many-body_2015} 
necessitate the knowledge of the exact 
equilibrium electronic density $n^{0}$. 
This means that the {\em ab initio} 
$e$-$ph$ Hamiltonian is {\em unknown} unless 
$n^{0}$ is calculated by other means, e.g., by
solving the original electron-nuclear problem. 
Even assuming that we could find $n^{0}$ somehow, we would still have 
to face a practical problem. All many-body techniques (including 
those based on GF)  can only be implemented in some approximation, 
for the exchange-correlation (xc) potential in 
Density Functional Theory (DFT), for the self-energy in GF theory, 
for the configuration state functions in the Multi-Configurational 
Hartree-Fock method,
for the  intermediate states in the Algebraic Diagrammatic Construction scheme, 
etc. An approximation in any of the available many-body methods
inevitably leads to an 
inconsistency if the {\em ab initio} $e$-$ph$ Hamiltonian is evaluated at 
the exact $n^{0}$. The inconsistency lies in the fact that the forces acting on the 
nuclei would not vanish in thermal equilibrium. As we shall see, the
{\em ab initio} $e$-$ph$ Hamiltonian must be determined {\em 
self-consistently} for it to be used in practice. Such self-consistent 
concept is completely general, i.e., it is not limited to GF 
approaches. Of course if an exact method is used than the 
self-consistent density $n^{0}$ coincides with the exact one.

The second issue is the extension of the Hedin-Baym equations 
at finite temperature and out-of-equilibrium. 
This is especially relevant in the light of the overwhelming 
and steadily increasing number 
of time-resolved spectroscopy experiments. 
We mention that at zero-temperature a nonequilibrium Green's 
function (NEGF) formulation has been put forward 
in terms of the nuclear-density correlation 
function~\cite{vanleeuwen_first-principles_2004,marini_functional_2018,harkonen_many-body_2020}. 
We here present the finite-temperature and nonequilibrium 
extension of the Hedin-Baym 
equations for the electronic GF and the displacement-displacement 
correlation function (or phononic GF)~\cite{feliciano_electron-phonon_2017}. 
We anticipate that the main differences are: 
(i) in all internal vertices the time arguments must be 
integrated over the $L$-shaped Konstantinov-Perel's 
contour~\cite{konstantinov1961diagram,Wagner_PhysRevB.44.6104,enzbook1992,sa-1.2004,svl-book}; (ii) the 
electronic GF satisfies a Dyson equation with 
an extra time-local self-energy, whose inclusion is fundamental to 
recover the Ehrenfest (mixed quantum-classical)
dynamics~\cite{horsfield_open-boundary_2004,horsfield_beyond_2004,li_ab-initio_2005,verdozzi_classical_2006,galperin_molecular_2007,galperin_the-non-linear_2008,dundas_current-driven_2009,hussein_semiclassical_2010,lu_blowing_2010,bode_scattering_2011,hubener_phonon_2018}. 
In the context of material science the Ehrenfest self-energy plays a 
key role in the description of 
polarons~\cite{galperin_the-non-linear_2008,lafuente_ab-initio_2022} 
and it is expected to be fundamental to detect
the phonon-induced coherent modulation of 
the excitonic resonances~\cite{trovatello_strongly_2020}. 

The Hedin-Baym equations have limited practical utility in 
nonequilibrium problems. Furthermore, the original derivation based 
on the source-field method allows for solving these equations  
only {\em iteratively}, starting from an approximation to the vertex.
The question of whether the iterative procedure converges towards the exact 
solution is still open. In most applications only one iteration step 
is performed since the equations resulting from the second interaction are 
already too complex. All these considerations bring us to the third issue, 
i.e., how to systematically improve  the 
approximations possibly preserving all 
fundamental conservation laws. We here present a {\em diagrammatic} 
derivation of the Hedin-Baym equations based on the skeletonic 
expansion of the electronic and phononic self-energies in terms of 
interacting electronic and phononic GF and the screened interaction.
We highlight three essential merits of the diagrammatic derivation:
(i) the possibility of including relevant scattering 
mechanisms through a proper selection of Feynman diagrams (to be 
converted into mathematical expressions through the Feynman rules 
which we provide); (ii) the possibility of using the $\F$-derivable 
criterion~\cite{baym_self-consistent_1962} to have a fully 
conserving dynamics; (iii) the possibility of closing the 
Kadanoff-Baym equations (KBE)~\cite{kadanoff1962quantum} 
through a skeletonic expansion of the 
self-energies in terms of only interacting GF.
The KBE are integro-differential equations for the electronic and 
phononic NEGF~\cite{danielewicz_quantum_1984,svl-book}, 
and they are definitely more practical for investigating the real-time 
evolution of systems driven out of equilibrium. Furthermore, through 
the so called Generalized Kadanoff-Baym Ansatz for fermions~\cite{lipavsky_generalized_1986} and 
bosons~\cite{karlsson_fast_2021} the KBE can be mapped onto a much 
simpler system 
of ordinary differential 
equations for a large number of self-energy 
approximations~\cite{schlunzen_achieving_2020,joost_g1-g2_2020,karlsson_fast_2021,pavlyukh_photoinduced_2021,pavlyukh_time-linear_2022,pavlyukh_interacting_2022,perfetto_real_2022}.

The paper is organized as follows. In Section~\ref{qselnusec} we 
derive the low-energy Hamiltonian for any system of electrons and 
nuclei, highlighting its dependence on the equilibrium electronic 
density and pointing out the necessity of a 
self-consistent procedure for its determination. In Section~\ref{inthamelphonsec}
we specialize the discussion to lattice 
periodic systems, introduce general time-dependent external perturbations 
and derive the $e$-$ph$ Hamiltonian on the $L$-shaped contour.
The equations of motion for the electronic and phononic field 
operators are derived in Section~\ref{eomsec}. In 
Section~\ref{dispG+MSHsec} we define the many-particle electronic and 
phononic GF on the contour and construct the Martin-Schwinger hierarchy that these 
GF satisfy. In Section~\ref{wicksec} we present the Wick's theorem 
as the solution of the noninteracting Martin-Schwinger hierarchy and 
in Section~\ref{exactexpsec} we provide the exact formula of the 
many-body expansion of the 
interacting GF in terms of the 
noninteracting ones. The many-body expansion is mapped onto a 
diagrammatic theory in Section~\ref{ebdiagexpsec} where we also 
introduce the notion of self-energies and 
skeleton diagrams. The skeletonic expansion of the self-energies in 
terms of the interacting GF and screened Coulomb 
interaction is shown to lead to the Hedin-Baym equations in 
Section~\ref{hedinsec}. The Hedin-Baym equations are applicable to 
systems in- and out-of equilibrium as well as at zero and finite 
temperature. To study the system evolution or the finite-temperature 
spectral properties the 
equations of motion for the GF are more convenient 
than the Hedin-Baym equations. These equations of motion are derived 
in Section~\ref{kbe-p+bsec}. In Section~\ref{conservingsec} we 
discuss the so called $\F$-derivable approximations to the 
self-energies. The GF satisfying the equations of 
motion with $\F$-derivable self-energies guarantee the fulfillment of all 
fundamental conservation laws. In Section~\ref{kbesec} we convert  
the equations of motion into a coupled system of 
integro-differential equations for the Keldysh components of the 
GF; these are the KBE. We first 
discuss the self-consistent solution of the equilibrium problem and 
then derive the real-time equations of motion 
to study the system evolution. We also present the formal solution of 
the KBE in the long-time limit and for 
steady-state conditions. The expansion of the phononic GF 
around the quasi-phonon energies reveals the possibility of 
a correlation-induced splitting of the phonon dispersion in materials 
with no time-reversal invariance.
The presented formalism can be extended to 
deal with much more general Hamiltonians than the $e$-$ph$ 
Hamiltonian. In Section~\ref{outlooksec} we provide a summary of the 
main results, illustrate possible
extensions and discuss their physical relevance.

\section{Quantum systems of electrons and nuclei}
\label{qselnusec}

In this section we lay down a quantum theory of electrons and 
nuclei which is suited whenever the
average nuclear positions remain  
close to the equilibrium values. Under this ``near-equilibrium 
hypothesis'' the nuclei stay away from each other and
they can be treated as quantum {\em distinguishable} particles. 
In fact, we can distinguish them 
by means of techniques like scanning tunneling microscopy or 
electron diffraction~\cite{feliciano_electron-phonon_2017}. 

Let $\hat{\q}(\blx=\blr\s)$ be the field operators that annihilate 
an electron in position $\blr$ with spin $\s$, hence they satisfy the 
anticommutation relations
$\{\hat{\q}(\blx),\hat{\q}^{\dag}(\blx')\}=
\d_{\s\s'}\d(\blr-\blr')\equiv \d(\blx-\blx')$, and 
$\hat{\blR}_{i}$, $\hat{\blP}_{i}$ be the position and momentum operators of 
the $i$-th nucleus, $i=1,\ldots N_{n}$,  
satisfying the  standard
commutation relations 
$[\hat{R}_{i,\a},\hat{P}_{j,\b}]_{-}=i \d_{ij}\d_{\a\b}$, with $\a$ 
and $\b$ running over the three components of the vectors.
The Hamiltonian describing an unperturbed system of electrons 
interacting with $N_{n}$ nuclei of charge 
$Z_{i}$ and mass $M_{i}$  reads (we use atomic units throughout the 
paper) 
\begin{align}
\hat{H}=\hat{H}_{e}+\hat{H}_{n}+\hat{H}_{e-n},
\label{elnucham}
\end{align}
where 
\begin{align}
\hat{H}_{e}&=\!\int \!d\blx \,
\hat{\q}^{\dag}(\blx)\Big[-\frac{\nabla^{2}}{2}\Big]\hat{\q}(\blx),
\nn\\
&+\frac{1}{2}\!\int \!d\blx 
d\blx'\hat{\q}^{\dag}(\blx)\hat{\q}^{\dag}(\blx')
v(\blr,\blr')\hat{\q}(\blx')\hat{\q}(\blx),
\label{he}
\end{align}
is the electronic Hamiltonian,
\begin{align}
\hat{H}_{n}=
\sum_{i=1}^{N_{n}}\frac{\hat{P}_{i}^{2}}{2M_{i}}+
\frac{1}{2}\sum_{i\neq j}^{N_{n}}Z_{i}Z_{j}
v(\hat{\blR}_{i},\hat{\blR}_{j}),
\label{hn}
\end{align}
is the nuclear Hamiltonian (with  
$\hat{P}_{i}^{2}=\hat{\blP}_{i}\cdot\hat{\blP}_{i}$) and 
\begin{align}
\hat{H}_{e-n}=-\int \!d\blx \,\hat{n}(\blx)\sum_{j=1}^{N_{n}}Z_{j}
v(\blr,\hat{\blR}_{j}),
\label{hen}
\end{align}
is the electron-nucleus interaction.	In Eqs.~(\ref{he}-\ref{hen}) the 
integral $\int \!d\blx\equiv \int \!d\blr\sum_{\s}$ signifies a 
spatial integral and a sum over spin, $v(\blr,\blr')=1/|\blr-\blr'|$ 
is the Coulomb interaction 
and $\hat{n}(\blx)\equiv \hat{\q}^{\dag}(\blx)\hat{\q}(\blx)$ is the 
density operator in $\blr$ for particles of spin $\s$.
For later purposes we find it convenient to collect all nuclear 
position and momentum operators into the vectors $\hat{\blR}=
(\hat{\blR}_{1},\ldots,\hat{\blR}_{N_{n}})$ and 
$\hat{\blP}=(\hat{\blP}_{1},\ldots,\hat{\blP}_{N_{n}})$. We also find 
it 
useful to define the nuclear potential energy
\begin{align}
E_{n-n}(\hat{\blR})\equiv \frac{1}{2}\sum_{i\neq j}^{N_{n}}Z_{i}Z_{j}
v(\hat{\blR}_{i},\hat{\blR}_{j}),
\end{align}
appearing in Eq.~(\ref{hn}),
and the electron-nuclear potential
\begin{align}
V(\blr,\hat{\blR})\equiv \sum_{j=1}^{N_{n}}Z_{j}
v(\blr,\hat{\blR}_{j}),
\end{align}
appearing in Eq.~(\ref{hen}).
The operators of $\hat{H}$ act on the direct-product space 
$\mathbb{F}\otimes \mathbb{D}_{N_{n}}$ where $\mathbb{F}$ is the 
electronic Fock space and $\mathbb{D}_{N_{n}}$ is the Hilbert space of 
the $N_{n}$ distinguishable nuclei. 

{\em Expansion around thermal equilibrium.--}
Consider the interacting system of electrons and nuclei in thermal 
equilibrium at a certain temperature. 
Under the ``near-equilibrium hypothesis'' we can 
approximate the full Hamiltonian by its 
second-order Taylor 
expansion around the equilibrium values of the nuclear positions, 
which we name $\blR^{0}=(\blR^{0}_{1},\ldots,\blR^{0}_{N_{n}})$, 
and around the equilibrium value of the
electronic density, which we name $n^{0}(\blx)$. 
In fact, also the 
electronic density must stay close to $n^{0}(\blx)$ for otherwise the 
forces acting on the nuclei would be strong enough to drive the nuclei
away from $\blR^{0}$.  
Notice that $\blR^{0}$ and $n^{0}(\blx)$ do in general 
depend on the temperature. We also observe that 
the existence of an inertial reference frame for the coordinates 
$\blR^{0}$ is supported by the macroscopic size of the system, i.e., 
$N_{n}\to\iif$. For finite systems, e.g., molecules or molecular 
aggregates, the choice of a suitable reference frame is more subtle, 
see 
Refs~\cite{eckart_some-studies_1935,howard_the-molecular_1970,Wilson_book,Bunker_book,sutcliffe_the-decoupling_2000,meyer_molecular_2002}.

We  introduce the 
displacement (or position fluctuation) 
operators $\hat{\blU}_{i}$ and the density fluctuation 
operator $\D\hat{n}$ according to
\begin{align}
\hat{\blU}_{i}=\hat{\blR}_{i}-\blR^{0}_{i},\quad\quad\quad\quad
\D\hat{n}(\blx)=\hat{n}(\blx)-n^{0}(\blx).
\label{displ+densfluct}
\end{align}
In the following we refer to these operators as the {\em fluctuation 
operators}.
Formally, the ``near-equilibrium hypothesis'' is equivalent to 
restrict full the space $\mathbb{F}\otimes 
\mathbb{D}_{N_{n}}$ to the subspace of states giving a small average of 
$\hat{\blU}_{i}$ and 
$\D\hat{n}$. 

The expansion of the nuclear potential energy around the equilibrium 
nuclear positions  yields to second order
\begin{align}
E_{n-n}(\hat{\blR})=
E_{n-n}(\blR^{0})&+\sum_{i\a}
\left.\frac{\de E_{n-n}(\blR)}{\de R_{i,\a}}
\right|_{\blR=\blR^{0}}\!
\hat{U}_{i,\a}
\nn\\
&+\frac{1}{2}\sum_{i\a,j\b}\left.\frac{\de^{2} E_{n-n}(\blR)}
{\de R_{i,\a}\de R_{j,\b}}
\right|_{\blR=\blR^{0}}\!
\hat{U}_{i,\a}\hat{U}_{j,\b}.
\label{nnintham3}
\end{align}
In the first term we recognize the electrostatic energy of a 
nuclear geometry $\blR^{0}$. As 
this term is only responsible for an overall energy shift we do not 
include it in the following discussion. 
Similarly, 
the expansion of the electron-nuclear potential
yields to second order
\begin{align}
V(\blr,\hat{\blR})=V(\blr)
+\sum_{i\a}g_{i,\a}(\blr)\,\hat{U}_{i,\a}
+\frac{1}{2}\sum_{i\a,j\b}
g^{\rm DW}_{i,\a;j,\b}(\blr)\,
\hat{U}_{i,\a}\hat{U}_{j,\b},
\label{elnuclint3}
\end{align}
where we define 
\begin{subequations}
\begin{align}
V(\blr)&\equiv V(\blr,\blR^{0}),
\label{V(r)}
\\
g_{i,\a}(\blr)&\equiv \left.\frac{\de V(\blr,\blR)}{\de 
R_{i,\a}}\right|_{\blR=\blR^{0}}=-Z_{i}\frac{\de}{\de 
r_{\a}}v(\blr,\blR^{0}_{i}),
\label{gia(r)}
\\
g^{\rm DW}_{i,\a;j,\b}(\blr)&\equiv \left.\frac{\de^{2} 
V(\blr,\blR)}{\de R_{i,\a}\de R_{j,\b}}
\right|_{\blR=\blR^{0}}=\d_{ij}Z_{i}\frac{\de^{2}}{\de 
r_{\a}\de r_{\b}}v(\blr,\blR^{0}_{i}).
\end{align}
\label{nucldenscoeff}
\end{subequations}
Inserting Eq.~(\ref{elnuclint3}) into Eq.~(\ref{hen}) we see that the first 
term gives rise to a purely electronic operator; it 
is the potential energy operator for electrons in the classical field 
generated by a nuclear geometry $\blR^{0}$. The second and 
third terms emerge when relaxing the 
infinite-mass approximation for the nuclei. The third 
term of Eq.~(\ref{elnuclint3})
is already quadratic in the displacements and it can therefore be 
multiplied by the equilibrium density, i.e., 
$\hat{n}\to n^{0}$~\cite{baym_field-theoretic_1961}. 
Going beyond the quadratic (or harmonic) approximation the replacement $\hat{n}\to n^{0}$
is no longer justified; in this case the third term gives rise to the 
so called Debye-Waller (DW) interaction~\cite{allen_theory_1976}.

{\em The low-energy Hamiltonian.--}
Inserting the expansion of $E_{n-n}$ and $V$ into Eqs.~(\ref{hn}) and 
(\ref{hen}) the total Hamiltonian becomes
\begin{align}
\hat{H}=\hat{H}_{0,e}+\hat{H}_{0,ph}+\hat{H}_{e-e}+\hat{H}'.
\label{el-phonham}
\end{align}
The first two terms describe an uncoupled system of noninteracting 
electrons and $N_{n}$ 
interacting nuclei in the electric field 
generated by a {\em frozen} electronic density $n^{0}(\blr)$
\begin{align}
\hat{H}_{0,e}=\!\int \!d\blx \,
\hat{\q}^{\dag}(\blx)
\Big[-\frac{\nabla^{2}}{2}+V(\blr)
\Big]\hat{\q}(\blx),
\label{h0e}
\end{align}
\begin{align}
\hat{H}_{0,ph}=\sum_{i=1}^{N_{n}}\frac{\hat{P}_{i}^{2}}{2M_{i}}
+\frac{1}{2}\sum_{i\a,j\b}\hat{U}_{i,\a} K_{i,\a;j,\b}\,\hat{U}_{j,\b},
\label{h0ph}
\end{align}
where we define the {\em elastic tensor}
\begin{align}
K_{i,\a;j,\b}\equiv\left.\frac{\de^{2} E_{n-n}(\blR)}
{\de R_{i,\a}\de R_{j,\b}}
\right|_{\blR=\blR^{0}}-
\int d\blx \,n^{0}(\blx)
g^{\rm DW}_{i,\a;j,\b}(\blr),
\label{elasticmat}
\end{align}
which is real and symmetric under the exchange $(i,\a)\leftrightarrow 
(j,\b)$. Already at this stage of the presentation a remark is due.
The eigenvalues $\w^{2}_{\l}$ of the tensor
$\overline{K}_{i,\a;j,\b}\equiv K_{i,\a;j,\b}/\sqrt{M_{i}M_{j}}$
are not physical and can even 
be {\em negative} such that $\hat{H}_{0,ph}$ does not have a proper 
ground state. It is therefore generally {\em not} possible to define 
annihilation and creation operators $\hat{b}_{\l}$ and 
$\hat{b}^{\dag}_{\l}$ to rewrite Eq.~(\ref{h0ph}) in the form 
$\hat{H}_{0,ph}=\sum_{\l}\w_{\l}(\hat{b}^{\dag}_{\l}\hat{b}_{\l}+\frac{1}{2})$. 
The {\em ab initio} low-energy Hamiltonian evaluated at vanishing 
coupling $g_{i,\a}$ [hence $\hat{H}'=0$, see Eq.~(\ref{el-phonintham2})]
does not contain physical quanta of vibrations (phonons in solids). These excitations can only 
emerge from a proper {\em nonperturbative} treatment, see 
Section~\ref{ephmatzprobsec}.

The third term in Eq.~(\ref{el-phonham})
is the electron-electron ($e$-$e$) interaction 
Hamiltonian
\begin{align}
\hat{H}_{e-e}=
\frac{1}{2}\!\int \!d\blx 
d\blx'\hat{\q}^{\dag}(\blx)\hat{\q}^{\dag}(\blx')
v(\blr,\blr')\hat{\q}(\blx')\hat{\q}(\blx),
\label{hee}
\end{align}
while the last term is the contribution linear in the nuclear 
displacements
\begin{align}
\hat{H}'=\sum_{i\a}\left[
\left.\frac{\de E_{n-n}(\blR)}{\de R_{i,\a}}
\right|_{\blR=\blR^{0}}
-\int \!d\blx \;g_{i,\a}(\blr)\hat{n}(\blx)\right]
\hat{U}_{i,\a}.
\label{el-phonintham}
\end{align}

The Hamiltonian in Eq.~(\ref{el-phonham}) with the four contributions as in 
Eqs.~(\ref{h0e}), (\ref{h0ph}), (\ref{hee}) and (\ref{el-phonintham}) is identical 
to that of Ref.~\cite{baym_field-theoretic_1961}. We here make a step further.  
Although it is not evident the operator $\hat{H}'$ is quadratic 
in the fluctuation operators.  To show it we consider 
the Heisenberg equation of 
motion  for 
the time-dependent average of the nuclear momentum operators.
Let $\hat{\callU}(t,t_{0})$ be the evolution operator from some 
initial time $t_{0}$ to time $t>t_{0}$ and $\hat{\callU}(t_{0},t)=
[\hat{\callU}(t,t_{0})]^{\dag}$. Henceforth any operator $\hat{O}(t)$
in the Heisenberg picture carries a subscript ``$H$'', i.e., 
$\hat{O}_{H}(t)=\hat{\callU}(t_{0},t)\hat{O}(t)\hat{\callU}(t,t_{0})$.
The time-dependent average $O(t)$ of the operator $\hat{O}(t)$ is 
defined according to 
\begin{align}
O(t)=\Tr[\hat{\r}\,\hat{O}_{H}(t)],
\label{tdaverage}
\end{align}
where 
\begin{align}
\hat{\r}=\frac{e^{-\b(\hat{H}-\m\hat{N}_{e})}}{\Tr[e^{-\b(\hat{H}-\m\hat{N}_{e})}]}
\end{align}
is the thermal 
density matrix, with $\b$ the inverse temperature, $\m$ the chemical 
potential and
$\hat{N}_{e}=\int d\blx \,\hat{n}(\blx)$ the 
operator for the total number of electrons. Using 
$i\frac{d}{dt}\hat{O}_{H}(t)=[\hat{O}_{H}(t),\hat{H}_{H}(t)]$ we find
\begin{align}
\frac{d P_{i,\a}(t)}{dt}&=- 
\left.\frac{\de E_{n-n}(\blR)}{\de R_{i,\a}}
\right|_{\blR=\blR^{0}}
+\int \!d\blx \;g_{i,\a}(\blr)n(\blx,t)
\nn\\
&-\sum_{j\b}K_{i,\a;j,\b}U_{j,\b}(t),
\label{eomfornuclp}
\end{align}
where $n(\blx,t)$ and $U_{j,\b}(t)$ are the time-dependent averages of 
the electronic density $\hat{n}(\blr)$ and nuclear displacement 
$\hat{U}_{j,\b}$. In thermal equilibrium the l.h.s. vanishes, 
$n(\blx,t)=n^{0}(\blx)$ and $U_{j,\b}=0$. Therefore
\begin{align}
\left.\frac{\de E_{n-n}(\blR)}{\de R_{i,\a}}
\right|_{\blR=\blR^{0}}=\int \!d\blx 
\;g_{i,\a}(\blr)n^{0}(\blx),
\label{idforlinearterm}
\end{align}
according to which we can rewrite Eq.~(\ref{el-phonintham}) as 
\begin{align}
\hat{H}'=-\sum_{i\a}
\int d\blx \;g_{i,\a}(\blr)
\D\hat{n}(\blx)
\,\hat{U}_{i,\a},
\label{el-phonintham2}
\end{align}
where $\D\hat{n}(\blr)$ is the density fluctuation operator defined 
in Eq.~(\ref{displ+densfluct}). In this form $\hat{H}'$ is 
manifestly quadratic in the fluctuation operators.

Inserting Eq.~(\ref{idforlinearterm}) in Eq.~(\ref{eomfornuclp}) 
we also see that the equation of motion for the momentum operators 
simplifies to
\begin{align}
\frac{d P_{i,\a}(t)}{dt}=
\int \!d\blx \;g_{i,\a}(\blr)\D n(\blx,t)
-\sum_{j\b}K_{i,\a;j,\b}U_{j,\b}(t).
\label{eomfornuclp2}
\end{align}
We can interpret the elastic tensor $K$ as the {\em nuclear-force 
tensor} of a system with frozen electronic density, i.e., with 
$\D n(\blx,t)=0$, or alternatively with vanishing coupling $g_{i,\a}$. 
In general $g_{i,\a}\neq 0$ and out of equilibrium $\D n(\blx,t)\neq 
0$, and the first 
term in Eq.~(\ref{eomfornuclp2}) significantly contributes to the nuclear 
forces.

The Hamiltonian in Eq.~(\ref{el-phonham}) is the low-energy 
approximation of the full Hamiltonian in Eq.~(\ref{elnucham}). The 
expansion around the equilibrium nuclear geometry and around the equilibrium 
density has inevitably made Eq.~(\ref{el-phonham}) to depend on 
these quantities. The scalar potential $V$ and the 
electron-nuclear coupling $g$ are determined from the sole knowledge 
of the equilibrium positions $\blR^{0}$ whereas the 
elastic tensor $K$ depends on both $\blR^{0}$ and $n^{0}$, see 
Eq.~(\ref{elasticmat}). 
Notice that the dependence of $\hat{H}$ on $n^{0}$ is through $K$ as well as 
$\D\hat{n}$, see Eq.~(\ref{el-phonintham2}). 
In the following we assume that $\blR^{0}$ is known 
and therefore that $V$ and $g$ are given. 
Strategies to obtain good approximations to the equilibrium 
nuclear geometry are indeed available, e.g., 
the Born-Oppenheimer approximation, see also discussion in 
Ref.~\cite{harkonen_many-body_2020}. 
Alternatively, $\blR^{0}$ can be taken from 
X-ray crystallographic measurements. 
The equilibrium density $n^{0}$ must 
instead be determined, and the proper way of doing it is 
{\em self-consistently}. Let us expand on this 
point. 

We write the dependence of $\hat{H}$ on $n^{0}$ explicitly: 
$\hat{H}=\hat{H}[n^{0}]$. For any given many-body treatment 
(whether exact or approximate)
a possible self-consistent strategy to obtain $n^{0}$ is: (i) Make an initial guess 
$n^{0}_{1}$ and (ii) 
Use the chosen many-body treatment to calculate the equilibrium density 
$n^{0}_{2}$ of $\hat{H}[n^{0}_{1}]$,
then the equilibrium density 
$n^{0}_{3}$ of $\hat{H}[n^{0}_{2}]$
and so on and so forth until convergence.
If the initial guess $n^{0}_{1}$ does already produce a good approximation 
to $K$ then a partial self-consistent scheme in which $K$ is not updated 
is also conceivable. Self-consistency is however unavoidable to determine $n^{0}$ in 
$\D\hat{n}$. In fact, it is only at self-consistency that the equilibrium value 
$\D n=0$, an essential requirement for the r.h.s. of Eq.~(\ref{eomfornuclp2}) 
to vanish and hence for the nuclear geometry to remain stationary.
In Section~\ref{ephmatzprobsec} we discuss how to implement the
self-consistent strategy using  
NEGF. In particular we show that $n^{0}$ 
can be obtained from the self-consistent 
solution of the Dyson equation for the 
Matsubara GF. 

For any given equilibrium geometry $\blR^{0}$  different 
scenarios are possible. If $\blR^{0}$ is too off target 
the self-consistent scheme may not converge, indicating that the 
nuclear geometry must be improved. If convergence is 
achieved then the self-consistent equilibrium state can 
be either stable or unstable. In the unstable scenario an infinitesimally 
small perturbation brings the nuclei away from $\blR^{0}$, indicating 
again that the nuclear geometry must be improved. 
Let us finally consider the stable scenario. 
The nuclear geometry can in this case be further optimized by minimizing 
the total energy (at zero temperature) or the grand-potential (at 
finite temperature) with respect to $\blR^{0}$. 
At the minimum the equilibrium geometry is the 
exact one only if an exact many-body treatment is used. Needless to 
say that the minimum is defined up to arbitrary overall shifts and 
rotations of the nuclear coordinates.

\section{Interacting Hamiltonian for electrons and phonons in- and 
out-of-equilibrium}
\label{inthamelphonsec}

Independently of the method chosen to find $\blR^{0}$ and of the 
self-consistent 
many-body treatment chosen to determine $n^{0}$ 
the low-energy Hamiltonian of a system of electrons and 
nuclei is given by Eq.~(\ref{el-phonham}). Let us discuss in detail the 
case of a crystal and introduce some notations.

In a crystal we can label the position of every nucleus 
with the vector (of integers) $\bln$ of 
the unit cell it belongs to and with the position $s$ relative to 
some point of the unit cell, i.e., 
$\blR^{0}_{i=\bln,s}=\blR^{0}_{\bln}+\blR^{0}_{s}$. 
If the unit cell contains $N_{u}$ nuclei then $s=1,\ldots,N_{u}$.
By definition the 
vector $\blR^{0}_{s}$ for the $s$-th nucleus is the same in all unit 
cells, and the mass $M_{i=\bln,s}=M_{s}$ and charge 
$Z_{i=\bln,s}=Z_{s}$ of the $s$-th nucleus
in cell $\bln$ is independent of $\bln$. 
The invariance of the crystal under discrete 
translations implies that the elastic tensor 
depends only on the difference between unit cell vectors, i.e.,
\begin{align}
K_{\bln s,\a;\bln's',\a'}=K_{s,\a;s',\a'}(\bln-\bln').
\end{align}
The periodicity of the crystal also implies an important property for 
the electron-nuclear coupling $g$. 
According to the definition in Eq.~(\ref{gia(r)})
we have 
$g_{\bln s,\a}(\blr)=-Z_{s}(\de/\de r_{\a})
v(\blr,\blR^{0}_{\bln s})$. The Coulomb interaction 
depends only on the relative 
coordinate and therefore $v(\blr,\blR^{0}_{\bln s})=
v(\blr+\blR^{0}_{\bln'},\blR^{0}_{\bln s}+\blR^{0}_{\bln'})=
v(\blr+\blR^{0}_{\bln'},\blR^{0}_{\bln+\bln' s})$ for all 
vectors $\blR^{0}_{\bln'}$. 
This implies that 
\begin{align}
g_{\bln s,\a}(\blr)=g_{\bln+\bln' s,\a}(\blr+\blR^{0}_{\bln'}).
\label{propginrs}
\end{align}

{\em Equilibrium Hamiltonian.--}
We consider a finite piece of the crystal with $N_{\a}$ cells 
along direction $\a=x,y,z$ and impose the Born-von Karman boundary 
conditions. The total number of cells is therefore $N=N_{x}N_{y}N_{z}$.
Accordingly the displacement and momentum operators 
can be expanded as
\begin{subequations}
\begin{align}
\hat{U}_{\bln s,\a}=\frac{1}{\sqrt{M_{s}N}}
\sum_{\blq}e^{i\blq\cdot\bln}\,
\sum_{\n}
e^{\n}_{s,\a}(\blq)
\;\hat{U}_{\blq \n},
\label{displdecnm}
\end{align}
\begin{align}
\hat{P}_{\bln s,\a}=\sqrt{\frac{M_{s}}{N}}
\sum_{\blq}e^{i\blq\cdot\bln}\,
\sum_{\n}
e^{\n}_{s,\a}(\blq)
\;\hat{P}_{\blq \n},
\label{momdecnm}
\end{align}
\label{momdisplexp}
\end{subequations}
where the sum over $\blq=(q_{x},q_{y},q_{z})$ 
runs over all vectors satisfying the 
property $q_{\a}N_{\a}=2\p m_{\a}$ with $m_{\a}$ integers, and 
$q_{\a}\in (-\p,\p]$ for $\a=x,y,z$. In Eqs.~(\ref{momdisplexp}) 
the vectors $\ble^{\n}(\blq)$ with components $e^{\n}_{s,\a}(\blq)$ 
form an orthonormal basis for each $\blq$, i.e., 
$\ble^{\n}(\blq)^{\dag}\cdot \ble^{\n'}(\blq)=\d_{\n\n'}$. 
In three dimensions the set of all vectors $\ble^{\n}(\blq)$ 
spans a $3N_{u}$ dimensional space for each $\blq$.
We refer to these vectors as the {\em normal modes}. 
As we shall see the most convenient choice of normal modes depends on 
the approximation made to treat the problem. 
A typical choice is the eigenbasis of the 
Hessian of the Born-Oppenheimer energy. At this stage of the 
presentation the set of normal 
modes is just a basis to expand the displacement and momentum 
operators.

The hermiticity of the operators $\hat{U}_{\bln s,\a}$ and $\hat{P}_{\bln s,\a}$
imposes the following constraints on the operators 
$\hat{U}_{\blq \n}$ and $\hat{P}_{\blq \n}$ and on the normal modes:
\begin{align}
	\hat{U}_{\blq \n}=\hat{U}_{-\blq \n}^{\dag}
	\quad,\quad
\hat{P}_{\blq \n}=\hat{P}_{-\blq \n}^{\dag}
\quad,\quad
\ble^{\n\ast}(-\blq)=\ble^{\n}(\blq).
\label{quasfinconst}
\end{align}
Inserting the expansions Eqs.~(\ref{momdisplexp}) into 
Eqs.~(\ref{h0ph}) and (\ref{el-phonintham2}) we obtain
\begin{align}
\hat{H}_{0,ph}=
\frac{1}{2}\sum_{\blq \n}\hat{P}^{\dag}_{\blq \n}
\hat{P}_{\blq \n}
+\frac{1}{2}\sum_{\blq\n\n'}
\hat{U}^{\dag}_{\blq \n}K_{\n\n'}(\blq)\,
\hat{U}_{\blq \n'},
\label{H0phondexp}
\end{align}
\begin{align}
\hat{H}'=-\sum_{\blq\n}\int d\blx \;g_{-\blq\n}(\blr)
\D\hat{n}(\blx)
\;\hat{U}_{\blq \n},
\label{el-phonintham3}
\end{align}
where
\begin{align}
K_{\n\n'}(\blq)&\equiv 
\sum_{\bln}e^{-i\blq\cdot\bln}
\sum_{s\a,s'\a'}
e^{\n\ast}_{s,\a}(\blq)\;
\frac{K_{s,\a;s'\b}(\bln)}{\sqrt{M_{s}M_{s'}}}
\;e^{\n'}_{s'\a'}(\blq)\nn\\&=K_{\n\n'}^{\ast}(-\blq)
=K_{\n'\n}^{\ast}(\blq),
\label{Kprop}
\end{align}
and
\begin{align}
g_{-\blq\n}(\blr)\equiv 
\sum_{\bln s \a}\frac{1}{\sqrt{M_{s}N}}
e^{i\blq\cdot\bln}\,e^{\n}_{s,\a}(\blq)\,
g_{\bln s,\a}(\blr)=g^{\ast}_{\blq\n}(\blr).
\label{proelphoncoup}
\end{align}
For crystals the Hamiltonian of Eq.~(\ref{el-phonham}) is known as 
the electron-phonon ($e$-$ph$) Hamiltonian.

{\em Nonequilibrium Hamiltonian.--}
We are interested in formulating a NEGF approach to deal with 
 systems described by the $e$-$ph$ Hamiltonian of 
Eq.~(\ref{el-phonham}) 
possibly driven out of equilibrium by external driving fields. 
Of course the external fields must be such that the nonequilibrium 
density and displacements are small enough to justify the harmonic 
approximation. As we shall see the developed formalism can accommodate 
many different kinds of drivings.  

Without any loss of generality we take 
the system in thermal equilibrium for times $t<t_{0}$ and 
then perturb it  by letting 
\begin{subequations}
\begin{align}	
\Big[-\frac{\nabla^{2}}{2}+V(\blr)
\Big]&\to h(\grad,\blr,t),
\label{h(z)}
\\
K_{\n\n'}(\blq)&\to K_{\n\n'}(\blq,t),
\label{K(z)}
\\
v(\blr,\blr')&\to v(\blr,\blr',t),
\label{v(z)}
\\
g_{\blq\n}(\blr)&\to g_{\blq\n}(\blr,t).
\label{g(z)}
\end{align}
\label{tdhampar}
\end{subequations}
Correspondingly $\hat{H}_{0,e}\to \hat{H}_{0,e}(t)$, 
$\hat{H}_{0,ph}\to \hat{H}_{0,ph}(t)$,
$\hat{H}_{e-e}\to \hat{H}_{e-e}(t)$, 
$\hat{H}'\to \hat{H}'(t)$ and hence $\hat{H}\to \hat{H}(t)$. 
The time-dependence of the Coulomb interaction $v$ and
$e$-$ph$ coupling $g$ may be due to, e.g., 
an adiabatic switching protocol
or a sudden quench of the interaction, 
whereas the time dependence of the one-particle
Hamiltonian $h$ and elastic tensor $K$ may 
be due to laser fields,
phonon drivings, etc. 

\begin{figure}[tbp]
    \centering
\includegraphics[width=0.48\textwidth]{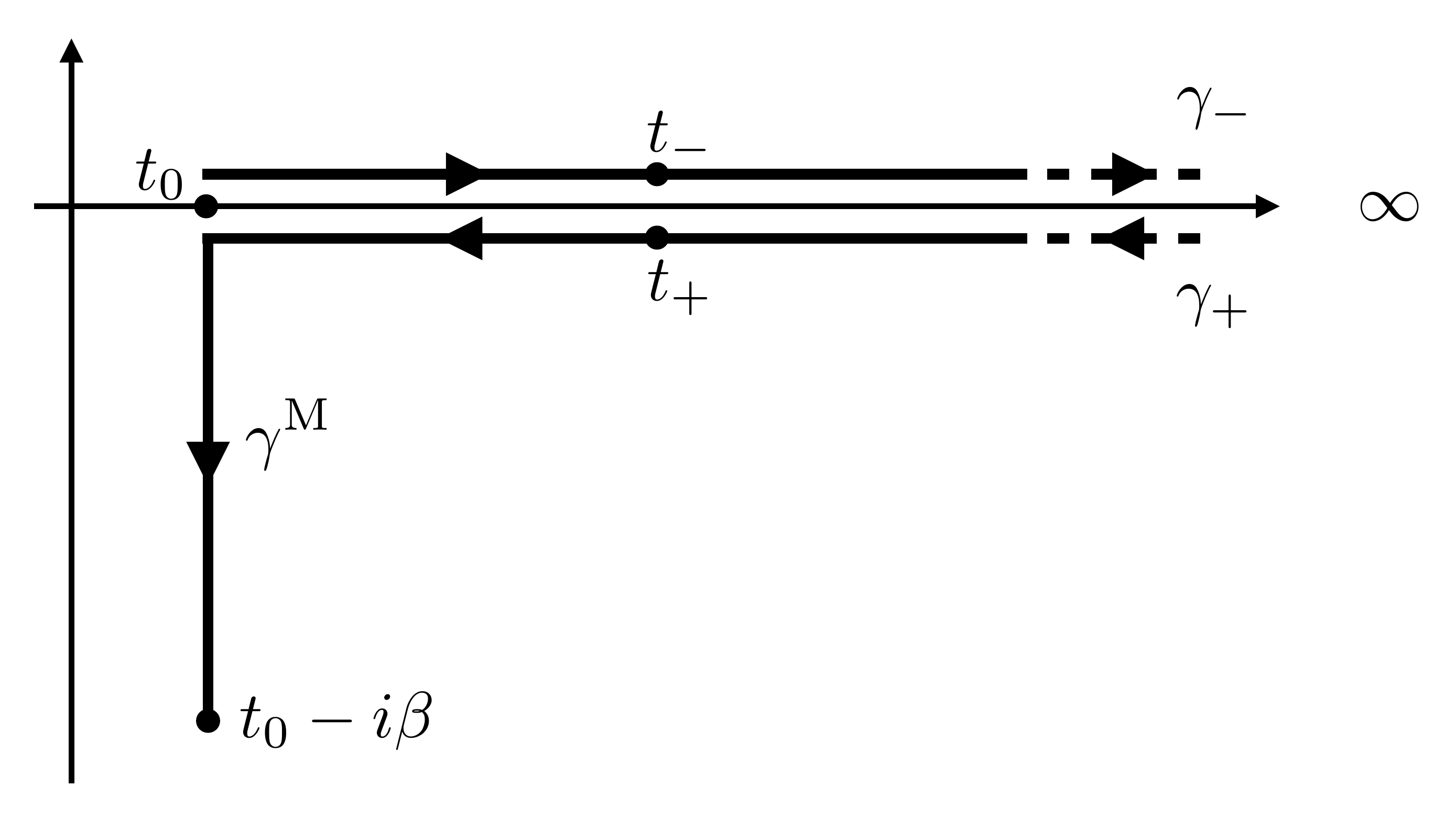}
\caption{$L$-shaped Konstantinov-Perel's contour.}
\label{contour}
\end{figure}

{\em Hamiltonian on the contour.--}
The explicit form of the evolution operators is 
$\hat{\callU}(t,t_{0})=T\big\{\exp
\big[-i\int_{t_{0}}^{t}d\bar{t}\,\hat{H}(\bar{t})\big]\big\}$, 
with $T$ the time-ordering operator, and 
$\hat{\callU}(t_{0},t)=\bar{T}
\big\{\exp\big[i\int_{t_{0}}^{t}d\bar{t}\,\hat{H}(\bar{t})\big]\big\}$, 
with $\bar{T}$ the anti-time-ordering operator. Therefore the 
time-dependent average in Eq.~(\ref{tdaverage}) can be 
written as~\cite{Wagner_PhysRevB.44.6104,svl-book}
\begin{align}
O(t)&=\Tr\big[\hat{\r}\,
\bar{T}
\big\{e^{i\int_{t_{0}}^{t}d\bar{t}\,\hat{H}(\bar{t})}\big\}\hat{O}(t)
\,T\big\{e^{-i\int_{t_{0}}^{t}d\bar{t}\,\hat{H}(\bar{t})}\big\}
\big]
\nn\\
&=\frac{\Tr\big[
\callT\big\{e^{-i\int_{\g}d\bar{z}\,\hat{H}(\bar{z})}\hat{O}(z)\big\}
\big]}{\Tr\big[
\callT\big\{e^{-i\int_{\g}d\bar{z}\,\hat{H}(\bar{z})}\big\}
\big]}.
\label{contformula}
\end{align}
In the second equality $z$ and $\bar{z}$ are contour-times running
on the $L$-shaped oriented contour $\g$~\cite{konstantinov1961diagram} consisting of a 
forward branch $\g_{-}$
going from the initial time $t_{0}$ to $\iif$, a backward 
branch $\g_{+}$
going from $\iif$ to $t_{0}$ and a vertical track on the complex 
plane $\g^{\rm M}$
going from $t_{0}$ to $t_{0}-i\b$, see Fig.~\ref{contour},
and $\callT$ is the contour 
ordering operator.
Henceforth we denote by $z=t_{\pm}$, $\bar{z}=\bar{t}_{\pm}$, 
$z'=t'_{\pm}$, etc.
contour times on $\g_{\pm}$.
For any quantity 
$q(z)$, being it a function or an operator, we define 
$q(t_{\pm})=q(t)$ and $q(t_{0}-i\t)=q$ independent of $\t\in 
(0,\b)$.
The only exception is $h(\grad,\blr,t_{0}-i\t)=h(\grad,\blr)-\m$. 
Thanks to this different definition for $h$ we have 
$\hat{H}(t_{\pm})=\hat{H}(t)$ and 
$\hat{H}(t_{0}-i\t)=\hat{H}-\m\hat{N}_{e}$.
Notice that the Hamiltonian $\hat{H}(z)$ depends on $z$ through the dependence of 
$h$, $K$, $v$ and $g$, see Eqs.~(\ref{tdhampar}). 

One remark about the 
dependence of  
$\hat{O}(z)$ on $z$. If the operator 
does not depend on $z$ we can safely 
write $\hat{O}(t)=\hat{O}$ in the first line of Eq.~(\ref{contformula}). 
However, if we do so 
in the second line then it is not clear where to place the operator 
$\hat{O}$ after the contour ordering. 
The reason to keep the contour argument even 
for operators that do not have an explicit time dependence (like the  
operators $\hat{\q}$, $\hat{U}$ and $\hat{P}$)
stems from the need of specifying 
their position along the contour, thus rendering unambiguous the 
action of $\callT$. Once the operators are ordered we can omit 
the time arguments if there is no time dependence.

To shorten the equations we gather the displacement and momentum 
operators into a two-dimensional vector of operators having components 
\begin{align}
\hat{\bgf}_{\blq\n}=\left(\begin{array}{c} 
\hat{\f}_{\blq\n}^{1}\\\hat{\f}_{\blq\n}^{2}
\end{array}\right)=
\left(\begin{array}{c}
\hat{U}_{\blq\n}\\\hat{P}_{\blq\n}
\end{array}\right).
\label{phicomp}
\end{align}
The commutation relations for the $\hat{\f}$-operators 
follow from the commutation relations $[\hat{U}_{\bln 
s,\a},\hat{P}_{\bln's',\a'}]=i\d_{\bln\bln'}\d_{ss'}\d_{\a\a'}$ after 
inserting the expansions in Eqs.~(\ref{momdisplexp}) and using the 
properties in Eq.~(\ref{quasfinconst}). We find
\begin{align}
[\hat{\f}_{\blq\n}^{i},\hat{\f}_{-\blq'\n'}^{i'}]=
[\hat{\f}_{\blq\n}^{i},\hat{\f}^{i'\,\dag}_{\blq'\n'}]
=\d_{\blq,\blq'}\a^{ii'}_{\n\n'}\;,
\label{commfifi}
\end{align}
where
\begin{align}
\a^{ii'}_{\n\n'}=\d_{\n\n'}\left(
\begin{array}{cc}
	0 & i \\ -i & 0 
\end{array}
\right)_{ii'}.
\end{align}

Let us express the contour Hamiltonian $\hat{H}(z)$, see 
Eqs.~(\ref{el-phonham}) and (\ref{tdhampar}), in terms of the 
$\hat{\f}$-operators. We first write the final result and then 
prove its correctness. We have
\begin{align}
\hat{H}(z)=\hat{H}_{0,e}^{s}(z)+\hat{H}_{0,ph}^{s}(z)+
\hat{H}_{e-e}(z)+\hat{H}_{e-ph}(z),
\label{el-phonham3}
\end{align}
where 
\begin{subequations}
\begin{align}
\hat{H}_{0,e}^{s}(z)=&\!\int \!\!d\blx \,
\hat{\q}^{\dag}(\blx)
\Big[h(\grad,\blr,z)-\!\sum_{\blq\n}
\bcallg_{\blq\n}^{\dag}(\blr,z)\cdot\bcalls_{\blq\n}(z)
\Big]\hat{\q}(\blx),
\label{h0es}
\\
\hat{H}_{0,ph}^{s}(z)&=\frac{1}{2}\sum_{\blq}\sum_{\n\n'}
\hat{\bgf}^{\dag}_{\blq\n}\;
Q_{\n\n'}(\blq,z)\;\hat{\bgf}_{\blq\n'}
\nn\\
&\quad+\sum_{\blq\n}\int \!d\blx\,n^{0}(\blx)\,
\bcallg^{\dag}_{\blq\n}(\blr,z)\cdot\hat{\bgf}_{\blq\n}\;,
\label{h0phs}
\\
\hat{H}_{e-e}(z)&=
\frac{1}{2}\!\int \!d\blx 
d\blx'\hat{\q}^{\dag}(\blx)\hat{\q}^{\dag}(\blx')
v(\blr,\blr',z)\hat{\q}(\blx')\hat{\q}(\blx),
\label{hees}
\\
\hat{H}_{e-ph}(z)&=-
\sum_{\blq\n}\int \!d\blx\,\hat{n}(\blx)\,
\bcallg^{\dag}_{\blq\n}(\blr,z)\cdot
\big(\hat{\bgf}_{\blq\n}-\bcalls_{\blq\n}(z)\big),
\label{hephs}
\end{align}
\label{hcomp}
\end{subequations}
and
\begin{subequations}
\begin{align}	
Q_{\n\n'}(\blq,z)&=\left(\begin{array}{cc}
	K_{\n\n'}(\blq,z) & 0 \\ 0 & \d_{\n\n'} ,
\end{array}\right)
\label{Qiipqnunup}
\\
\bcallg_{\blq\n}(\blr,z)&=\left(\begin{array}{c}
g_{\blq\n}(\blr,z) \\ 0 
\end{array}\right).
\label{giqnu}
\end{align}
\end{subequations}
In Eqs.~(\ref{h0es}) and (\ref{hephs}) we introduce a (two-dimensional 
vector) shift $\bcalls_{\blq\n}(z)$ with the purpose of 
simplifying the NEGF treatment, see below. The exact form of the shift is 
irrelevant for the time being as the terms containing $\bcalls_{\blq\n}(z)$
cancel out in the sum of Eq.~(\ref{el-phonham3}), 
thereby $\hat{H}(z)$ is independent of this quantity.
The shifted electronic Hamiltonian 
$\hat{H}_{0,e}^{s}(z)$ evaluated at zero shift, i.e., 
$\bcalls_{\blq\n}(z)=0$,
is the same as $\hat{H}_{0,e}(z)$, see 
Eqs.~(\ref{h0e}) and (\ref{h(z)}).
The first term of the shifted phononic Hamiltonian 
$\hat{H}_{0,ph}^{s}(z)$ is the same as $\hat{H}_{0,ph}(z)$, see 
Eqs.~(\ref{H0phondexp}) and (\ref{K(z)}), 
whereas the second term coincides with the 
contribution to $\hat{H}'$ coming from $n^{0}$ in $\D\hat{n}$, see 
Eqs.~(\ref{el-phonintham3}) and (\ref{g(z)}). The $e$-$e$ 
interaction Hamiltonian $\hat{H}_{e-e}(z)$ has not changed, see 
Eqs.~(\ref{el-phonham}) and (\ref{v(z)}). Finally, the 
$e$-$ph$ interaction Hamiltonian $\hat{H}_{e-ph}(z)$ 
evaluated at zero shift coincides with the 
contribution to $\hat{H}'$ coming from $\hat{n}$ in $\D\hat{n}$.
We conclude that Eq.~(\ref{el-phonham3}) is identical to the contour 
Hamiltonian for any two-dimensional vector $\bcalls_{\blq\n}(z)$.

\section{Equations of motion for operators in the contour Heisenberg picture}
\label{eomsec}

The {\em contour Heisenberg picture} is an extremely useful concept to 
develop the NEGF formalism. We define the contour evolution operators 
according to
\begin{subequations}
\begin{align}
\hat{\callU}(z,t_{0})&=\callT
\big\{e^{-i\int_{t_{0}}^{z}d\bar{z}\,\hat{H}(\bar{z})}\big\}
\\
\hat{\callU}(t_{0},z)&=\bar{\callT}
\big\{e^{i\int_{t_{0}}^{z}d\bar{z}\,\hat{H}(\bar{z})}\big\}
\end{align}
\label{contevolop}
\end{subequations}
where $\bar{\callT}$ is the anti-contour-ordering operator. These 
operators are unitary for $z=t_{\pm}$ and hermitian for $z=t_{0}-i\b$.
For all $z\in\g$ we have the property 
$\hat{\callU}(t_{0},z)\hat{\callU}(z,t_{0})=1$. We define an operator 
in the contour Heisenberg picture as 
\begin{align}
\hat{O}_{H}(z)\equiv 
\hat{\callU}(t_{0},z)\hat{O}(z)\hat{\callU}(z,t_{0}).
\end{align}
It is straightforward to verify that $O_{H}(t_{\pm})=O_{H}(t)$. The 
equation of motion for operators in the contour Heisenberg picture 
follows directly from the definitions in Eqs.~(\ref{contevolop})
\begin{align}
i\frac{d\hat{O}_{H}(z)}{dz}=\hat{\callU}(t_{0},z)\Big(
[\hat{O}(z),\hat{H}(z)]+
i\frac{d O(z)}{dz}\Big)\hat{\callU}(z,t_{0}).
\label{eomO}
\end{align}
As the Hamiltonian is written in terms of $\hat{\q}(\blx)$ and 
$\hat{\bgf}_{\blq\n}$ it is 
clear that the equation of motion for these operators plays a crucial 
role in the following derivation.

{\em Equations of motion.--}
Using Eq.~(\ref{eomO}) the equation of motion for the 
$\hat{\f}$-operators reads
\begin{align}
\sum_{\n'}\Big[i
\frac{d}{dz}\a_{\n\n'}&-Q_{\n\n'}(\blq,z)\Big]
\hat{\bgf}_{\blq\n',H}(z)
\nn\\
&=
-\int\!
d\blx 
\,\bcallg_{\blq\n}(\blr,z)\big(\hat{n}_{H}(\blx,z)-n^{0}(\blx)\big).
\label{eomphixi}
\end{align}
To remove the inhomogeneous term with $n^{0}$  
we define the $2\times 2$
GF $D_{0,\blq\n\n'}$ as the solution of
\begin{align}
\sum_{\bar{\n}}\Big[i
\frac{d}{dz}\a_{\n\bar{\n}}&-Q_{\n\bar{\n}}(\blq,z)\Big]	
D_{0,\blq\bar{\n}\n'}(z,z')=\mathbb{1}\d_{\n\n'}	
\d(z,z'),
\label{eomford01}
\end{align}
and satisfying the periodic Kubo-Martin-Schwinger (KMS) boundary conditions along 
the contour $\g$. We use $D_{0}$ to define the shift in Eqs.~(\ref{hcomp})
\begin{align}
\bcalls_{\blq\n'}(z)=
\sum_{\bar{\n}}\int \!d\bar{z}\,D_{0,\blq\n'\bar{\n}}(z,\bar{z})
\int\!d\blx\, \bcallg_{\blq\bar{\n}}(\blr,\bar{z})n^{0}(\blx).
\end{align}
Then the equation of motion for the (time-dependent) shifted displacement
\begin{align}
\hat{\bgvf}_{\blq\n}(z)\equiv 
\hat{\bgf}_{\blq\n}-\bcalls_{\blq\n}(z)
\label{phisdef}
\end{align}
reads
\begin{align}
\sum_{\n'}\Big[i
\frac{d}{dz}\a_{\n\n'}-Q_{\n\n'}&(\blq,z)\Big]
\hat{\bgvf}_{\blq\n',H}(z)
\nn\\
&=-\int\!d\blx 
\,\bcallg_{\blq\n}(\blr,z)\,\hat{n}_{H}(\blx,z),
\label{eomqpartbosphi}
\end{align}
which is a homogeneous equation in the field operators. Notice 
that the shift is proportional to the 
identity operator and therefore 
$\bcalls_{\blq\n,H}(z)=\bcalls_{\blq\n}(z)$. Such proportionality 
also implies 
that the operators $\hat{\bgvf}(z)$ and $\hat{\bgf}(z')$ 
[these operators are not in the contour-Heisenberg picture and in particular 
$\hat{\bgf}(z')$ does not depend on $z'$]
satisfy the 
same commutation relations for any $z$ and $z'$ : 
\begin{align}
[\hat{\vf}_{\blq\n}^{i}(z),\hat{\vf}^{i'\,\dag}_{\blq'\n'}(z')]	=
[\hat{\vf}_{\blq\n}^{i}(z),\hat{\f}^{i'\,\dag}_{\blq'\n'}(z')]	
=[\hat{\f}_{\blq\n}^{i}(z),\hat{\f}^{i'\,\dag}_{\blq'\n'}(z')]	.
\label{commrelfvf}
\end{align}
	
From the commutator $[\hat{\q}(\blx),\hat{H}(z)]$ we can easily 
derive the equation of motion for the electronic field operators
\begin{align}
\Big[
i\frac{d}{dz}-h^{s}(\grad,\blr,z)\Big]&
\hat{\q}_{H}(\blx,z)
\nn\\
&=\int 
d\blx'v(\blr,\blr',z)\hat{n}_{H}(\blx',z)\hat{\q}_{H}(\blx,z)
\nn\\
&-\sum_{\blq\n}\hat{\q}_{H}(\blx,z)\,
\bcallg_{\blq\n}^{\dag}(\blr,z)
\cdot\hat{\bgvf}_{\blq\n,H}(z),
\label{eomqpartbos}
\end{align}
where we define the shifted one-particle Hamiltonian 
\begin{align}
h^{s}(\grad,\blr,z)\equiv h(\grad,\blr,z)-\!\sum_{\blq\n}
\bcallg_{\blq\n}^{\dag}(\blr,z)\cdot\bcalls_{\blq\n}(z), 
\label{oneparths}
\end{align}
see Eq.~(\ref{h0es}),  and in the last term of the r.h.s. we  
recognize that the density operator in Eq.~(\ref{hephs}) multiples the 
shifted displacement $\hat{\bgvf}$ defined in Eq.~(\ref{phisdef}).

\section{Green's functions and  Martin-Schwinger hierarchy}
\label{dispG+MSHsec}

The building blocks of the NEGF formalism are the electronic and 
phononic GF.
Let us introduce 
a notation which is used throughout the reminder of the paper.
We denote the position, spin and contour-time coordinates of an 
electronic field operator with the collective indices
\begin{align}
k&=\blx_{k},z_{k},\;\;\;\;j=\blx_{j},z_{j},\;\;\;\ldots
\nn\\
k'&=\blx_{k}',z_{k}',\;\;\;\;j'=\blx_{j}',z_{j}',\;\;\;\ldots
\label{elnot}
\end{align}
etc.. Thus, for instance, $\hat{\q}(1)=\hat{\q}(\blx_{1},z_{1})$ and $
\hat{\q}(2')=\hat{\q}(\blx_{2}',z_{2}')$. We recall that 
the electronic field operators have no explicit dependence on the 
contour-time, i.e., $\hat{\q}(\blx,z)=\hat{\q}(\blx)$.
Without any risk of ambiguity we use the same notation to 
denote the momentum, branch, component and contour-time coordinates of a 
phononic field operator
\begin{align}
k&=\blq_{k},\n_{k},i_{k},z_{k},\;\;\;\;j=\blq_{j},\n_{j},i_{j},z_{j},\;\;\;\ldots
\nn\\
k'&=\blq'_{k},\n'_{k},i'_{k},z'_{k},\;\;\;\;j'=\blq'_{j},\n'_{j},i'_{j},z'_{j},\;\;\;\ldots
\end{align}
etc.. Thus, for instance, $\hat{\vf}(1)=\hat{\vf}_{\blq_{1}\n_{1}}^{i_{1}}(z_{1})$ and 
$\hat{\vf}(2')=\hat{\vf}_{\blq'_{2}\n'_{2}}^{i'_{2}}(z'_{2})$.
We also use the superscript star `` $^{\ast}$ '' to denote the 
composite index with reversed momentum, e.g., 
$k^{\ast}=-\blq_{k},\n_{k},i_{k},z_{k}$.
Of course starring twice is the same as no starring, i.e.,
$k^{\ast\ast}=k$. 
We then define 
\begin{subequations}
\begin{align}
\overrightarrow{D}_{0}^{-1}(1,2)&\equiv
\Big[i
\frac{d}{dz_{1}}\a^{i_{1}i_{2}}_{\n_{1}\n_{2}}-
Q^{i_{1}i_{2}}_{\n_{1}\n_{2}}(\blq_{1},z_{1})\big]
\d_{\blq_{1}\blq_{2}}\d(z_{1},z_{2})
\\
\overrightarrow{G}_{0}^{-1}(1;2)&\equiv
\Big[
i\frac{d}{dz_{1}}-h^{s}(\grad_{1},\blr_{1},z_{1})\Big]
\d(\blx_{1}-\blx_{2})\d(z_{1},z_{2}),
\label{G0-1}
\\
g(1;2)&\equiv\d(z_{1},z_{2})
g^{i_{2}}_{\blq_{2}\n_{2}}(\blr_{1},z_{1}),
\label{g(1;2)}
\\
v(1;2)&\equiv \d(z_{1},z_{2})v(\blr_{1},\blr_{2},z_{1}),
\label{v(1;2)}
\end{align}
\end{subequations}
where $\d(z_{1},z_{2})$ is the Dirac-delta on the 
contour~\cite{svl-book}.
Accordingly, the equations of motion for the field operators, i.e., 
Eqs.~(\ref{eomqpartbosphi}) and (\ref{eomqpartbos}), are 
shortened as 
\begin{subequations}
\begin{align}
\int \!d\tilde{2}\,\overrightarrow{D}_{0}^{-1}(1,\tilde{2})
\hat{\vf}_{H}(\tilde{2})&=-\int\!d\bar{2}\,
g(\bar{2};1)\hat{\q}^{\dag}_{H}(\bar{2})
\hat{\q}_{H}(\bar{2}),
\label{eomphishort}
\\
\int 
\!d\bar{2}\,\overrightarrow{G}_{0}^{-1}(1;\bar{2})\,
\hat{\q}_{H}(\bar{2})&=
\int 
\!d\bar{2}\,v(1;\bar{2})\hat{\q}^{\dag}_{H}(\bar{2})
\hat{\q}_{H}(\bar{2})\hat{\q}_{H}(1)
\nn\\
&-\int\!d\tilde{2}\,g(1;\tilde{2}^{\ast}) \hat{\q}_{H}(1) 
\hat{\vf}_{H}(\tilde{2}).
\label{eompsishort}
\end{align}
\label{eomshort}
\end{subequations}
To distinguish the integration variables of the electronic operators $\hat{\q}$ from 
those of the phononic operators $\hat{\vf}$ we use a bar for the 
former and a tilde for the latter, thus $\int d\bar{1}\equiv \int 
d\bar{\blx}_{1}d\bar{z}_{1}$ and 
$\int d\tilde{1}\equiv \sum_{\tilde{q}_{1}\tilde{\n}_{1}\tilde{i}_{1}}
\int d\tilde{z}_{1}$. Setting $v=g=0$ in the r.h.s. of 
Eqs.~(\ref{eomshort}) we obtain the equations of motion of the field 
operators governed by the Hamiltonian 
\begin{align}
\hat{H}_{0}^{s}(z)\equiv \hat{H}_{0,e}^{s}(z)+\hat{H}_{0,ph}^{s}(z).
\label{H0s}
\end{align}
We shall use this observation in the next section.

{\em Green's functions.--}
The phononic and electronic GF are contour-ordered 
correlators of strings of phononic and electronic field operators.
The $m$-particle phononic GF is defined as 
\begin{widetext}
\begin{align}
\widetilde{D}_m&(1,2,...,2m)\equiv  \frac{1}{i^{m}}
\Tr  \left[ \hat{\r}\, \callT
\left\{\hat{\vf}_{H}(1)\hat{\vf}_{H}(2)
...\hat{\vf}_{H}(2m)\right\} 
\right]
= \frac{1}{i^{m}} \frac{ \Tr  \left[ 
\callT \left\{ e^{-i 
\int_{\g} d\bar{z} \hat{H} (\bar{z})} \hat{\vf}(1)\hat{\vf}(2)
...\hat{\vf}(2m) \right\} \right]}
{\Tr \left[ \callT \left\{  e^{-i \int_{\g} d\bar{z} \hat{H} 
(\bar{z})} \right\} \right]} .
\label{ndispGF}
\end{align}
In this definition $m$ can also be a {\em half-integer}; in 
particular $\widetilde{D}_{1/2}(1)=\vf(1)/i^{1/2}$.
The phononic GF $\widetilde{D}_{m}$
differs from that in 
Refs.~\cite{sakkinen_many-body_2015-2,karlsson_non-equilibrium_2020} 
as it is defined in terms of the shifted operators $\hat{\vf}$ 
instead of $\hat{\f}$.  
We observe that the operators appearing in the 
second equality are not in the contour Heisenberg picture [compare 
with Eq.~(\ref{contformula})]. 
The phononic GF is totally symmetric 
under an arbitrary permutation of its arguments $1,2,...,2m$.
Similarly, we define the $n$-particle electronic GF as 
\begin{align}
G_n& (1, ..., n; 1', ..., n') = \frac{1}{i^{n}} 
\Tr  \left[ \hat{\r} \,\callT \big\{
\hat{\psi}_{H} (1) ... 
\hat{\psi}_{H} (n) \hat{\psi}^{\dag}_{H} (n')   
...  \hat{\psi}^{\dag}_{H}(1')
\big\} \right]
= \frac{1}{i^{n}} \frac{\Tr\left[\callT\left\{e^{-i 
\int_{\g} d\bar{z} \hat{H} (\bar{z})} \hat{\psi} (1) ... 
\hat{\psi} (n) \hat{\psi}^{\dag} (n')   
...  \hat{\psi}^{\dag}(1') \right\} \right]}
{\Tr \left[ \callT \left\{  e^{-i \int_{\g} d\bar{z} \hat{H} 
(\bar{z})} \right\} \right]} .
\nonumber \\
\label{manyGF1}
\end{align}
The electronic GF is totally antisymmetric 
under an arbitrary permutation of the arguments $1,2,...,n$ and 
$1',2',...,n'$.

{\em Martin-Schwinger hierarchy.--} Taking into account that the 
commutation relations for the operators $\hat{\vf}$ are 
identical to the commutation relations for the operators $\hat{\f}$, 
see Eq.~(\ref{commrelfvf}),
the equations of motion for $\widetilde{D}_m$ with $m=1/2,1,3/2,...$ read
\begin{align}
\int d\tilde{1}\,\overrightarrow{D}_{0}^{-1}(k,\tilde{1})
\widetilde{D}_{m}(1,...,\tilde{1},...,2m)
&=-\frac{1}{i^{m}}\int 
d\bar{1}\,g(\bar{1};k)\,
\Tr\Big[\hat{\r}\;\callT\Big\{
\hat{\vf}_{H}(1),...,\underbrace{\hat{\q}^{\dag}_{H}(\bar{1})
\hat{\q}_{H}(\bar{1})}_{k-{\rm th\,place}},...,
\hat{\vf}_{H}(2m)\Big\}\Big]
\nn\\
&+
\sum_{\substack{j=1\\j\neq k}}^{2m}\d(k,j^{\ast})
\widetilde{D}_{m-1}(1,...,\stackrel{\sqcap}{k},...,\stackrel{\sqcap}{j},...,2m),
\label{intmshdm}
\end{align}
where the variable $\tilde{1}$ in the l.h.s. is at place  $k$ 
and we define $\widetilde{D}_{0}=1$ and $\widetilde{D}_{-1/2}=0$.
The first term in the r.h.s. originates from the equation of motion 
Eq.~(\ref{eomphishort}). The last term in the r.h.s. originates
from the derivative of the Heaviside step-functions implicit in 
the contour ordering~\cite{svl-book}. These derivatives generate quantities like 
$\d(z_{k},z_{j})[\hat{\vf}^{i_{k}}_{\blq_{k}\n_{k},H}(z_{k}),
\hat{\vf}^{i_{j}}_{\blq_{j}\n_{j},H}(z_{j})]=\d(z_{k},z_{j})
\d_{\blq_{k},-\blq_{j}}\a^{i_{k}i_{j}}_{\n_{k}\n_{j}}$, 
see Eq.~(\ref{commfifi}), which multiplied by $\a$ lead to 
$\d(k,j^{\ast})\equiv\d(z_{k},z_{j})
\d_{\blq_{k},-\blq_{j}}\d_{\n_{k}\n_{j}}\d_{i_{k}i_{j}}$.
To shorten the equations we 
have also introduced
the symbol `` $\sqcap$ '' above an index to indicate that the index is 
missing from the list. 
In a similar way we can derive the equations of motion for the 
$n$-particle GF
and find for $n=1,2,3,\ldots$
\begin{align}
\int d\bar{1}\,
\overrightarrow{G}_{0}^{-1}(k;\bar{1})
G_n(1,...,\bar{1},...,n;1',..., n')  
&=- \,i \int d\bar{1} \, v(k;\bar{1}) 
\, G_{n+1} (1, ..., n, \bar{1} ; 1', ... ,n',\bar{1}^+)
\nn\\
&-\frac{1}{i^{n}}\int 
 d\tilde{1}\,g(k;\tilde{1}^{\ast})
 \Tr\Big[\hat{\r}\;\callT\Big\{
\hat{\q}_{H}(1),...,\underbrace{
\hat{\q}_{H}(k)\hat{\vf}_{H}(\tilde{1})
}_{k-{\rm th\,place}},...,\hat{\q}_{H}(n)\hat{\q}_{H}(n'),...,
\hat{\q}^{\dag}_{H}(1')\Big\}\Big]
\nn \\
&
+\sum_{j=1}^n \, (-)^{k+j} \,\delta (k;j') \, G_{n-1} (1, 
..., \stackrel{\sqcap}{k},  ..., n; 1', ..., 
\stackrel{\sqcap}{j'}, ..., n' ) , 
\label{intmshgn}
\end{align}
\end{widetext}
where the variable $\bar{1}$ in the l.h.s. is at place  $k$ and 
we define $G_{0}=1$. The 
first two terms in the r.h.s. originate from the equation of motion 
Eq.~(\ref{eompsishort}). The last term in the r.h.s. originates
from the the derivative of the Heaviside step-functions implicit in 
the contour ordering. In the 
electronic case $\d(k;j)\equiv \d(\blx_{k}-\blx_{j})\d(z_{k},z_{j})$. We 
further notice that the last argument of the GF 
$G_{n+1}$ is $\bar{1}^{+}$. We use the superscript `` $+$ '' to 
indicate that the contour time is infinitesimally later than 
$\bar{z}_{1}$. This infinitesimal shift guarantees that 
the creation operator $\hat{\q}^{\dag}(\bar{1})$ in $G_{n+1}$ ends up to 
the left of the annihilation operator $\hat{\q}(\bar{1})$ when the 
operators are contour ordered.

The equation of motion for $G_{n}$ with derivative with respect to 
the primed arguments can be worked out similarly.
All equations of motion must be solved with 
KMS boundary conditions, i.e., $\widetilde{D}_{m}$ must be 
periodic on the contour with respect to all times $z_{1},\ldots,z_{2m}$ 
and $G_{n}$ must be antiperiodic on the contour 
with respect to all  times 
$z_{1},\ldots,z_{n},z'_{1},\ldots,z'_{n}$. 

If the $e$-$ph$ 
coupling $g=0$ then $\widetilde{D}_{m}$ couples only to $\widetilde{D}_{m-1}$.
In the electronic sector things are different. For $g=0$ the equations 
of motion reduce to the Martin-Schwinger hierarchy for a system of 
only electrons, and $G_{n}$ couples to $G_{n-1}$  and $G_{n+1}$ 
through the Coulomb interaction $v$. For $G_{n}$
to couple
only to $G_{n-1}$ the Coulomb interaction has to vanish too.

When both $e$-$e$ and $e$-$ph$  
interactions are present, $G_{n}$  
couples to $G_{n-1}$,  $G_{n+1}$ as well as  to mixed 
GF consisting of a mix string of $\hat{\q}$, $\hat{\q}^{\dag}$, 
and $\hat{\vf}$ operators, see second line in the equations of motion. Likewise, 
$\widetilde{D}_{m}$ couples to $\widetilde{D}_{m-1}$ but also to mixed GF,
see first term in the r.h.s.  of the equations of motion. The equations 
of motion for the mixed GF can  be derived in precisely the same 
way, see also Ref.~\cite{sakkinen_thesis}.
We refer to the full set of equations as the Martin-Schwinger 
hierarchy for electron-phonon systems.
In the next sections we lay down a perturbative method to calculate 
all GF.

\section{Wick's theorem for the many-particle Green's functions}
\label{wicksec}

The Wick theorem provides the 
solution of the Martin-Schwinger hierarchy with r.h.s evaluated at 
$g=v=0$. This is the same as solving the Martin-Schwinger hierarchy
for a system of electrons and phonons 
governed by the Hamiltonian $\hat{H}_{0}^{s}$, see comment above 
Eq.~(\ref{H0s}). As $\hat{H}_{0}^{s}$ depends on $g$ explicitly, 
setting $g=v=0$ in the r.h.s. of Eqs.~(\ref{intmshdm})
and (\ref{intmshgn}) is 
not the same as solving the Martin-Schwinger 
hierarchy with $\hat{H}|_{g=v=0}$ (noninteracting hierarchy). 
Henceforth we name the GF governed by  
$\hat{H}_{0}^{s}$ as the {\em independent} GF and we 
denote them by $D_{0,m}$ and $G_{0,n}$.
We then have

\begin{widetext}
\begin{subequations}	
\begin{align}
\int \!d\tilde{1}\,\overrightarrow{D}_{0}^{-1}(k,\tilde{1})
D_{0,m}(1,...,\tilde{1},...,2m)
&=\sum_{\substack{j=1\\j\neq k}}^{2m}\d(k,j^{\ast})
D_{0,m-1}(1,...,\stackrel{\sqcap}{k},...,\stackrel{\sqcap}{j},...,2m),
\label{nintmshdm}
\\
\int d\bar{1}\,
\overrightarrow{G}_{0}^{-1}(k;\bar{1})
G_{0,n}(1,...,\bar{1},...,n;1',..., n')  
&=\sum_{j=1}^n \, (-)^{k+j} \,\delta (k;j') \, G_{0,n-1} (1, 
..., \stackrel{\sqcap}{k},  ..., n; 1', ..., 
\stackrel{\sqcap}{j'}, ..., n' ) , 
\label{nintmshgn}
\end{align}
\end{subequations}
\end{widetext}
and the like with time derivatives with respect to the primed 
arguments. The independent GF satisfy two independent 
hierarchies. Despite the similarities the phononic and 
electronic hierarchies present 
important differences.
The sum in the r.h.s. runs over all arguments of $D_{0,m}$ 
in Eq.~(\ref{nintmshdm}) whereas it runs only over the unprimed 
arguments of $G_{0,n}$ in Eq.~(\ref{intmshgn}).
Moreover, in the phononic case $m$ can also be a half-integer.
From Eq.~(\ref{nintmshdm}) 
we see that the integer $m$ connects with the integer $m-1$ and 
the half-integer $m$ connects with the half-integer $m-1$. Therefore, 
we have two separate hierarchies of equations for the phononic 
GF.

{\em Wick's theorem for phonons.--}
The proof of Wick's theorem for $D_{0,m}$ goes along the same lines 
as in 
Ref.~\cite{karlsson_non-equilibrium_2020}. The phononic GF 
$D_{0,m} = 0$ for all half-integers $m$. This can easily be proven 
 by considering the average of the equation of motion 
Eq.~(\ref{eomqpartbosphi}) with $g=0$. 
By definition this average is proportional to 
$D_{0,1/2}$ and the only solution satisfying the KMS boundary 
conditions is $D_{0,1/2}=0$. Consider now $m = 3/2$: 
\begin{align}
\int d\tilde{1}\,&\overrightarrow{D}_{0}^{-1}(1,\tilde{1})
D_{0,3/2}(\tilde{1},2,3)
\nn\\
&=\d(1,2^{\ast})D_{0,1/2}(3)+\d(1,3^{\ast})D_{0,1/2}(2)=0,
\end{align}
and the like for the variables $2$ and $3$. We see that
$D_{0,3/2}=0$ is a solution satisfying the KMS boundary 
conditions. By induction, $D_{0,m}=0$ 
for half-integers. Henceforth we only consider integers $m$ in the 
noninteracting case.     

For $m=1$ we have the equation of motion for $D_{0,1}$:
\begin{align}
\int d\tilde{1}\,\overrightarrow{D}_{0}^{-1}(1,\tilde{1})
D_{0,1}(\tilde{1},2)
=\d(1,2^{\ast}),
\label{nintmshd1}
\end{align}
Comparing with Eq.~(\ref{eomford01}) we realize that 
\begin{align}
D_{0,1}(1,2)=\d_{\blq_{1},-\blq_{2}}
D_{0,\blq_{1}\n_{1}\n_{2}}^{i_{1}i_{2}}(z_{1},z_{2}).
\label{connctD0D0}
\end{align} 
Without 
any risk of ambiguity we denote $D_{0,1}(1,2)$ simply by $D_{0}(1,2)$ 
in the remainder of the paper. For $D_{0,m}$ with $m>1$ the solution 
of Eq.~(\ref{nintmshdm}) is given by the so-called {\em 
 hafnian}~\cite{caianiellobook,izbook,sakkinen_thesis}. The  hafnian can be 
defined recursively starting from any of the arguments in $D_{0,m}$. 
Choosing for instance the argument $k$ we have
\begin{align}
D_{0,m}(1,...,2m)=\sum_{\substack{j=1\\j\neq 
k}}^{2m}D_{0}(k,j)D_{0,m-1}(1,...,\stackrel{\sqcap}{k},...,\stackrel{\sqcap}{j},...,2m).
\label{haffnian}
\end{align}
Using again Eq.~(\ref{haffnian}) for $D_{0,m-1}$ and then for 
$D_{0,m-2}$ and so on and so forth we obtain an 
expansion of $D_{0,m}$ in terms of products of $m$ $D_{0}$'s.
A compact way to write this expansion is 
\begin{align}
D_{0,m}(1,...,2m)=\frac{1}{2^{m}m!}\sum_{P}&
D_{0}(P(1),P(2))\ldots 
\nn\\&\times D_{0}(P(2m-1),P(2m)),
\end{align}
where the sum runs over all permutations of the indices 
$1,2,\ldots,2m$.
The recursive form of Eq.~(\ref{haffnian}) 
makes it clear that $D_{0,m}$ satisfies Eq.~(\ref{nintmshdm}) and the KMS 
boundary conditions.

{\em Wick's theorem for electrons.--} The solution of  
Eq.~(\ref{intmshgn}) has been discussed at length in 
Refs.~\cite{vLS.2012,svl-book}. We here  write the final result for 
completeness. For $n=1$ Eq.~(\ref{intmshgn}) yields
\begin{align}
\int d\bar{1}\,
\overrightarrow{G}_{0}^{-1}(1;\bar{1})
G_{0,1}(\bar{1};2)  
=\delta (1;2),
\label{eomG0}
\end{align}
to be solved with KMS boundary conditions. Like for the phononic case 
we shorten the notation and write 
$G_{0,1}(1;2)=G_{0}^{s}(1;2)$. The superscript `` $s$ '' reminds 
us that this GF depends on $g$ through $h^{s}$.
For $G_{0,n}$ with $n>1$ the solution of Eq.~(\ref{intmshgn}) is 
again defined recursively choosing either an unprimed or a primed 
argument
\begin{align}
G_{0,n} (1 ,..., n; 1' ,..., n') 
&= \sum_{k=1}^n \, (-)^{k+j} G_0^{s} (k;j') 
\nn\\ &\times G_{0,n-1} (1 ,..., \stackrel{\sqcap}{k}, ..., n; 1' 
,..., \stackrel{\sqcap}{j'}, ..., n' )   
\nn
\nn\\&= \sum_{j=1}^n (-)^{k+j} G_0^{s} 
(k;j') \nn\\ &\times G_{0,n-1} (1, ... ,\stackrel{\sqcap}{k}, ... ,n; 1' ,
...,\stackrel{\sqcap}{j'}, ..., n' ) . 
\label{G0nwick}
\end{align}
Using again Eq.~(\ref{G0nwick}) for $G_{0,n-1}$ and then for 
$G_{0,n-2}$ and so on and so forth we obtain an 
expansion of $G_{0,n}$ in terms of products of $n$ $G^{s}_{0}$'s.
A compact way to write this expansion is the determinant
\begin{align}
G_{0,n} (1, ..., n; 1' ,... ,n') &= \left|
\begin{array}{ccc}
 G_0^{s} (1;1') & \ldots  &G_0^{s}(1;n')    \\
 \vdots & & \vdots \\
 G_0^{s} (n;1') & \ldots  & G_0^{s}(n;n')
\end{array}
\right|
\nn\\
=\sum_{P}(-)^{P}&G_0^{s} (P(1);1')\ldots G_{0}^{s}(P(n);n')
\end{align}
where the sum runs over all permutations of the indices $1, 2, ... , 
n$ and $(-)^{P}$ is the sign of the permutation.

The recursive form of the Wick's theorem 
highlights the differences between the phononic and the electronic 
case. For $G_{0,n}$ 
we need to connect unprimed arguments to primed 
arguments in all possible ways. For $D_{0,m}$  there is no 
such distinction, and we need to 
connect all arguments in all possible ways. 

\section{Exact Green's functions from Wick's theorem}
\label{exactexpsec}

The interacting GF $\widetilde{D}_{m}$ and $G_{n}$ can be expanded 
in powers of the $e$-$e$ interaction $v$ and  
$e$-$ph$ coupling $g$. 
In this section we focus on the one-particle electronic GF 
$G\equiv G_{1}$, the one-particle phononic GF
$\widetilde{D}\equiv \widetilde{D}_{1}$ and the half-particle phononic GF 
$\widetilde{D}_{1/2}$. The final results are Eqs.~(\ref{gabexppb}), 
(\ref{dabexppb}) and (\ref{phiaexppb}).
In Section~\ref{hedinsec} we show that the perturbative expansion 
leads to a closed system of equations for these quantities.
Higher order Green's functions as well as 
mixed Green's functions (relevant for linear response theory) 
can be investigated along the same lines, see Ref.~\cite{svl-book}.

The starting point is the  Hamiltonian written in the form 
of Eq.~(\ref{el-phonham3}), i.e., 
$\hat{H}=\hat{H}_{0}^{s}+\hat{H}_{e-e}+\hat{H}_{e-ph}$ with 
$\hat{H}_{0}^{s}$ defined in Eq.~(\ref{H0s}).
Inside the contour-ordering  the Hamiltonians $\hat{H}_{0}^{s}$, 
$\hat{H}_{e-e}$ and $\hat{H}_{e-ph}$ can be treated as 
commuting operators
and hence the
exponential of their sum  can be separated into the product of three exponentials. 
It is then natural to
define the independent averages as
\begin{align}
\bra\callT\big\{\ldots\big\}\ket^{s}_{\rm 0}\equiv
\Tr\Big[\callT\big\{e^{-i\int_{\g}d\bar{z}\hat{H}^{s}_{0}(\bar{z})}\ldots\big\}\Big].
\end{align}
We emphasize again that the independent averages are not the same as the 
noninteracting averages, i.e., the averages with $v=g=0$, 
since both $\hat{H}_{0,e}^{s}$ and $\hat{H}_{0,ph}^{s}$ 
depend on $g$, see Eqs.(\ref{h0es}) and (\ref{h0phs}).

{\em One-particle electronic Green's function.--} The 
GF $G\equiv G_{1}$ is defined in Eq.~(\ref{manyGF1}). 
We have 
\begin{widetext}
\begin{align}
G(a;b)&=\frac{1}{i} \frac{\Tr\left[\callT\left\{e^{-i \int_{\gamma} d\bar{z}\, \hat{H}_{0}^{s} (\bar{z})}  
e^{-i \int_{ \gamma} d\bar{z}\, \hat{H}_{e-e} (\bar{z})}
e^{-i \int_{ \gamma} d\bar{z}\, \hat{H}_{e-ph} (\bar{z})}
\hat{\psi} (a) \hat{\psi}^{\dag}(b) \right\} \right]}
{\Tr \left[ \callT \left\{   
e^{-i \int_{\gamma} d\bar{z}\, \hat{H}_{0}^{s} (\bar{z})}  
e^{-i \int_{ \gamma} d\bar{z}\, \hat{H}_{e-e} (\bar{z})}
e^{-i \int_{ \gamma} d\bar{z}\, \hat{H}_{e-ph} (\bar{z})} 
\right\} \right]},  
\label{Gabexpprelpb}
\end{align}
where $a=\blx_{a},z_{a}$ and $b=\blx_{b},z_{b}$ in accordance with 
the notation of Eq.~(\ref{elnot}).
The denominator in this equation is the interacting partition 
function $\callZ=\Tr[e^{-\b(\hat{H}-\m\hat{N}_{e})}]$. Expanding the exponentials containing 
$\hat{H}_{e-e}$ and $\hat{H}_{e-ph}$ we find
\begin{align}
G(a;b)&=\frac{1}{i} \frac{ 
\sum_{k,p=0}^{\iif}
\frac{(-i)^{k+p}}{k!\,p!}
\int_{\g}\!dz_{1}...dz_{k} d\tilde{z}_{1}...d\tilde{z}_{p}
\bra\callT\Big\{\!\hat{H}_{e-e} (z_{1})...\hat{H}_{e-e} (z_{k})
\hat{H}_{e-ph}(\tilde{z}_{1})...\hat{H}_{e-ph}(\tilde{z}_{p})
\hat{\psi} (a) \hat{\psi}^{\dag}(b)\!\Big\}\ket_{0}^{s}}
{ 
\sum_{k,p=0}^{\iif}
\frac{(-i)^{k+p}}{k!\,p!}
\int_{\g}\!dz_{1}...dz_{k} d\tilde{z}_{1}...d\tilde{z}_{p}
\bra\callT\Big\{\!\hat{H}_{e-e} (z_{1})...\hat{H}_{e-e} (z_{k})
\hat{H}_{e-ph}(\tilde{z}_{1})...\hat{H}_{e-ph}(\tilde{z}_{p})
\!\Big\}\ket_{0}^{s}}.  
\label{Gabexpprelpb2}
\end{align}	
To facilitate the identification of the expansion terms 
we use a tilde for the contour times of the $e$-$ph$
interaction Hamiltonian.
Let us write the integrated Hamiltonians in terms of $g(i;j)$ 
and $v(i;j)$, see Eqs.~(\ref{g(1;2)}) and (\ref{v(1;2)}). We have
\begin{subequations}
\begin{align}
\int d\tilde{z}_{j} \hat{H}_{e-ph}(\tilde{z}_{j})&=
-\int d\bar{j}d\tilde{j} \,
g(\bar{j};\tilde{j}^{\ast})\,
\hat{\q}^{\dag}(\bar{j}^{+})\hat{\q}(\bar{j})
\,\hat{\vf}(\tilde{j}),
\\
\int dz_{j} \hat{H}_{e-e}(z_{j})&=
\frac{1}{2}\int dj \,dj' v(j;j')\hat{\q}^{\dag}(j^{+})\hat{\q}^{\dag}(j'^{+})
\hat{\q}(j')\hat{\q}(j).
\end{align}
\label{hinthfraka}
\end{subequations}
The infinitesimal shift in the contour-times of the electronic 
creation operators guarantees that these operators end up to the left 
of the annihilation operators calculated at the same contour-times
after the contour reordering.
Inserting Eqs.~(\ref{hinthfraka}) into Eq.~(\ref{Gabexpprelpb2}) we 
are left with the evaluation of 
contour-ordered strings like 
$\bra\callT\{\hat{\q}\ldots\hat{\q}\hat{\q}^{\dag}\ldots\hat{\q}^{\dag}
\hat{\vf}\ldots\hat{\vf}\}
\ket_{0}^{s}$ with an arbitrary number of operators.
We observe that 
$\hat{H}_{0,e}^{s}$ acts on the Fock space $\mathbb{F}$ of 
the electrons and $\hat{H}_{0,ph}^{s}$ acts on the Hilbert space 
$\mathbb{D}_{N_{n}}$ of $N_{n}$ distinguishable nuclei. 
As such, the eigenkets of
$\hat{H}_{0}^{s}=\hat{H}_{0,e}^{s}+\hat{H}_{0,ph}^{s}$ 
factorize into tensor products of 
kets in $\mathbb{F}$ and kets in $\mathbb{D}_{N_{n}}$. 
Therefore, the partition function for independent electrons and 
phonons
\begin{align}\
\callZ_{0}^{s}&=\Tr\big[\callT\big\{e^{-i 
\int_{\gamma} d\bar{z} \,\hat{H}_{0}^{s} (\bar{z})} \big\}\big]
=
\Tr\big[\callT\big\{e^{-i 
\int_{\gamma} d\bar{z} \,\hat{H}_{0,e}^{s} (\bar{z})} \big\}\big]
\times \Tr\big[\callT\big\{e^{-i 
\int_{\gamma} d\bar{z}\, \hat{H}_{0,ph}^{s}(\bar{z})} 
\big\}\big]=\callZ_{0,e}^{s}\callZ_{0,ph}^{s}
\end{align}
factorizes into electron and phonon 
contributions. The same type of factorization allows us to simplify 
the independent average of any string of operators  as
\begin{align}
\bra\callT\{\hat{\q}\ldots\hat{\q}\hat{\q}^{\dag}\ldots\hat{\q}^{\dag}
\hat{\vf}\ldots\hat{\vf}\}\ket_{0}^{s}=
\bra\callT\{\hat{\q}\ldots\hat{\q}\hat{\q}^{\dag}\ldots\hat{\q}^{\dag}
\}\ket_{0,e}^{s}\times
\bra\callT\{\hat{\vf}\ldots\hat{\vf}\}\ket_{0,ph}^{s}\;,
\label{facteb}
\end{align}
where the average $\bra\ldots\ket_{0,e}^{s}$ is performed with 
$\hat{H}_{0,e}^{s}$ and the average $\bra\ldots\ket_{0,ph}^{s}$ is 
performed with $\hat{H}_{0,ph}^{s}$.

To order $k$ in $v$ and to order $p$ in $g$ the numerator of the 
GF contains the independent average of the following string
\begin{align}
&\bra\callT\Big\{\underbrace{\hat{\q}^{\dag}(1^{+})\hat{\q}^{\dag}(1'^{+})
\hat{\q}(1')\hat{\q}(1)\ldots}_{4k\;{\rm operators}}
\underbrace{
\hat{\q}^{\dag}(\bar{1}^{+})\hat{\q}(\bar{1})
\hat{\vf}(\tilde{1})\ldots}_{3p\;{\rm operators}}\hat{\psi} (a) 
\hat{\psi}^{\dag}(b)\Big\}
\ket_{0}^{s}
\nn\\
&=(-)^{p}\bra\callT\Big\{\hat{\psi}(a)
\underbrace{\hat{\q}(1)\hat{\q}(1')\ldots}_{2k\;\hat{\q}}
\underbrace{\hat{\q}(\bar{1})\ldots}_{p\;\hat{\q}}
\underbrace{\ldots \hat{\q}^{\dag}(\bar{1}^{+})}_{p\;\hat{\q}^{\dag}}
\underbrace{
\ldots\hat{\q}^{\dag}(1'^{+})\hat{\q}^{\dag}(1^{+})}_{2k\;\hat{\q}^{\dag}}
\hat{\psi}^{\dag}(b)
\Big\}
\ket_{0,e}^{s}
\times\bra\callT\Big\{
\underbrace{\hat{\vf}(\tilde{1})\ldots}_{p \;\hat{\vf}}\Big\}
\ket_{0,ph}^{s}
\nn\\
&=(-)^{p}\callZ_{0,e}^{s}\;i^{2k+p+1}
G_{0,2k+p+1}(a,1,1',...,\bar{1},...;b,1^{+},1'^{+},...,\bar{1}^{+},...)
\times \callZ_{0,ph}^{s}\; i^{p/2}D_{0,p/2}(\tilde{1},...).
\end{align}
With similar 
manipulations we can work out the expansion of the partition 
function. Taking into account that $D_{0,p/2}$ is nonvanishing only 
for even integers $p$ the expansion of the interacting GF reads
\begin{align}
G(a;b)&=\frac{\callZ_{0}^{s}}{\callZ}\sum_{k,p=0}^{\iif}
\frac{i^{k+p}}{2^{k}k!\,(2p)!}\int d1d1'...dkdk' v(1;1')...v(k;k')\int d\tilde{1}
d\bar{1}...d(\widetilde{2p})
d(\overline{2p})\, g(\bar{1};\tilde{1}^{\ast})...g(\overline{2p};\widetilde{2p^{\ast}})
\nn\\
&\times 
G_{0,2k+2p+1}(a,1,1',...,\bar{1},\ldots;b,1^{+},1'^{+},...,\bar{1}^{+},...)
D_{0,p}(\tilde{1},...),
\label{gabexppb}
\end{align}
with
\begin{align}
\frac{\callZ}{\callZ_{0}^{s}}&=	\sum_{k,p=0}^{\iif}
\frac{i^{k+p}}{2^{k}k!\,(2p)!}\int d1d1'...dkdk' v(1;1')...v(k;k')\int d\tilde{1}
d\bar{1}...d(\widetilde{2p})
d(\overline{2p})\, g(\bar{1};\tilde{1}^{\ast})...g(\overline{2p};\widetilde{2p^{\ast}})
\nn\\
&\times 
G_{0,2k+2p}(1,1',...,\bar{1},\ldots;1^{+},1'^{+},...,\bar{1}^{+},...)
D_{0,p}(\tilde{1},...).
\label{Zexppb}
\end{align}	
The zero-th order term in the expansion of Eq.~(\ref{gabexppb}) 
($k=p=0$) is the GF
$G_{0,1}=G_{0}^{s}$ calculated from Eq.~(\ref{eomG0}) where 
$\overrightarrow{G}_{0}^{-1}$ is defined in Eq.~(\ref{G0-1}).

Using Wick's theorem for $G_{0,n}$ and $D_{0,m}$ 
Eqs.~(\ref{gabexppb}) and (\ref{Zexppb}) provide an exact expansion in terms of 
the one-particle electronic GF $G_{0}^{s}$ and phononic
GF $D_{0}$. 

{\em One-particle phononic Green's function.--}
The interacting GF $\widetilde{D}\equiv \widetilde{D}_{1}$
is defined in Eq.~(\ref{ndispGF}). Writing the exponential 
like in Eq.~(\ref{Gabexpprelpb}), expanding with respect to
$\hat{H}_{e-e}$ and $\hat{H}_{e-ph}$, and 
using  Eqs.~(\ref{hinthfraka}) we find
\begin{align}
\widetilde{D}(a,b)&= \frac{\callZ_{0}^{s}}{\callZ}
\sum_{k,p=0}^{\iif}
\frac{i^{k+p}}{2^{k}k!\,(2p)!}\int d1d1'...dkdk' v(1;1')...v(k;k') 
\int d\tilde{1}d\bar{1}... d(\widetilde{2p})
d(\overline{2p})
g(\bar{1};\tilde{1}^{\ast})...g(\overline{2p};\widetilde{2p}^{\ast})
\nn\\
&\times G_{0,2k+2p}(1,1',...,\bar{1},...;1^{+},1'^{+},...,\bar{1}^{+},...)
D_{0,p+1}(a,b,\tilde{1},...),
\label{dabexppb}
\end{align}
where we take into account that $D_{0,m}$ vanishes for half-integers $m$.
In Eq.~(\ref{dabexppb}) the arguments $a=\blq_{a},\n_{a},i_{a},z_{a}$ 
and $b=\blq_{b},\n_{b},i_{b},z_{b}$.

Using Wick's theorem for $G_{0,n}$ and $D_{0,m}$ we  have 
an exact expansion of the interacting one-particle phononic GF
in terms of $G_{0}^{s}$ and $D_{0}$. The 
zero-th order term ($k=p=0$) is the GF $D_{0}$
since $G_{0,0}=1$.

{\em Half-particle phononic Green's function.--}
The interacting GF $\widetilde{D}_{1/2}$ is proportional to 
the time-dependent average of the field operator $\hat{\vf}$, i.e., 
$\widetilde{D}_{1/2}(a)=\frac{1}{i^{1/2}}\,\vf(a)$.
Proceeding along the same lines as for the derivation of the 
expansion Eq.~(\ref{dabexppb}) we find
\begin{align}
\vf(a)&= \frac{\callZ_{0}^{s}}{\callZ}
\sum_{k,p=0}^{\iif}
\frac{i^{k+p+1}}{2^{k}k!\,(2p+1)!}\int d1d1'...dkdk' v(1;1')...v(k;k') 
\int d\tilde{1}d\bar{1}... d(\widetilde{2p+1})
d(\overline{2p+1})
g(\bar{1};\tilde{1}^{\ast})...g(\overline{2p+1};\widetilde{2p+1}^{\ast})
\nn\\
&\times G_{0,2k+2p+1}(1,1',\ldots,\bar{1},\ldots;1^{+},1'^{+},\ldots,\bar{1}^{+},\ldots)
D_{0,p+1}(a,\tilde{1},\ldots),
\label{phiaexppb}
\end{align}
where we take into account that $D_{0,m}$ vanishes for half-integers $m$.
The average $\vf$ vanishes for $g=0$, in agreement with 
the equation of motion Eq.~(\ref{eomqpartbosphi}).

\end{widetext}

\section{Diagrammatic theory}
\label{ebdiagexpsec}

The expansions of $G$, $\widetilde{D}$ and $\vf$ contain the GF
$G_{0}^{s}$ and $D_{0}$, the Coulomb interaction $v$ and 
the $e$-$ph$ coupling $g$. Let us assign a 
graphical object to these quantities. We use an   
oriented line from $2$ to $1$ to represent 
$G_{0}^{s}(1;2)$, and a wiggly line between $1$ and $2$ to represent 
$v(1;2)=v(2;1)$. 
For the noninteracting phononic GF
$D_{0}(1,2)=D_{0}(2,1)$ we use
a spring from $1$ to $2$. The $e$-$ph$ coupling $g(1;2^{\ast})$
is instead represented by a square, half black and half white, 
where $1$ is attached to the black vertex and $2$ is attached 
to the white vertex. To summarize
\begin{align}
    \raisebox{-6pt}{\includegraphics[width=0.26\textwidth]{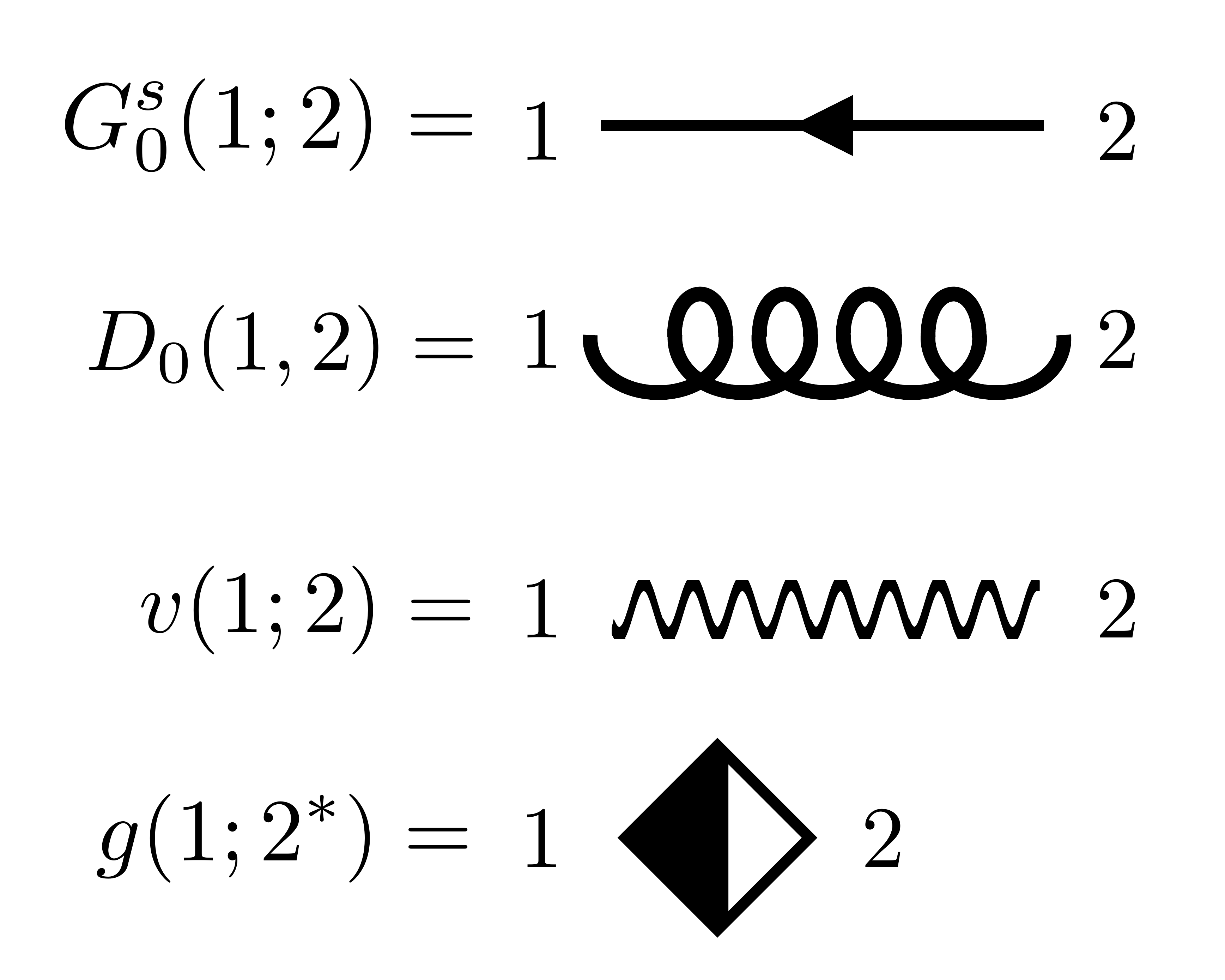}} \;\;,
    \label{Phitilde}
\end{align}
We can now represent every term of the expansions with diagrams. 
The diagrams for $G$ are either 
connected or products of a connected diagram and a vacuum 
diagram. In a connected diagram for $G(a;b)$ all internal vertices are 
connected to both $a$ and $b$ through $G_{0}^{s}$, $D_{0}$, $v$ 
and $g$. Thus a disconnected $G$-diagram is characterized 
by a subset of internal vertices that are not connected to either $a$ 
or $b$ and hence they form a vacuum diagram.
Similarly, the diagrams for $\vf(a)$ fall into two 
main classes: those with all internal vertices connected to 
$a$ and those where a subset of internal vertices is 
disconnected, thus forming a vacuum diagram. The diagrams for
$\widetilde{D}(a,b)$ can instead be grouped 
into three different classes: (c1) all internal vertices 
connected to both $a$ and $b$ (c2) a subset of 
internal vertices connected only to $a$ and the complementary set 
connected only to $b$ and (c3) diagrams where a subset of internal vertices is 
not connected to either $a$ or $b$, thus forming
a vacuum diagram. In all cases the 
contributions containing vacuum diagrams factorize and 
cancel with the expansion of the partition function $\callZ$, see 
Eq.~(\ref{Zexppb}). Furthermore, 
many connected diagrams are topologically equivalent 
and it is therefore enough to consider only the topologically 
inequivalent diagrams. The number of 
topologically equivalent diagrams  
cancel the combinatorial factor $2^{k}k!\,(2p)!$ in Eqs.~(\ref{gabexppb}) 
and~(\ref{dabexppb}) and $2^{k}k!\,(2p+1)!$ in Eq.~(\ref{phiaexppb}). 
The proof of these statements goes along the same 
lines as the proof for only electrons and we refer to 
Refs.~\cite{svl-book,sakkinen_thesis} for more details. 
The resulting formulas for $G$, 
$\widetilde{D}$ and $\vf$ become 

\begin{widetext}
\begin{align}
G(a;b)&=\sum_{k,p=0}^{\iif}
i^{k+p}\int d1d1'...dkdk' v(1;1')...v(k;k')\int d\tilde{1}
d\bar{1}...d(\widetilde{2p})
d(\overline{2p})\, g(\bar{1};\tilde{1}^{\ast})...g(\overline{2p};\widetilde{2p^{\ast}})
\nn\\
&\times \left.
G_{0,2k+2p+1}(a,1,1',...,\bar{1},...;b,1^{+},1'^{+},...,\bar{1}^{+},...)
D_{0,p}(\tilde{1},...)\right|_{\substack{c\;\;\; \\ t.i.}},
\label{gabexppb2}
\end{align}
\begin{align}
\widetilde{D}(a,b)&=
\sum_{k,p=0}^{\iif}
i^{k+p}\int d1d1'...dkdk' v(1;1')...v(k;k') 
\int d\tilde{1}d\bar{1}... d(\widetilde{2p})
d(\overline{2p})
g(\bar{1};\tilde{1}^{\ast})...g(\overline{2p};\widetilde{2p}^{\ast})
\nn\\
&\times\left.G_{0,2k+2p}(1,1',...,\bar{1},...;1^{+},1'^{+},...,\bar{1}^{+},...)
D_{0,p+1}(a,b,\tilde{1},...)\right|_{\substack{c\;\;\; \\ t.i.}},
\label{dabexppb2}
\end{align}
and 
\begin{align}
\vf(a)&= \sum_{k,p=0}^{\iif}
i^{k+p+1}\int d1d1'...dkdk' v(1;1')...v(k;k') 
\int d\tilde{1}d\bar{1}... d(\widetilde{2p+1})
d(\overline{2p+1})
g(\bar{1};\tilde{1}^{\ast})...g(\overline{2p+1};\widetilde{2p+1}^{\ast})
\nn\\
&\times \left.G_{0,2k+2p+1}(1,1',\ldots,\bar{1},\ldots;1^{+},1'^{+},\ldots,\bar{1}^{+},\ldots)
D_{0,p+1}(a,\tilde{1},\ldots)\right|_{\substack{c\;\;\; \\ t.i.}},
\label{phiaexppb2}
\end{align}
\end{widetext}	
where the labels `` $c$ '' and `` $t.i.$ '' indicate
that when expanding $G_{0,n}$ 
in determinants and $D_{0,m}$ in  hafnians only 
connected and 
topologically inequivalent diagrams are retained. In particular 
the expansion Eq.~(\ref{dabexppb2}) for $\widetilde{D}$ contains all diagrams in 
classes (c1) and (c2).

\begin{figure*}[tbp]
    \centering
\fbox{\includegraphics[width=0.9\textwidth]{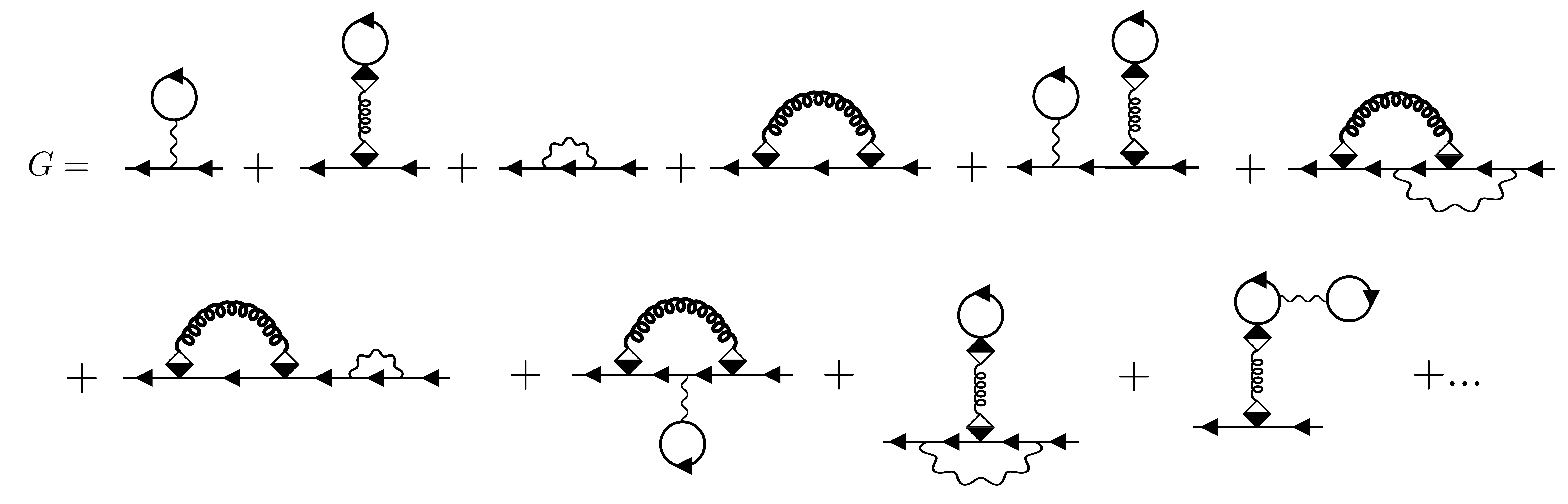}}
\caption{Low-order diagrams 
in the expansion of the interacting electronic Green's function $G$.}
\label{Gexpansionfig}
\end{figure*}

{\em Diagrammatic expansion for $G$.--}
In Fig.~\ref{Gexpansionfig} we show a few low-order Feynman diagrams
for $G$. The Feynman rules to convert 
the diagrams into a mathematical expression are

\begin{itemize}
    \item Number all vertices and assign an interaction  
    $v(i;j)$ to a wiggly line connecting $i$ and $j$, 
	an $e$-$ph$ coupling $g(i;j^{\ast})$ to a 
	square with white vertex in $j$ and black vertex $i$, 
	a GF $G_{0}^{s}(i;j^{+})$ to an oriented line from $j$ to 
	$i$, a GF $D_{0}(i,j)$ 
	to a spring connecting $i$ and $j$.
    \item
    Integrate over all internal vertices and multiply by 
    $i^{k+p}(-)^{l}$ where $l$ is the number of electronic
	loops, $k$ is the number of wiggly lines and $2p$ is the number of 
	squares.
\end{itemize}

There are diagrams (5-th and 7-th diagrams in the second row 
of Fig.~\ref{Gexpansionfig}) that are one-$G_{0}^{s}$-line reducible, 
i.e., they can be disconnected into two pieces by cutting an 
internal $G_{0}^{s}$-line. 
We define the irreducible 
self-energy\index{self-energy} $\S$ as the set of all 
one-$G_{0}^{s}$-line irreducible diagrams 
with the external (ingoing and outgoing) 
$G_{0}^{s}$-line removed.
Then $G$ can be written as a geometric series 
\begin{align}
G=G_{0}^{s}+G_{0}^{s}\S G_{0}^{s}+G_{0}^{s}\S G_{0}^{s}\S 
G_{0}^{s}+\ldots=
G_{0}^{s}+G_{0}^{s}\S G.
\label{dysongeb}
\end{align}
Each product in this formula stand for a space-spin-time convolution.
The self-energy $\S=\S[G_{0}^{s},D_{0},v,g]$
is an infinite sum of irreducible diagrams with $G_{0}^{s}$-lines and 
$D_{0}$-lines connected through $v$ and $g$. 

Among the self-energy diagrams there are some with
self-energy insertions, i.e., 
diagrams that can be disconnected into two pieces 
by cutting two $G_{0}^{s}$-lines. 
We say that a diagram is $G$-skeletonic if it does not contain 
self-energy insertions.
Then the full set of $\S$-diagrams is obtained 
by dressing the $G_{0}^{s}$-lines of the skeleton 
diagrams with all possible self-energy insertions. 
This amounts to evaluate the 
skeleton diagrams with the interacting GF $G$ instead of 
$G_{0}^{s}$~\cite{svl-book}. Denoting by $\S_{1\rm skel}$ the sum of only 
$G$-skeleton diagrams we can write
\begin{align}
\S=\S[G_{0}^{s},D_{0},v,g]=\S_{1\rm skel}[G,D_{0},v,g].
\end{align}

\begin{figure*}[t]
    \centering
\fbox{\includegraphics[width=0.9\textwidth]{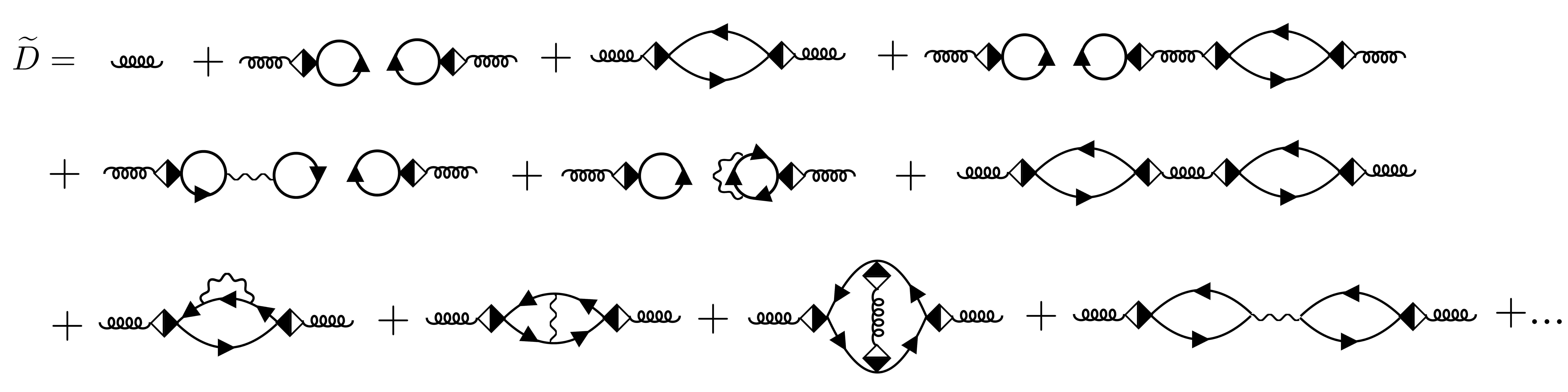}}
\caption{Low-order diagrams 
in the expansion of the interacting phononic Green's function
$\widetilde{D}$.}
\label{Dexpansionfig}
\end{figure*}

{\em Diagrammatic expansion for $\widetilde{D}$.--}
Let us now consider the one-particle phononic GF
$\widetilde{D}$. Expanding $G_{0,n}$ and $D_{0,m}$ in 
Eq.~(\ref{dabexppb2}) according to Wick's theorem and representing 
every term of the expansion with a diagram we obtain the  
diagrammatic expansion of $\widetilde{D}$.
The Feynman rules for the $\widetilde{D}$-diagrams are the 
same as for the $G$ diagrams. In Fig.~\ref{Dexpansionfig} we 
illustrate a few low-order diagrams.

\begin{figure*}[tbp]
    \centering
\fbox{\includegraphics[width=0.9\textwidth]{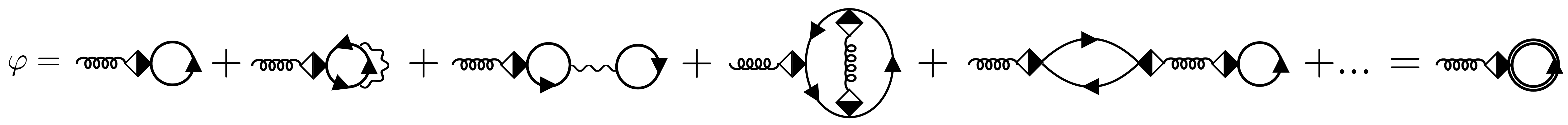}}
\caption{Resummation of the tadpole diagrams.}
\label{tadpoleexpansionfig}
\end{figure*}

We can clearly distinguish the diagrams belonging to class (c2). These are 
the double tadpole diagrams (see 2-nd, 4-th, 
5-th and 6-th diagrams of 
Fig.~\ref{Dexpansionfig}). 
The single tadpole diagrams constitute the diagrammatic 
expansion of the half-particle phononic GF 
$\widetilde{D}_{1/2}=\vf/i^{1/2}$. 
In Fig.~\ref{tadpoleexpansionfig} we show the diagrammatic expansion 
of $\vf$.  The full set of diagrams can be 
easily summed up, see r.h.s. in
Fig.~\ref{tadpoleexpansionfig} where the oriented double line
denotes the interacting GF $G$. 
The Feynman rules for the $\vf$-diagrams are the same as for the $G$ 
diagrams and $D$ diagrams except that the prefactor is 
$i^{k+p+1}(-)^{l+1}$ where $l$ is the number of electronic loops, 
$k$ is the number of wiggly lines and $2p+1$ is the number of 
squares, see Eq.~(\ref{phiaexppb2}).
We thus have
\begin{align}
 \vf(a)= i \int d\tilde{1}d\bar{1}
 D_{0}(a,\tilde{1})g(\bar{1};\tilde{1}^{\ast})
 G(\bar{1},\bar{1}^{+}),
\label{resumoftadpole}
\end{align}
a result that could alternatively be found by direct integration of 
the equation of motion Eq.~(\ref{eomqpartbosphi}) after
taking into account that the electronic density 
$n(1)=-iG(1;1^{+})$. We can therefore write 
the expansion of $\widetilde{D}$ as
\begin{align}
\widetilde{D}(a,b)&=\widetilde{D}_{1/2}(a)
\widetilde{D}_{1/2}(b)
\nn\\&+\left[D_{0}+
D_{0}\P D_{0}+D_{0}\P D_{0}\P D_{0}+...\right](a,b),
\end{align}
in which the products in this formula stand for 
momentum-mode-component-time convolutions.
The phononic self-energy
$\P=\P[G_{0}^{s},D_{0},v,g]$ is the set of all diagrams
that, after the removal of the ingoing and outgoing $D_{0}$-lines, 
cannot be cut into two pieces by cutting an internal $D_{0}$-line 
(one-$D_{0}$-line irreducible diagrams).

The phononic Green's function $\widetilde{D}(a,b)$ does 
{\em not} fulfill a Dyson 
equation but the fully connected phononic GF 
\begin{align}
D(a,b) &\equiv \widetilde{D}(a,b) - \widetilde{D}_{1/2}(a)
\widetilde{D}_{1/2}(b)
\nn\\&=\widetilde{D}(a,b)-
\frac{1}{i}\vf(a)\vf(b)
\end{align}
does.
The GF $D(a,b)$ can alternatively 
be written in terms of 
the fluctuation operators $\D\hat{\vf}(a)\equiv 
\hat{\vf}(a)-\vf(a)=
\hat{\f}(a)-\f(a)=\D\hat{\f}(a)$. We have 
\begin{align}
D(a,b) &=\frac{1}{i}\Tr\Big[
\hat{\r}\;\callT\Big\{\D\hat{\vf}_{H}(a)
\D\hat{\vf}_{H}(b)\Big\}\Big]
\nn\\&=\frac{1}{i}\Tr\Big[
\hat{\r}\;\callT\Big\{\D\hat{\f}_{H}(a)
\D\hat{\f}_{H}(b)\Big\}\Big].
\label{connectedDcdef}
\end{align}
The GF 
$D(a,b)$ fulfills a Dyson equation since
\begin{align}
D(a,b)&=\left[D_{0}+
D_{0}\P D_{0}+D_{0}\P D_{0}\P D_{0}+...\right](a,b)
\nn\\
&=D_{0}(a,b)+
(D_{0}\P
D)(a,b).
\label{dysonDeb}
\end{align}
Like for the electronic Green's function we can express the phononic 
self-energy in terms of $G$-skeleton diagrams.
We remove all diagrams with 
$\S$-insertions inside the $\P$-diagrams 
and then
replace $G_{0}^{s}$ by the full GF $G$, thus obtaining 
the functional $\P_{1\rm skel}$:
\begin{equation}
\P=\P[G_{0}^{s}, D_{0},v,g]=\P_{1\rm skel}[G, D_{0},v,g].
\end{equation}

The topological idea of the skeletonic expansion in $G$ 
is completely general and it can 
be extended to the phononic GF $D$ and 
the screened Coulomb interaction $W$. This is done in the next 
section.

\section{From the skeletonic expansion in $D$ and $W$ to the Hedin-Baym equations}
\label{hedinsec}

\begin{figure}[tbp]
    \centering
\fbox{\includegraphics[width=0.4\textwidth]{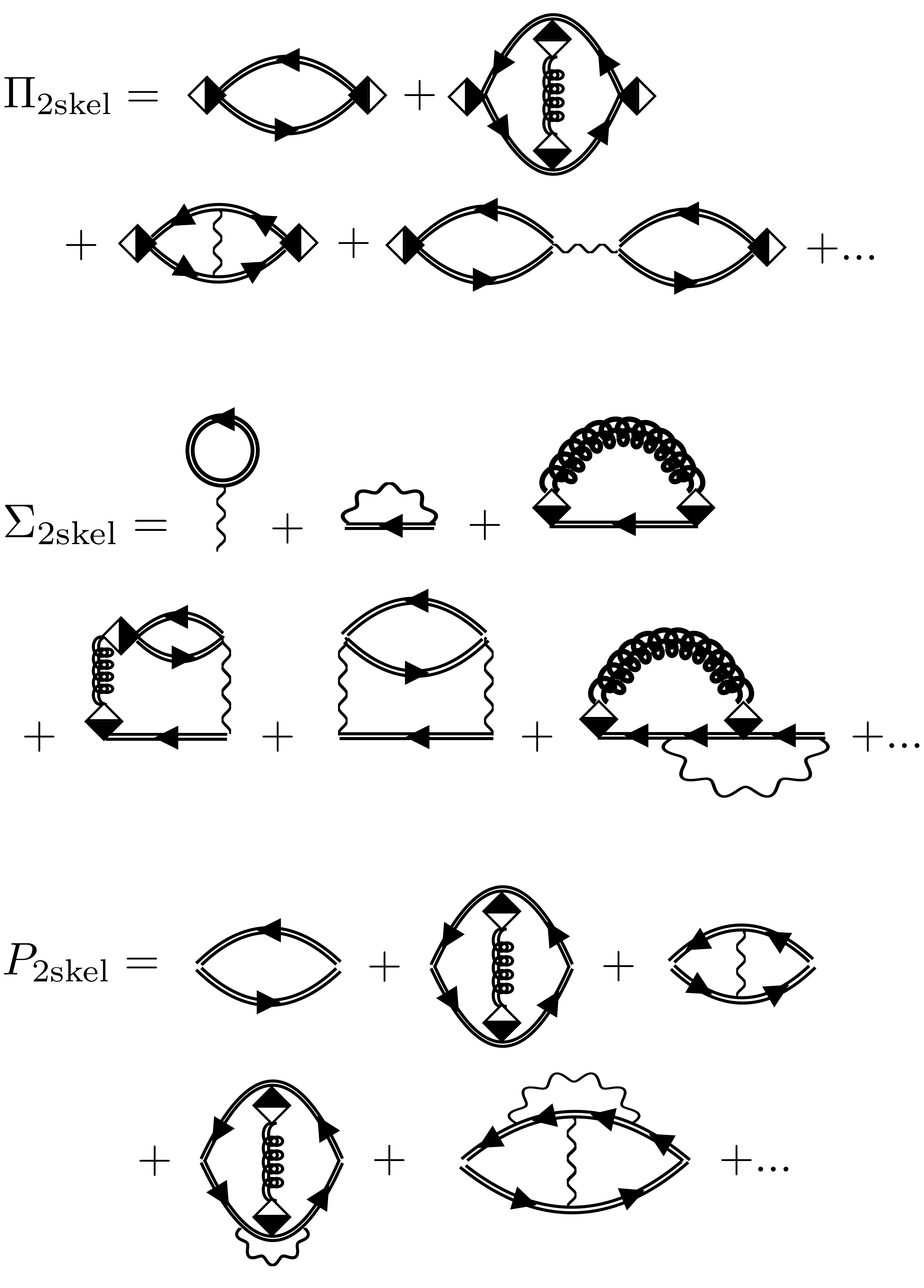}}
\caption{Expansion of the phononic (top) and electronic (middle) 
self-energies and of the polarization (bottom) in $G$-skeleton 
diagrams and $D$-skeleton diagrams.}
\label{se-expansionssfig}
\end{figure}

{\em Skeletonic expansion in $D$.--}
If we remove all phononic 
self-energy insertions inside the $\P_{1\rm skel}$-diagrams and then 
replace $D_{0}$ with $D$ we can write 
\begin{align}
\P=\P_{1\rm skel}[G,D_{0},v,g]=\P_{2\rm skel}[G,D,v,g] .
\end{align}
Here $\P_{2\rm skel}$ 
contains all doubly skeletonic self-energy diagrams, i.e., all 
those diagrams that do not 
contain either $\S$-insertions or $\P$-insertions.
Examples of $\P_{2\rm skel}$-diagrams are shown in 
Fig.~\ref{se-expansionssfig} (top) where the double spring represents 
$D$.
A similar procedure can be applied to the electronic self-energy  
except that for $\S$ we must exclude 
the time-local diagrams. In fact, 
a $\P$-insertion is here equivalent to a $\S$-insertion and it would 
therefore lead to a double 
counting. Therefore
\begin{align}
\S=\S_{1\rm skel}[G,D_{0},v,g]=\S_{\rm Eh}[G,D_{0},g]
+\S_{2\rm skel}[G,D,v,g],
\end{align}
where $\S_{\rm Eh}$ is the self-energy in the second diagram of 
Fig.~\ref{Gexpansionfig} with $G_{0}^{s}\to G$. Using the Feynman rules 
\begin{align}
\S_{\rm Eh}(1;2)&=-i\d(1;2)\!\int\!\! d\tilde{1}
d\tilde{2}d\bar{1}\,g(1;\tilde{1}^{\ast})
\nn\\&\quad\quad\quad\quad\quad\times D_{0}(\tilde{1},\tilde{2})
g(\bar{1};\tilde{2}^{\ast})G(\bar{1};\bar{1}^{+})
\nn\\
&=-\d(1;2)\int 
d\tilde{1}\,g(1;\tilde{1}^{\ast})\vf(\tilde{1}),
\label{EhrenfestSigma}
\end{align}
where in the last equality we use Eq.~(\ref{resumoftadpole}).
The self-energy $\S_{\rm Eh}$ is known as the {\em Ehrenfest
self-energy}.
The Ehrenfest approximation consists in including the phononic 
feedback on the electrons through $\S_{\rm Eh}$, the electronic 
feedback on the phonons through the density, see 
Eq.~(\ref{resumoftadpole}), and in setting $\P=0$.
In this approximation 
the nuclei are therefore treated as classical particles since they are 
described in terms of displacements and momenta only, i.e., the 
components of 
$\vf$~\cite{horsfield_open-boundary_2004,horsfield_beyond_2004,li_ab-initio_2005,verdozzi_classical_2006,galperin_molecular_2007,galperin_the-non-linear_2008,dundas_current-driven_2009,hussein_semiclassical_2010,lu_blowing_2010,bode_scattering_2011,hubener_phonon_2018}. 
The importance of the Ehrenfest diagram in the description of 
polarons has been pointed 
out in 
Refs.~\cite{galperin_the-non-linear_2008,lafuente_ab-initio_2022}.
We expect that the Ehrenfest diagram is also crucial to capture 
the phonon-induced coherent modulation of 
the excitonic resonances~\cite{trovatello_strongly_2020}. 
Examples of $\S_{2\rm skel}$ diagrams are shown in 
Fig.~\ref{se-expansionssfig} (middle).

{\em Skeletonic expansion in $W$.--}
Like in the case of only electrons~\cite{hedin_new_1965} we can further reduce the number 
of diagrams by removing all those diagrams containing a polarization 
insertion, which we here define as a piece that can be cut away by cutting two 
$v$-lines {\em and at the same time} it does not break into two disjoint 
pieces by cutting one $v$-line or one $D$-line, an example is the 5-th diagram in 
the $\S$-expansion of
Fig.~\ref{se-expansionssfig} (middle). 
The polarization diagrams are therefore one-$v$-line irreducible {\em and} 
one-$D$-line irreducible. 
To the best of our knowledge the diagrammatic definition of the 
polarization $P$ in systems of electrons and phonons is given here for 
the first time.
In Fig.~\ref{se-expansionssfig} (bottom) we show a few low-order diagrams 
for  $P=
P_{2\rm skel}[G,D,v,g]$ which are 
both $G$-skeletonic and $D$-skeletonic. The polarization 
$P$  contains both electronic and phononic contributions.
For later purposes we also define the one-$v$-line irreducible,
one-$D$-line irreducible and two-$G$-lines reducible
kernel $\tilde{K}_{r}$ from the polarization according to
\begin{align}
    \raisebox{-6pt}{\includegraphics[width=0.42\textwidth]{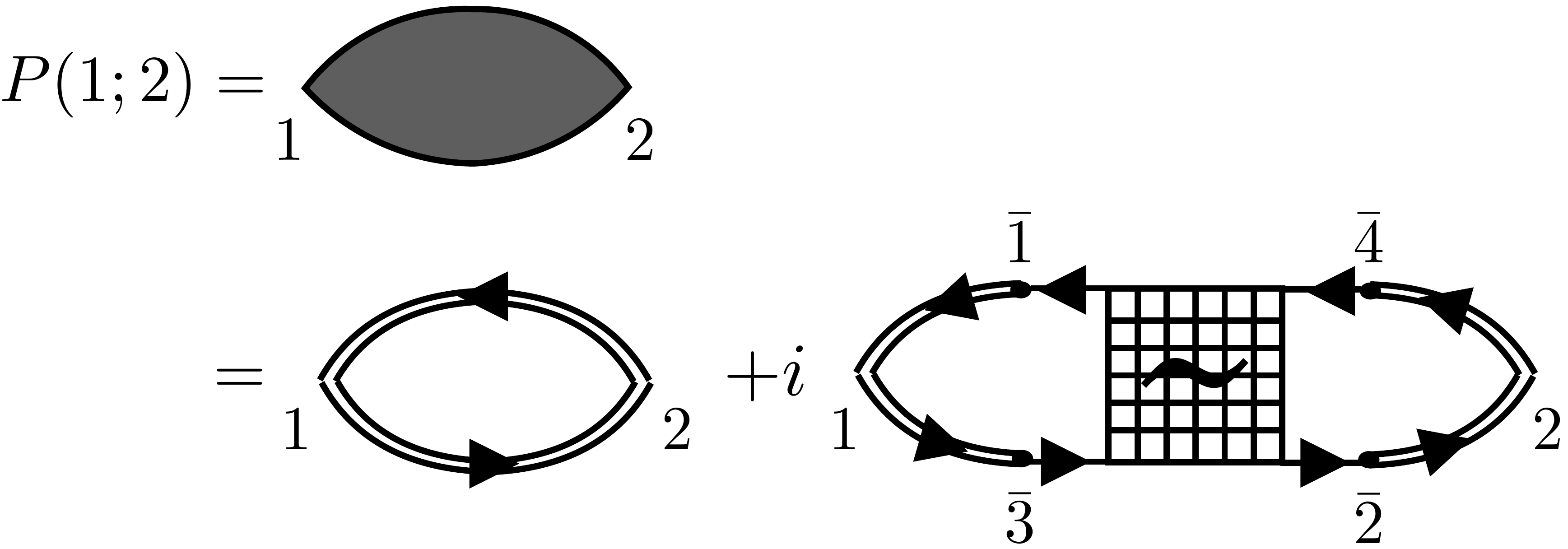}} \;\;,
    \label{ptildektilde}
\end{align}
where the dark-grey bubble represents $P$ and 
the square grid represents  
$\tilde{K}_{r}(\bar{1},\bar{2};\bar{3},\bar{4})
=\tilde{K}_{r}(\bar{2},\bar{1};\bar{4},\bar{3})$.
This is the same definition used in the case of systems of only 
electrons. In fact, the so called vertex function $\G=\d-\tilde{K}_{r}GG$ 
relates to $P$ through the well known formula $P=-iGG\G$ (the  
factor of `` $i$ '' in the second diagram of Eq.~(\ref{ptildektilde}) 
comes from the Feynman rules for the polarization diagrams). 
Following the same strategy as in Ref.~\cite{svl-book} one can show 
that the kernel $\tilde{K}_{r}$ satisfies the Bethe-Salpeter equation 
\begin{align}
    \raisebox{-6pt}{\includegraphics[width=0.45\textwidth]{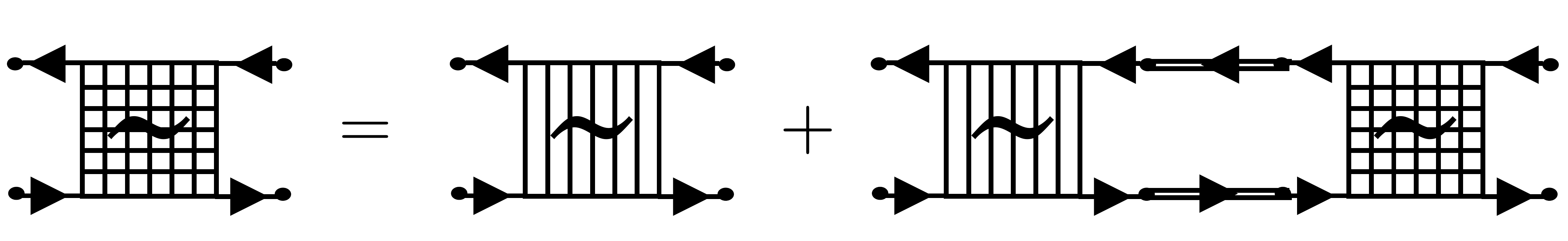}} \;\;,
    \label{BSE-kernel}
\end{align}
where the kernel $\tilde{K}$, represented by the square with only vertical 
lines, is one-$v$-line irreducible,
one-$D$-line irreducible and two-$G$-lines irreducible.

From the polarization diagrams we can construct the 
dynamically screened interaction in the usual manner 
\begin{align}
    \raisebox{-6pt}{\includegraphics[width=0.4\textwidth]{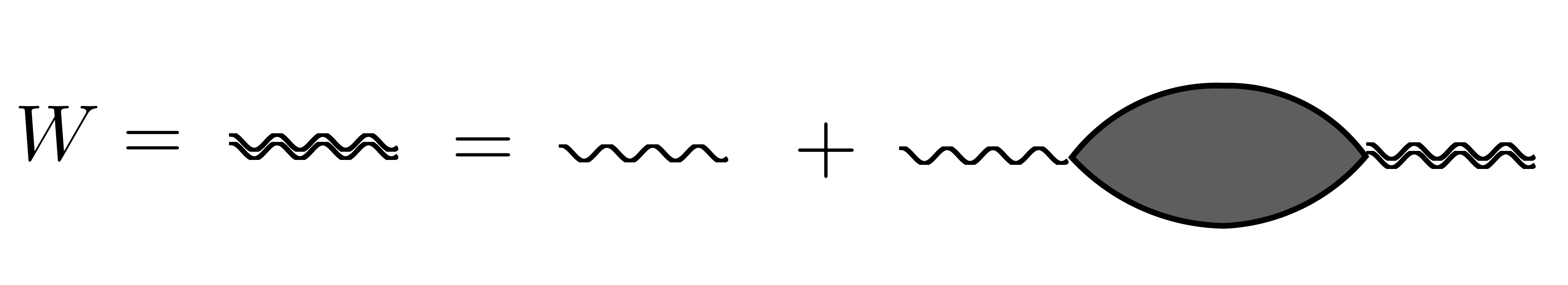}} \;\;.
    \label{screenedW}
\end{align}
We say that a 
diagram is $W$-skeletonic if it does not contain  
$P$-insertions.\index{diagrams!self-energy!skeleton in $W$}
\index{diagrams!displacement self-energy!skeleton in $W$}
Then the desired expression for $\P$ and $\S$ is obtained by discarding all 
those diagrams which are not $W$-skeletonic, 
and then replacing $v$ with $W$. 

\begin{figure*}[tbp]
    \centering
\fbox{\includegraphics[width=0.9\textwidth]{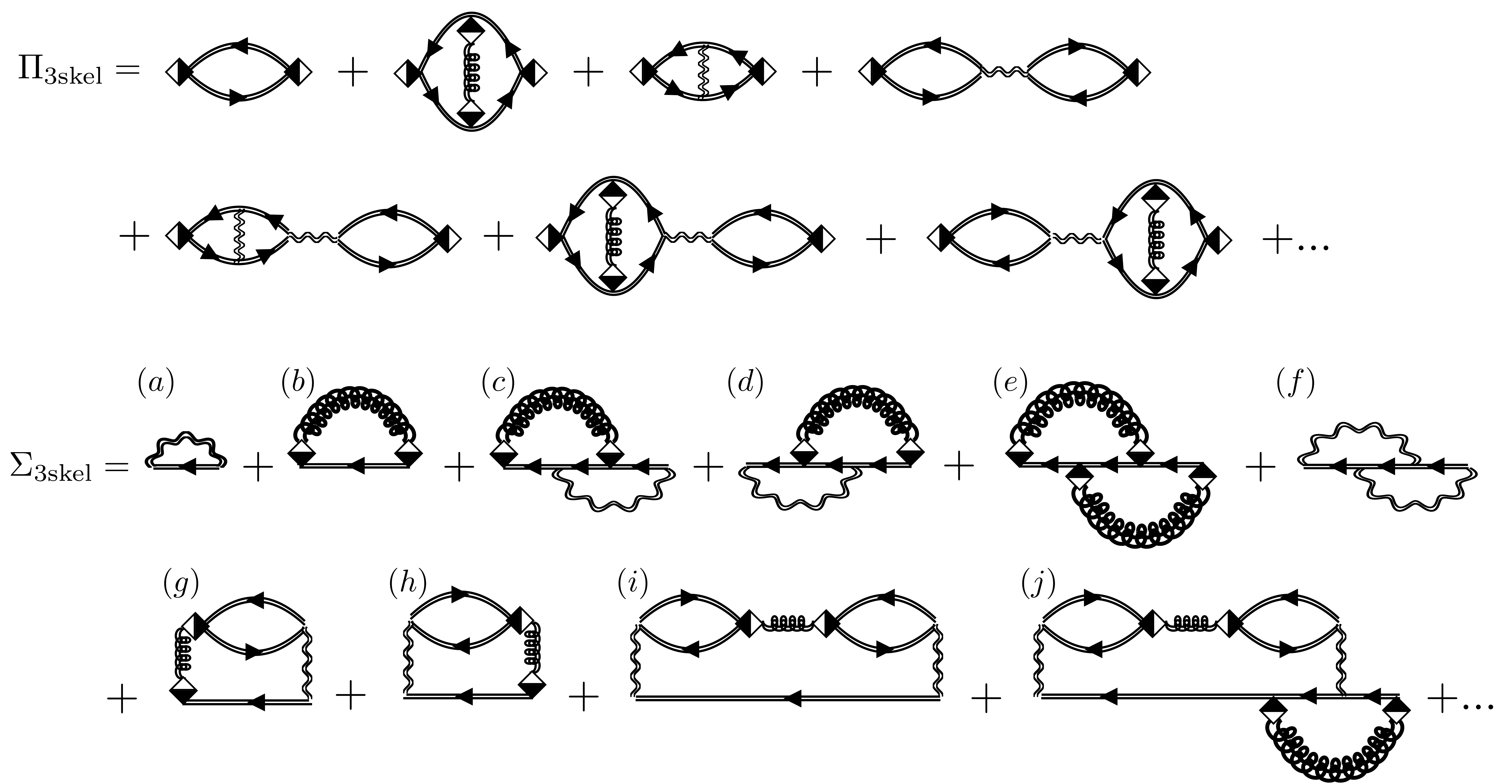}}
\caption{Expansion of the phononic (top) and electronic (bottom)
self-energy in $G$-skeleton diagrams, $D$-skeleton diagrams and 
$W$-skeleton diagrams. To facilitate the discussion in the main text 
we have labelled the $\S$-diagrams.}
\label{selfexpansionsss}
\end{figure*}

{\em  Phononic self-energy.--}
For the phononic 
self-energy we  get
\begin{align}
\P=\P_{2\rm skel}[G,D,v,g]=\P_{3\rm skel}[G,D,W,g],
\end{align}
where $\P_{3\rm skel}$ is the sum of all the triply skeletonic self-energy 
diagrams, i.e., all those diagrams that do not contain either 
$\S$-insertions, $\P$-insertions or $P$-insertions. 
In Fig.~\ref{selfexpansionsss} (top) we show a few low-order diagrams of the triply 
skeletonic expansion for $\P_{3\rm skel}$.
They are one-$G$-line irreducible, one-$D$-line irreducible and 
two-$W$-lines irreducible, i.e., they 
cannot break into disjoint pieces by cutting two  
$W$-lines for otherwise they would contain a polarization 
insertion. We then have either one-$W$-line 
reducible diagrams or one-$W$-line irreducible diagrams.
We can write $\P$ in a compact form using the polarization $P$. 
We define the {\em dressed} (or {\em screened}) 
$e$-$ph$ coupling $g^{d}$ as
\begin{align}
    \raisebox{-10pt}{\includegraphics[width=0.32\textwidth]{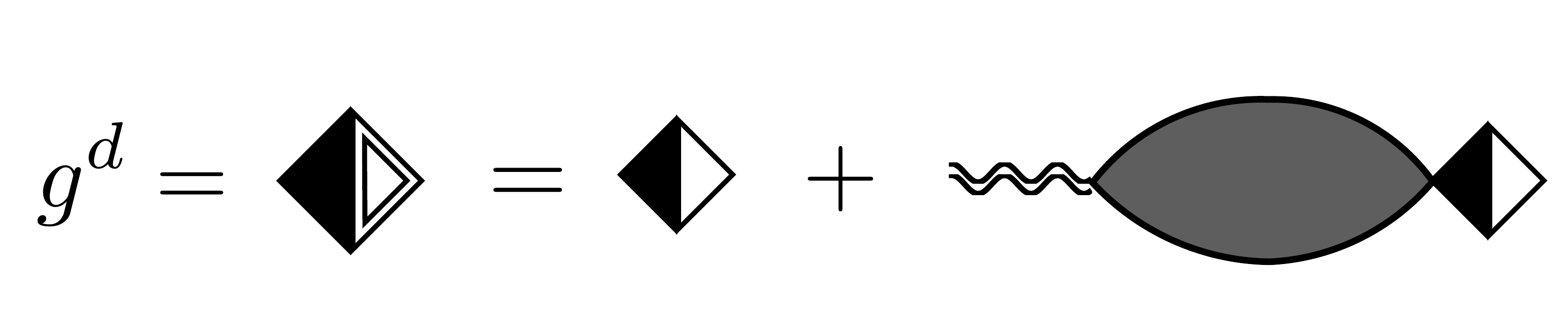}} 
    \nn\\ =g+WPg\;\;,\quad\quad\quad\quad\quad\;
	\label{dressedvertexeq}
\end{align}
In terms of the dressed $e$-$ph$ coupling the phononic self-energy can be 
represented as
\begin{align}
    \raisebox{-6pt}{\includegraphics[width=0.2\textwidth]{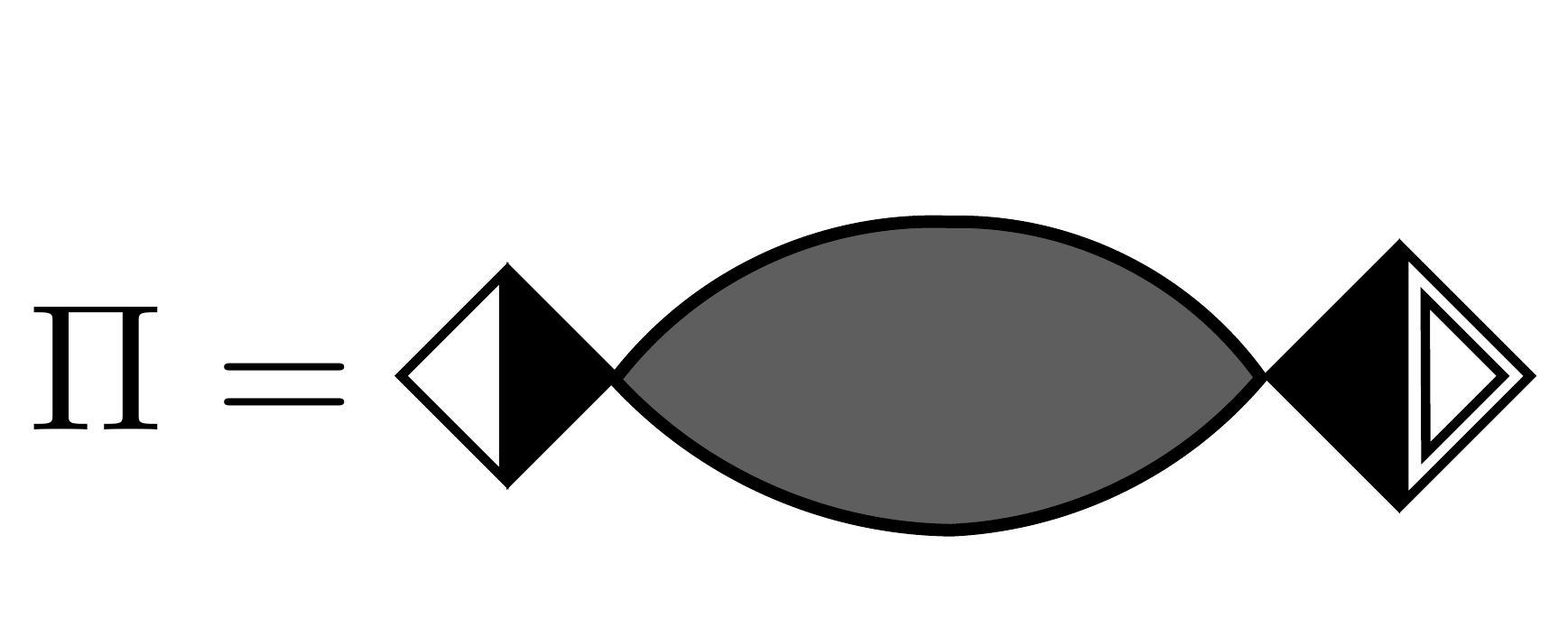}} \;\;,
    \nn\\ =gPg^{d}\;\;,\quad\quad\quad\quad\quad\quad
	\label{Pi=gPgd}
\end{align}
in agreement with the field-theoretic approach~\cite{feliciano_electron-phonon_2017}.
We see from this expression that one $e$-$ph$ coupling 
is bare whereas the other is dressed. As pointed out in 
Refs.~\cite{feliciano_electron-phonon_2017,marini_equilibrium_2023}
this structure has to be properly taken into account for the 
calculation of phonons, although a symmetrized form where both $e$-$ph$ 
couplings are dressed has been shown to give satisfactory results~\cite{calandra_adiabatic_2010}, 
which has been attributed to error 
cancellations~\cite{berges_phonon_2023}. 

{\em Electronic self-energy.--}
For the electronic self-energy $\S$ the only diagram for which 
we should not proceed with the 
replacement is the Hartree diagram [first diagram in  
Fig.~\ref{se-expansionssfig} (middle)] 
since here every polarization insertion is equivalent 
to a self-energy insertion, hence  $v\ra W$ would lead to 
a double counting. Therefore
\begin{align}
\S&=\S_{\rm Eh}[G,D_{0},g]+\S_{2\rm skel}[G,D,v,g]
\nn\\&=\S_{\rm Eh}[G,D_{0},g]+\S_{\rm H}[G,v]+
\S_{3\rm skel}[G,D,W,g],
\label{sigmasss}
\end{align}
where $\S_{\rm H}$ is the Hartree diagram.
The self-energy $\S_{3\rm skel}$ is also called the 
exchange-correlation (xc) self-energy. 
Henceforth we shall 
equivalently write $\S_{3\rm skel}$ or $\S_{\rm xc}$.
The diagrammatic expansion of the xc self-energy plays a crucial role 
in diagrammatic theory since the Bethe-Salpeter kernel 
$\tilde{K}(1,2;3,4)=-\d\S_{\rm xc}(1;3)/\d G(4;2)$, see 
Eq.~(\ref{BSE-kernel}). This statement can be proven along the same 
lines as in Ref.~\cite{svl-book}.
We illustrate in Fig.~\ref{selfexpansionsss} (bottom)  
a few low-order diagrams 
of the triply skeletonic expansion of $\S_{3\rm skel}$.
Diagrams like the last two in the figure  must be included since 
by cutting the two $W$-lines we get a piece that 
is neither a polarization 
insertion nor a $\P$-insertion.

Like the phononic self-energy also the electronic self-energy $\S$ 
can be written in 
a compact form using the polarization $P$, or more precisely the kernel
$\tilde{K}_{r}$. Let us consider the 
admissible ``effective'' interactions that can 
sprout from, e.g., the left vertex of $\S_{\rm xc}$. 
Keeping an eye on Fig.~\ref{selfexpansionsss} (bottom) 
we realize that we can have $W$, see diagrams $(a)$, $(d)$ and $(f)$,
and $[gDg]$, see diagrams $(b)$, $(c)$ and $(e)$.
We can also have $WP[gDg]$, see 
diagram $(h)$, but we cannot have 
$WPW$ since the corresponding diagram would contain a $P$-insertion. 
We can further have $[gDg]PW$, see  
diagram $(g)$, but we cannot have 
$[gDg]P[gDg]$ since 
the corresponding diagram would contain a  
$\P$-insertion. We can finally have 
$WP[gDg]PW$, see
diagrams $(i)$ and $(j)$. 
All other structures are non-skeletonic: 
$WP[gDg]P[gDg]$
and $[gDg]PWP[gDg]$ 
contain a $\P$-insertion whereas 
$[gDg]PWPW$ contains a polarization 
insertion. We conclude that the total ``effective'' interaction 
sprouting from the left vertex is 
\begin{align}
\widetilde{W}&=W\!+gDg+W\!PgDg+gDgPW
+W\!PgDgPW
\nn\\
&=W+ (g+ WPg)
D(g+gPW)
\nn\\
&=W+ g^{d}Dg^{d}.
\end{align}
To make these graphical considerations rigorous we define the 
phonon-mediated $e$-$e$ interaction from the  
dressed $e$-$ph$ coupling according to
\begin{align}
    \raisebox{-6pt}{\includegraphics[width=0.3\textwidth]{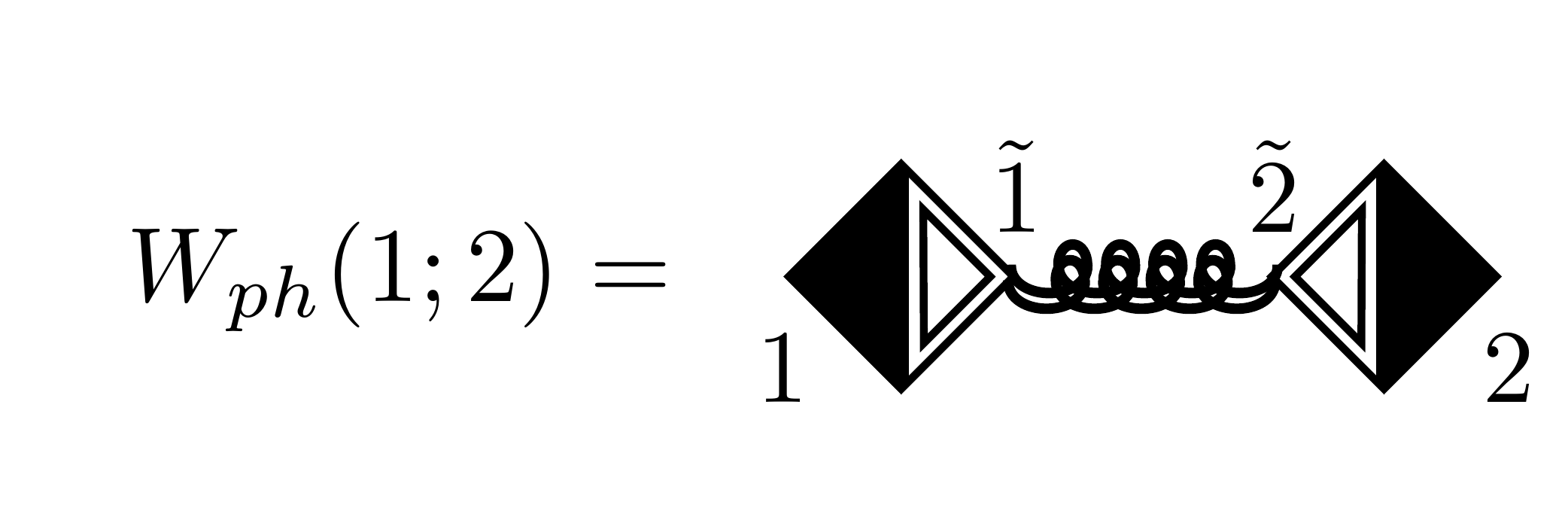}} \;\;,
    \label{Wdispl}
\end{align}
or in formulas
\begin{align}
W_{ph}(1;2)=\int d\tilde{1}d\tilde{2}\,
g^{d}(1;\tilde{1}^{\ast})D(\tilde{1},\tilde{2})
g^{d}(2;\tilde{2}^{\ast}),
\end{align}
and the total screened interaction $\widetilde{W}\equiv W+ W_{ph}$ according to
\begin{align}
    \raisebox{-6pt}{\includegraphics[width=0.45\textwidth]{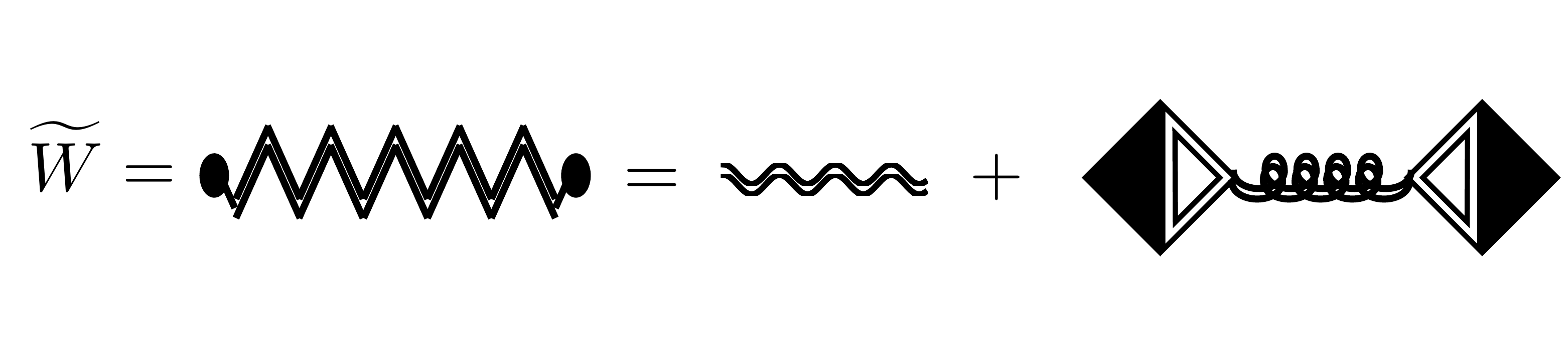}} \;\;,
    \label{Wtot}
\end{align}
or in formulas
\begin{align}
\widetilde{W}(1;2)=W(1;2)+ \,W_{ph}(1;2).
\end{align}
The total electronic self-energy can then be written as
\begin{align}
    \raisebox{-6pt}{\includegraphics[width=0.45\textwidth]{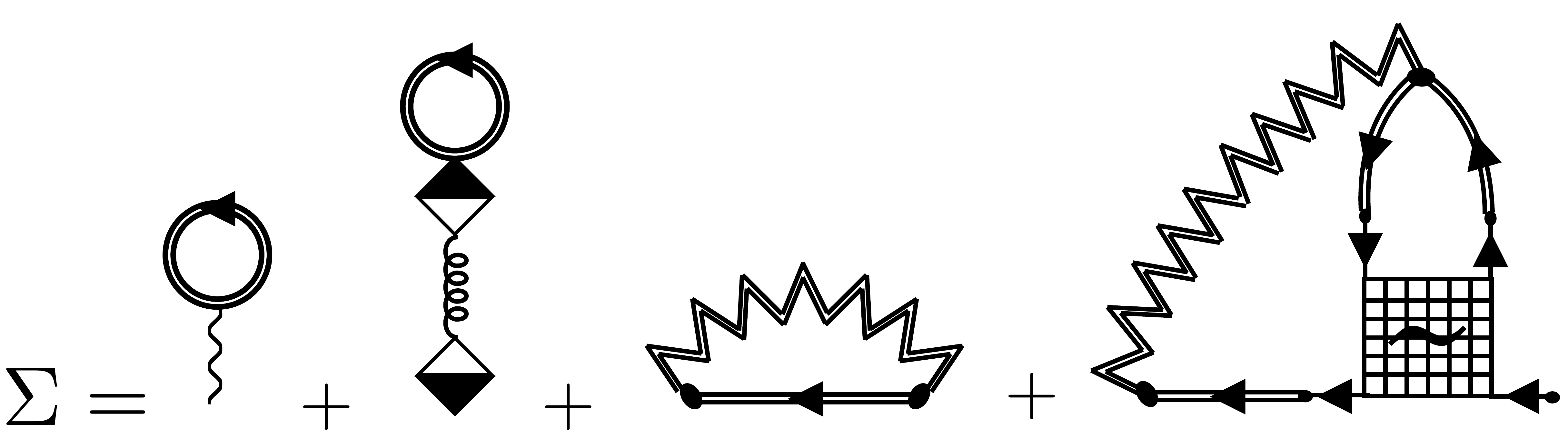}} \;\;.
    \label{SigmawithWtotandGamma}
\end{align}

\begin{table*}[tbp]
    \centering
    \begin{tabular}{|c|c|}
	\hline
	& \\
	$\begin{array}{c}\vspace{-1.2cm}\\G=G^{s}_{0}+G^{s}_{0}\S G  \end{array}$ & 
	\includegraphics[width=8.cm]{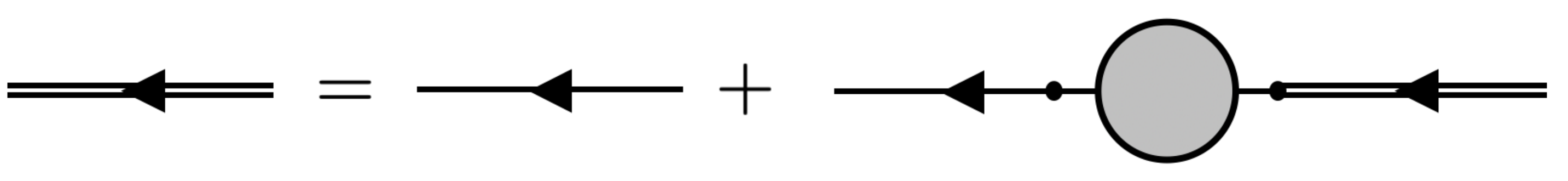} 
	\\\hline
	$\begin{array}{c}\vspace{-1.4cm} \\D=D_{0}+D_{0}\P D \end{array}$ & 
	\includegraphics[width=6.5cm]{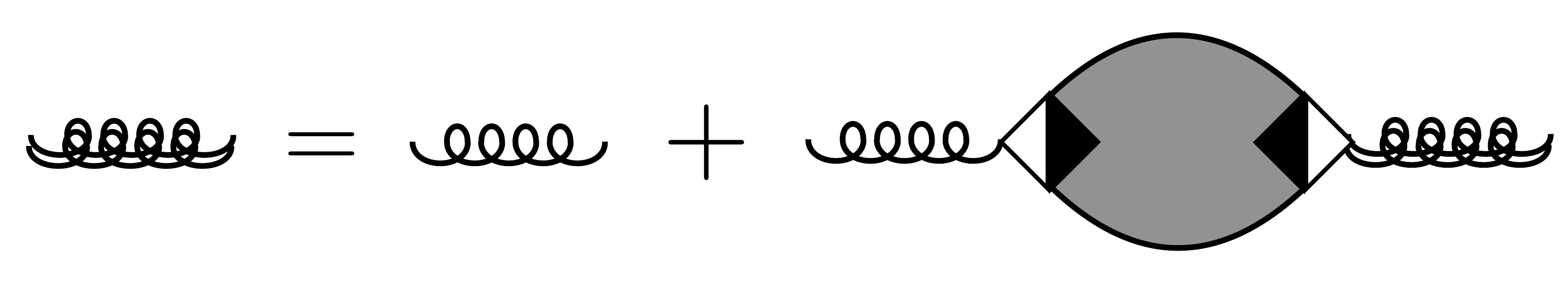} \\ 
	\hline
	$\begin{array}{l}\vspace{-1.cm}\\\S=\S_{\rm Eh}+\S_{\rm H}
	                 +i G\widetilde{W}(\d-GG\tilde{K}_{r}) \\ \end{array}$ & 
	\includegraphics[width=10.cm]{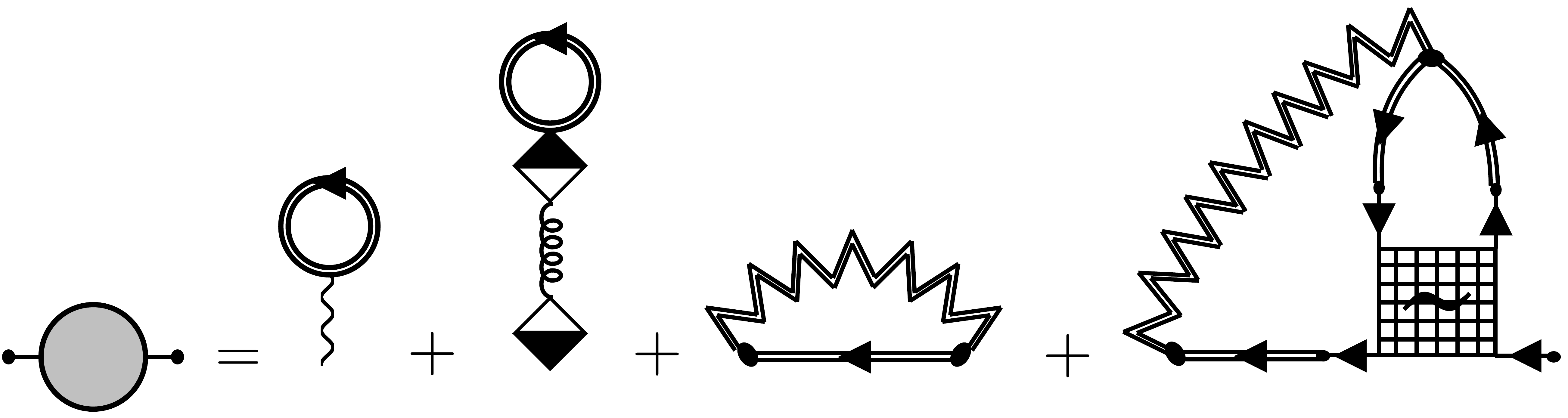} \\\hline
	$\begin{array}{c}\vspace{-1.4cm} \\ \P=gPg^{d} \end{array}$ & 
	\includegraphics[width=5cm]{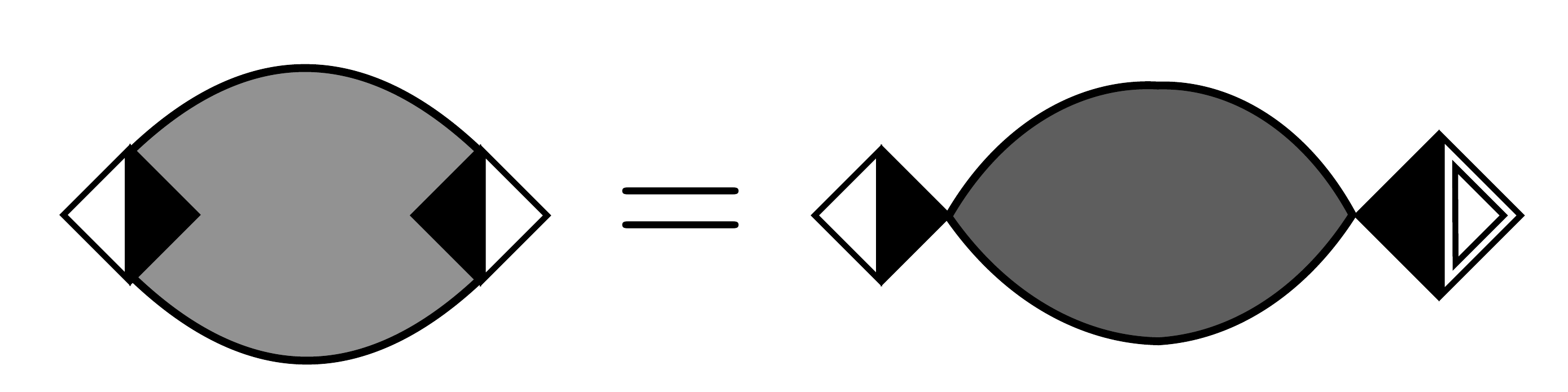} \\\hline
	$\begin{array}{c}\vspace{-1.6cm} \\ \widetilde{W}\equiv W+ 
	g^{d}Dg^{d} \end{array}$ & 
	\includegraphics[width=6cm]{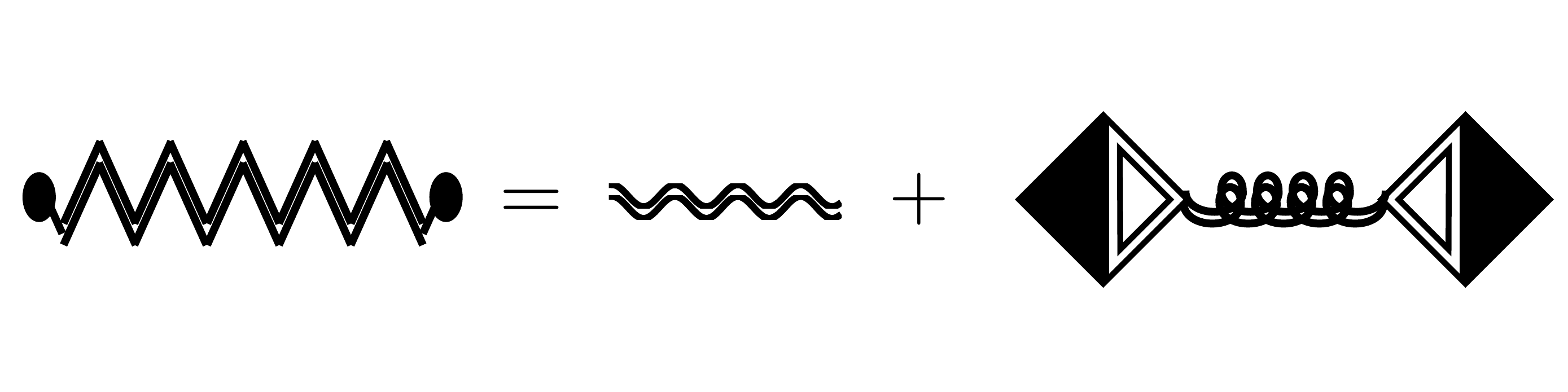} \\\hline
	$\begin{array}{c}\vspace{-1.6cm} \\ W=(\d+WP)v \end{array}$ & 
	\includegraphics[width=7cm]{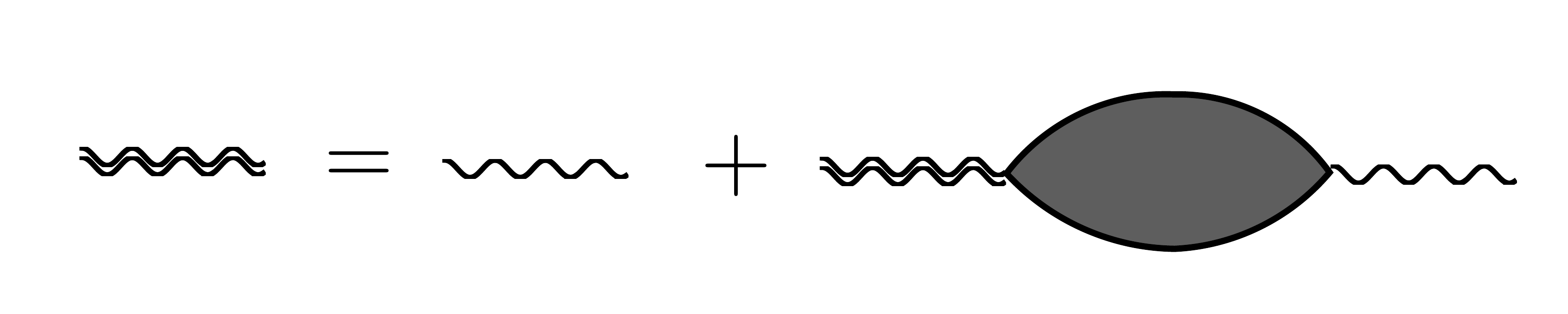} \\\hline
	$\begin{array}{c}\vspace{-1.3cm} \\ P=-i GG(\d-\tilde{K}_{r}GG)  \end{array}$ & 
	\includegraphics[width=7cm]{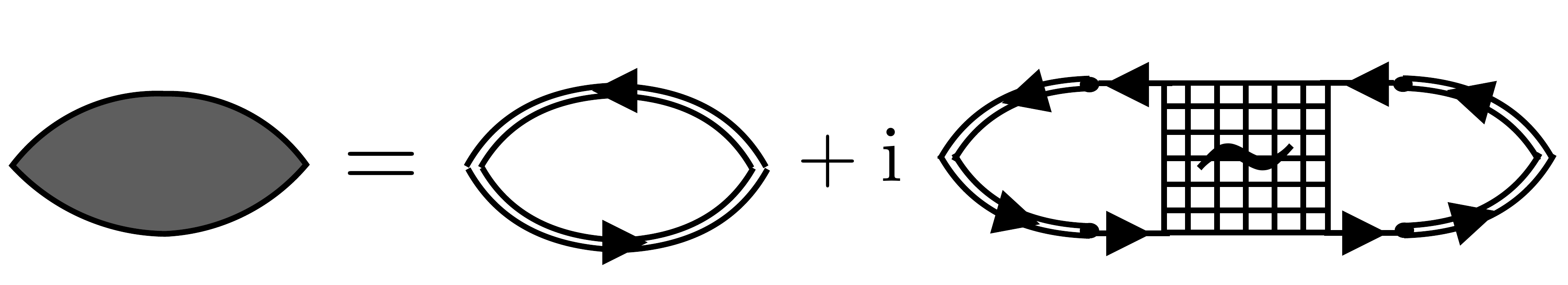} 
	\\\hline
	$\begin{array}{c}\vspace{-1.7cm} \\g^{d}=(\d+WP)g \end{array}$ & 
	\includegraphics[width=7cm]{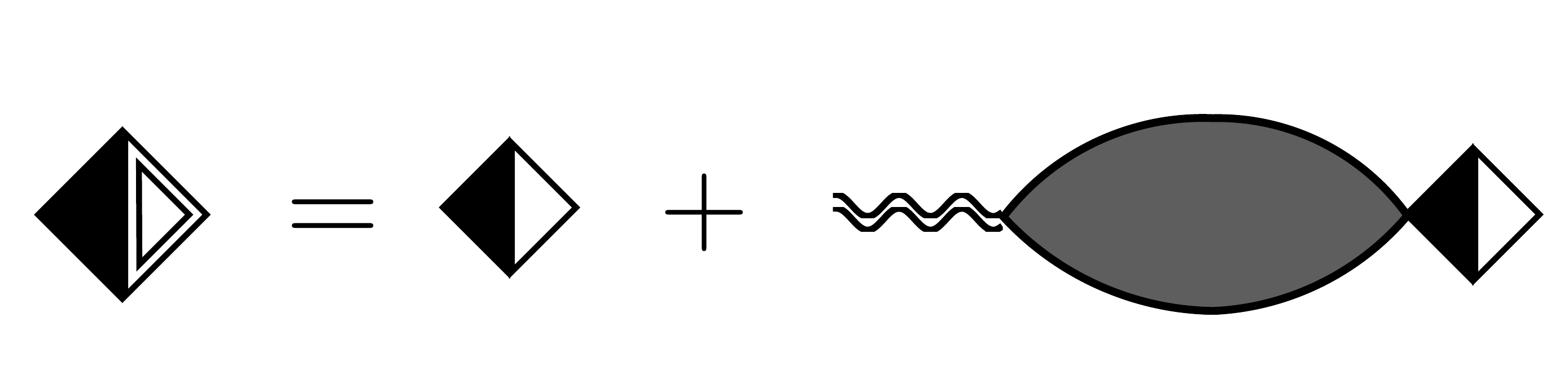} 
	\\\hline
	$\begin{array}{c}\vspace{0.cm}\\ \tilde{K}_{r}=-\frac{\mbox{$\d\S_{\rm 
	xc}$}}{\mbox{$\d G$}}
	(\d- GG\tilde{K}_{r}) \end{array}$ & $\begin{array}{c} 
	\vspace{0.1cm}\\
	\includegraphics[width=7.cm]{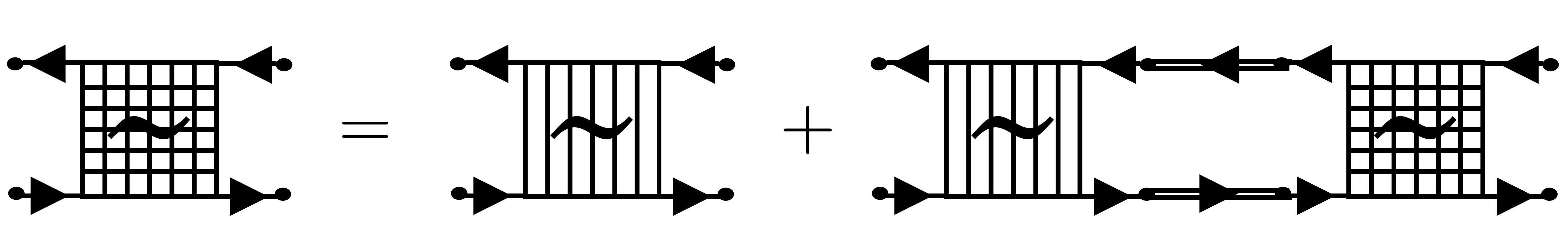} \end{array}$\\
	& \\
	\hline
    \end{tabular}
    \caption{Mathematical expression (left column) and diagrammatic 
    representation (right column) of the Hedin-Baym equations for systems 
	of interacting electrons and phonons.}
    \label{Hedinebtab}
\end{table*}

{\em Hedin-Baym equations.--} We summarize in Table~\ref{Hedinebtab} 
the fundamental equations that relate the various many-body 
quantities, i.e., $G$, $D$, $\S$, $\P$, $\widetilde{W}$, $W$, 
$P$, $g^{d}$ and $\tilde{K}_{r}$. 
We here align with Ref.~\cite{feliciano_electron-phonon_2017} and 
call to the full set of equations in Table~\ref{Hedinebtab}
the {\em Hedin-Baym equations}
for electrons and phonons.
The Hedin-Baym equations provide a closed system of equations 
for any diagrammatic approximation to the xc self-energy through the irreducible kernel 
$\tilde{K}=\d\S_{\rm xc}/\d G$.  They are equations 
on the contour and can therefore be used to study systems in 
equilibrium at any temperature as well as systems driven away from 
equilibrium by external fields. For systems in equilibrium at zero 
temperature the {\em adiabatic} assumption in conjunction with the 
assumption of a {\em nondegenerate ground state} allows for deforming the 
contour into a single branch going from $-\iif$ to $+\iif$, i.e., 
the real axis~\cite{svl-book}. In this case the contour Green's 
functions become the more familiar time-ordered Green's functions 
and the Hedin-Baym equations 
reduce to those presented in 
Ref.~\cite{feliciano_electron-phonon_2017}. We emphasize that no such 
shortcut is possible at finite temperature. 
One way to avoid the use of the $L$-shaped contour for equilibrium systems at 
finite temperature is the analytic continuation (from Matsubara to 
retarded to time-ordered) which may however be 
rather cumbersome in the presence of singularities or branch cuts, 
although notable progresses have been recently 
made~\cite{fei_nevanlinna_2021,fei_analytical_2021,kosuke_bosonic_2023}.

Like for the Hedin equations for only electrons the Hedin-Baym 
equations can be 
iterated to obtain an expansion of $\S$ and $\P$ in terms of $G$, 
$W$, $D$ and $g$. If 
we start with $\tilde{K}_{r}=0$, hence 
\begin{align}
P\simeq \chi^{0}=-iGG,
\end{align}
the electronic self-energy $\S$ 
is approximated by
\begin{align}
\S=\S_{\rm Eh}+\S_{\rm H}+\S_{\rm GW}+\S_{\rm FM},
\label{GW+FM}
\end{align}
where $\S_{\rm GW}=i GW$ is the well known  $GW$ 
self-energy with RPA screened interaction 
$W=v+v\chi^{0}W$\index{screened interaction!RPA 
approximation}, and 
\begin{align}\S_{\rm 
FM}=i g^{d}GDg^{d}
\end{align}
is the so called {\em Fan-Migdal 
self-energy}~\cite{fan_temperature_1951,migdal_interaction_1958}
with dressed 
electron-phonon coupling 
\begin{align}
g^{d}=(\d+ W\chi^{0})g.
\end{align}
The phononic self-energy for $\tilde{K}_{r}=0$ 
is simply 
\begin{align}
\P=g\,\chi^{0}\,g^{d}=
g(\chi^{0}+\chi^{0}W\chi^{0})g.
\label{aGGad}
\end{align}
We remark that the response function $\chi^{0}$ appearing in $W$ and 
$\P$ cannot, in general, be built with a quasi-particle GF $G$
since both $\S_{\rm GW}$ and $\S_{\rm FM}$ are 
non-local in time. Although the time-local Coulomb-hole plus Screened 
Exchange (COH-SEX) version of $\S_{\rm GW}$ often provides a good 
compromise in the trade-off between accuracy and computational cost 
we are not aware of a similar time-local version of $\S_{\rm FM}$. 
In Section~\ref{kbe-p+bsec} we show that 
the approximation in Eqs.~(\ref{GW+FM}) and (\ref{aGGad}) is
conserving, i.e., the resulting GF satisfy all fundamental conservation laws.

Inserting an approximation for the self-energies $\S$ and $\P$ into
the Dyson equations $G=G_{0}^{s}+G_{0}^{s}\S G$ and 
$D=D_{0}+D_{0}\P D$ we obtain a closed system of equations 
for $G$ and $D$ for any  $G_{0}^{s}$ and $D_{0}$.
The GF $D_{0}$ depends only on the parameters of the 
Hamiltonian, see also Appendix~\ref{nonintDsec}, whereas the GF $G_{0}^{s}$ depends also on the nuclear displacements 
through $h^{s}$. The Hedin-Baym equations must therefore be 
coupled to Eq.~(\ref{resumoftadpole}).

The Hedin-Baym equations can alternatively be derived using the 
field-theoretic 
approach~\cite{feliciano_electron-phonon_2017,vanleeuwen_first-principles_2004,melo_unified_2016,harkonen_many-body_2020,marini_equilibrium_2023}.
We emphasize that the field-theoretic 
approach prescinds from any diagrammatic notion, i.e., 
it does not tell us
how to expand the various many-body quantities diagrammatically.

\section{Equations of motion for the Green's functions}
\label{kbe-p+bsec}

An alternative route to solving the Hedin-Baym 
equations is the solution of the equations of motion for the GF.
This second route is more convenient for systems at 
finite temperature or out of equilibrium. 
We here consider the self-energies as functionals of $G$ 
and $D$ (doubly skeletonic expansion). Using the equation of 
motion for $D_{0}$, see Eq.~(\ref{nintmshd1}), 
and for $G_{0}^{s}$, see  Eq.~(\ref{eomG0}),
we can convert the Dyson equations 
Eqs.~(\ref{dysongeb}) and~(\ref{dysonDeb}) into integro-differential 
equations on the contour
\begin{subequations}
\begin{align}
\int d\bar{1}\;\overrightarrow{G}_{0}^{-1}(1;\bar{1})G(\bar{1};2)&=
\d(1;2)+\int d\bar{1}\,\S(1;\bar{1})G(\bar{1};2),
\label{eomGongamma}\\
\int 
d\tilde{1}\;\overrightarrow{D}_{0}^{-1}(1,\tilde{1})D(\tilde{1},2^{\ast})&=
\d(1,2)+\int 
d\tilde{1}\,\P(1^{\ast},\tilde{1})D(\tilde{1},2^{\ast}).
\label{eomDongamma}
\end{align}
\end{subequations}

{\em Equation of motion for $G$.--}
We separate 
the electronic self-energy into a time-local (singular) contribution 
$\S^{\d}(1;2)\propto \d(z_{1},z_{2})$ and a rest 
$\S_{\rm c}(1;2)$
which is called the {\em correlation 
self-energy}:
\begin{align}
\S=\S^{\d}+\S_{\rm c}.
\label{S=Sd+Sc}
\end{align}
The singular contribution is given by the sum of the Hartree-Fock (HF) self-energy 
$\S_{\rm HF}(1;2)=\d(z_{1},z_{2})V_{\rm 
HF}(\blx_{1},\blx_{2},z_{1})$, with $V_{\rm HF}$ the spatially 
nonlocal HF potential,
and the Ehrenfest self-energy
$\S_{\rm Eh}(1;2)=-\d(1;2)
\sum_{\blq\n}\bcallg_{\blq\n}^{\dag}(\blr_{1},z_{1})\cdot \bgvf_{\blq\n}(z_{1})$, 
see Eq.~(\ref{EhrenfestSigma}).
Then Eq.~(\ref{eomGongamma}) can 
be rewritten as 
\begin{align}
	\int d\bar{1}\;\overrightarrow{G}_{\rm mf}^{-1}(1;\bar{1})G(\bar{1};2)&=
\d(1;2)+\int d\bar{1}\,\S_{\rm c}(1;\bar{1})G(\bar{1};2),
\label{eomGmnc}
\end{align}
where the mean-field operator $\overrightarrow{G}_{\rm 
mf}^{-1}(1;2)\equiv \overrightarrow{G}_{0}^{-1}(1;2)-\S^{\d}(1;2)$. 
Taking into account the definition of $\overrightarrow{G}_{0}^{-1}$ 
in Eq.~(\ref{G0-1}) and the definition of the shifted one-particle 
Hamiltonian $h^{s}$ in Eq.~(\ref{oneparths}) one finds
\begin{align}
\overrightarrow{G}_{\rm mf}^{-1}(1;2)&=
\Big[i\frac{d}{dz_{1}}-h(\grad,\blr,z)
+\!\sum_{\blq\n}
g_{-\blq\n}(\blr,z)U_{\blq\n}(z)\Big]
\d(1;2)
\nn\\&-V_{\rm HF}(\blx_{1},\blx_{2},z_{1})\d(z_{1},z_{2})
\end{align}
where we also take into account 
Eq.~(\ref{phisdef}) and the fact that electrons are coupled to the 
phonons only through the displacements, see Eq.~(\ref{giqnu}).

For simplicity we specialize the discussion to the 
relevant case of external perturbations that 
do not break the lattice periodicity of the crystal, albeit the 
developed formalism is far more general. 
Let us write the coordinates $\blr=(x,y,z)$ of any point in space
as the sum of the vector $\blR^{0}_{\bln}$ of the unit cell 
the point $\blr$ belongs to and a displacement $\blu$ 
spanning the unit cell centered at the origin, i.e., 
$\blr=\blR^{0}_{\bln}+\blu$. 
We introduce a generic one-electron Bloch basis, e.g., the Kohn-Sham 
basis, 
$\Q_{\blk\m}(\blx=\blr\s)=e^{i\blk\cdot\bln}
u_{\blk\m }(\blu\s)/\sqrt{N}$ where the vector $\blk$ takes the same 
values as the vector $\blq$ defined below Eqs.~(\ref{momdisplexp}).
The wavefunctions $\Q_{\blk\m}(\blx)$ can be thought of as the 
one-electron eigenfunctions in some potential, e.g., the Kohn-Sham 
potential, with the 
periodicity of our lattice; hence the index $\m$ can be thought of as 
a band index. The matrix element of 
 any two-point correlator 
with the same periodicity, calculated by sandwiching 
with  $\Q^{\ast}_{\blk_{1}\m_{1}}$ and 
$\Q_{\blk_{2}\m_{2}}$,  is proportional to 
$\d_{\blk_{1},\blk_{2}}$.
If we then multiply Eq.~(\ref{eomGmnc}) by  
$\Q^{\ast}_{\blk\m_{1}}(\blx_{1})$ from the left and by 
$\Q_{\blk\m_{2}}(\blx_{2})$ from the right 
and we integrate over $\blx_{1}$ and $\blx_{2}$ we find (in matrix 
form)
\begin{align}
\Big[i
\frac{d}{dz_{1}}-h_{\rm HF}(\blk,z_{1})&+
\sum_{\n}\tilde{g}_{\n}(\blk,z_{1})U_{{\mathbf 0}\n}(z_{1})\Big]
G_{\blk}(z_{1},z_{2})
\nn \\
&=\d(z_{1},z_{2})+\!\!
\int_{\g}\!d\bar{z}\, \S_{\rm c,\blk}(z_{1},\bar{z})
G_{\blk}(\bar{z},z_{2}) ,
\label{contourkbeGephon2}
\end{align}
where
\begin{align}
&h_{\rm HF,\m_{1}\m_{2}}(\blk,z_{1})=
\int d\blx_{1} d\blx_{2}\Q^{\ast}_{\blk\m_{1}}(\blx_{1})
\nn\\
&\times
\Big(h(\grad_{1},\blr_{1},z_{1})\d(\blx_{1}-\blx_{2})+V_{\rm 
HF}(\blx_{1},\blx_{2},z_{1})\Big)\Q_{\blk\m_{2}}(\blx_{2})
\end{align}
and the like for the matrix elements $G_{\blk\m_{1}\m_{2}}$ 
and $\S_{\rm c,\blk\m_{1}\m_{2}}$. The term proportional to 
$U_{{\mathbf 0}\n}$ in Eq.~(\ref{contourkbeGephon2})
 originates from 
\begin{align}
\int d\blx_{1} \Q^{\ast}_{\blk\m_{1}}(\blx_{1})
g_{-\blq\n}(\blr_{1},z_{1})
\Q_{\blk\m_{2}}(\blx_{1})=\d_{\blq,{\mathbf 
0}}\,\tilde{g}_{\n,\m_{1}\m_{2}}(\blk,z_{1}),
\end{align}
which implicitly defines the matrix $\tilde{g}_{\n}(\blk,z_{1})$.
The Kronecker-delta on the r.h.s. follows from the property 
Eq.~(\ref{propginrs}) of the $e$-$ph$ coupling which in turns 
implies [see  Eq.~(\ref{proelphoncoup})] 
\begin{align}
g_{-\blq\n}(\blr+\blR^{0}_{\bln},z)=e^{-i\blq\cdot\bln}
g_{-\blq\n}(\blr,z).
\label{gpropcrystal}
\end{align}
Equation~(\ref{contourkbeGephon2}) 
is consistent with the fact that the only displacements activated
by an external perturbation preserving the lattice periodicity 
are the uniform ones.
Of course, this 
does not mean that only zero-momentum phonons are emitted or 
absorbed, see below. 

{\em Equation of motion for displacements and momenta.--}
Writing Eq.~(\ref{eomphixi}) component-wise we find
\begin{subequations}
\begin{align}
\frac{d P_{\blq\n}(z)}{dz}&=	
\int \!d\blx\,g_{\blq\n}(\blr,z)\D n(\blx,z)
-\sum_{\n'}K_{\n\n'}(\blq,z)U_{\blq\n'}(z),
\\
\frac{d U_{\blq\n}(z)}{dz}&=P_{\blq\n}(z) , 
\end{align}
\label{eomupelphon}
\end{subequations}
where $P_{\blq\n}$ is the average of 
$\hat{P}_{\blq\n}=\hat{\f}^{2}_{\blq\n}$ and $U_{\blq\n}$ is the average of 
$\hat{U}_{\blq\n}=\hat{\f}^{1}_{\blq\n}$.
The first of these equations agrees with Eq.~(\ref{eomfornuclp2}) whereas 
the second equation establishes that $P_{\blq\n}$ is the conjugate 
momentum of $U_{\blq\n}$. Alternatively, $U_{\blq\n}$ and 
$P_{\blq\n}$ can be calculated from 
$\f^{i}_{\blq\n}=\vf^{i}_{\blq\n}+s^{i}_{\blq\n}$, see 
Eq.~(\ref{phisdef}), where $\vf^{i}_{\blq\n}$ is given by Eq.~(\ref{resumoftadpole}). 
It is straightforward to find
\begin{subequations}
\begin{align}
U_{\blq\n}(z)&=-\sum_{\n'}\int d\bar{\blx}d\bar{z}\,
D_{0,\blq\n\n'}^{11}(z,\bar{z})\,g_{\blq\n'}(\bar{\blr})\D 
n(\bar{\blx},\bar{z}),
\\
P_{\blq\n}(z)&=-\sum_{\n'}\int d\bar{\blx}d\bar{z}\,
D_{0,\blq\n\n'}^{21}(z,\bar{z})g_{\blq\n'}(\bar{\blr})\D 
n(\bar{\blx},\bar{z}),
\end{align}
\label{uknpknexpl}
\end{subequations}
where have taken into account Eq.~(\ref{connctD0D0}).
In equilibrium $\D n=0$ and therefore $U_{\blq\n}=P_{\blq\n}=0$.
Under the hypothesis that the external perturbation does not 
break the lattice periodicity we also have 
$U_{\blq\n}(z)=P_{\blq\n}(z)=0$ for all $\blq\neq 0$.

{\em Equation of motion for $D$.--} We have already observed that 
$D_{0}(1,2)\propto \d_{\blq_{1},-\blq_{2}}$, see Eq.~(\ref{connctD0D0}). 
Let us now investigate 
the mathematical structure of the phononic self-energy. Combining 
Eqs.~(\ref{dressedvertexeq}) and~(\ref{Pi=gPgd}) we can write 
\begin{align}
\P=gP(g+WPg)=g(P+PWP)g=g\chi g
\end{align}
where $\chi=P+PWP=P+P v\chi$ is the density-density response 
function. Spelling out 
the space-spin-time convolutions 
\begin{align}
\P(1,2)&=\P(\blq_{1}\n_{1},i_{1},z_{1},\blq_{2}\n_{2},i_{2},z_{2})
=
\d_{i_{1},1}\d_{i_{2},1}
\int \!d\bar{\blx}_{1}d\bar{\blx}_{2}
\nn\\ &\times\,g_{-\blq_{1}\n_{1}}(\bar{\blr}_{1},z_{1})
\chi(\bar{\blx}_{1},z_{1};\bar{\blx}_{2},z_{2})
g_{-\blq_{2}\n_{2}}(\bar{\blr}_{2},z_{2}).
\label{elphonexPi}
\end{align}	
Working again under the hypothesis that the external perturbation does not 
break the lattice periodicity $\chi$ is invariant under a 
simultaneous translations of its spatial coordinates by an arbitrary 
lattice vector $\blR^{0}_{\bln}$. Therefore 
\begin{align}
\P(1,2)=\d_{\blq_{1},-\blq_{2}}
\P_{-\blq_{1}\n_{1}\n_{2}}^{i_{1}i_{2}}(z_{1},z_{2}).
\label{Pidiagq}
\end{align}
As both $D_{0}(1,2)$ and $\P(1,2)$ are proportional to 
$\d_{\blq_{1},-\blq_{2}}$ the interacting phononic GF
$D(1,2)=[D_{0}+D_{0}\P D_{0}+D_{0}\P D_{0}\P 
D_{0}+\ldots](1,2)$ is also proportional to 
$\d_{\blq_{1},-\blq_{2}}$, i.e.,
\begin{align}
D(1,2)=\d_{\blq_{1},-\blq_{2}}
D_{\blq_{1}\n_{1}\n_{2}}^{i_{1}i_{2}}(z_{1},z_{2}).
\label{Ddiagq}
\end{align}
Notice that the phononic self-energy $\P$ in the r.h.s. of 
Eq.~(\ref{Pidiagq})  is defined with a 
$-\blq_{1}$ whereas the phononic GF $D$
in the r.h.s. of Eq.~(\ref{Ddiagq}) is defined with a $+\blq_{1}$. 
These different definitions are chosen to have a more elegant 
equation of motion. Indeed if we insert these expressions into 
Eq.~(\ref{eomDongamma})
 we obtain (in matrix form)
\begin{align}
\Big[i
\frac{d}{dz_{1}}\,\a &-Q(\blq,z_{1})
\Big]D_{\blq}(z_{1},z_{2})
\nn\\
&=\d(z_{1},z_{2})+\int_{\g}d\tilde{z}\,
\P_{\blq}(z_{1},\tilde{z})D_{\blq}(\tilde{z},z_{2}).
\label{contourkbeDephon}
\end{align}

Equations~(\ref{contourkbeGephon2}), (\ref{contourkbeDephon}) and 
their counterparts with derivatives with respect to $z_{2}$ along with 
the equations of motion Eqs.~(\ref{eomupelphon}) for the displacements and 
momenta form a closed system of 
equations for any approximate functional $\S_{{\rm c},2\rm 
skel}[G,D,v,g]$ 
and $\P_{2\rm skel}[G,D,v,g]$.

\section{Conserving approximations}
\label{conservingsec}

\begin{figure}[tbp]
\fbox{\includegraphics[width=0.46\textwidth]{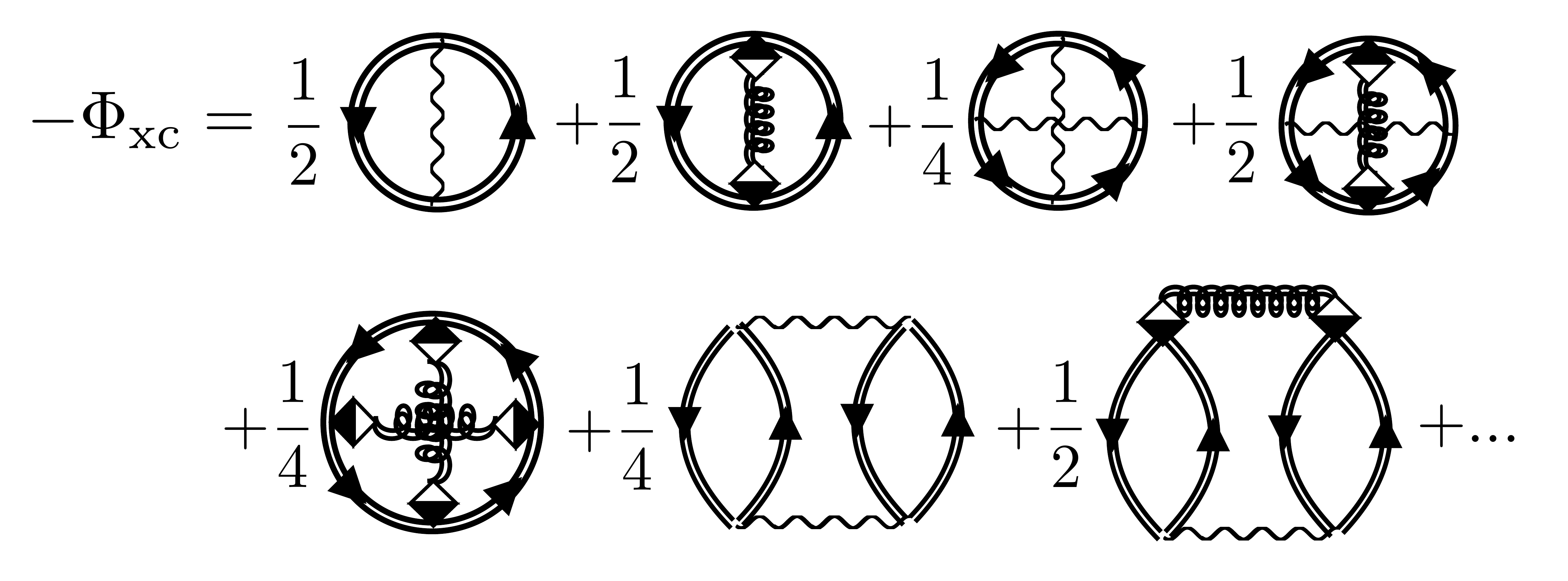}}
\caption{Expansion of the $\F_{\rm xc}$ functional in $G$-skeleton 
diagrams and 
$D$-skeleton diagrams.}
\label{vacdiageb}
\end{figure}

If the self-energies are 
$\F$-derivable~\cite{baym_self-consistent_1962,almbladh_variational_1999,karlsson_partial_2016} then the 
GF resulting from the solution 
of Eqs.~(\ref{contourkbeGephon2}), (\ref{contourkbeDephon}) 
and~(\ref{eomupelphon}) 
satisfy all fundamental conservation laws.
To define this properly in the context of electrons and phonons we
split off the Ehrenfest-Hartree 
part of the self-energy like in Eq.~(\ref{sigmasss}). 
In the doubly skeletonic expansion the remainder is the xc 
self-energy $\S_{\rm xc}[G,D,v,g]=\S_{3\rm skel}[G,D,W,g]$.
We then construct the functional $\F_{\rm xc}[G, D,v,g]$ 
using the same rules as for a system of only 
electrons~\cite{svl-book}: (i) close each skeleton diagram 
for $\S_{\rm xc}$ with a $G$-line, thereby producing a set of vacuum 
diagrams (ii) retain only the topologically inequivalent vacuum 
diagrams and (iii) multiply every diagram by the corresponding 
symmetry factor $1/N_{sym}$, where $N_{sym}$ is the number of 
equivalent $G$ lines  yielding the same self-energy diagram by their 
respective removal.
The lowest order diagrams of the expansion are shown in Fig.~\ref{vacdiageb}.      
The additional minus sign is due to the fact that the removal 
of a $G$-line from a vacuum diagram changes the number of electronic loops 
by one. By construction the $\F_{\rm xc}$ functional has the property that 
\begin{subequations}
\begin{align}
\S_{\rm xc}(1;2)&=\frac{\d\F_{\rm xc}}{\d G(2;1)},
\\
\P(1,2)&=\left.\frac{\d\F_{\rm xc}}{\d D(1,2)}\right|_{S},
\end{align}
\label{fderpeb}
\end{subequations}
where the subscript `` $S$ '' refers to the 
symmetrized derivative $[\d/\d D(1,2)+\d/\d D(2,1)]/2$.
The Hartree self-energy is obtained from the functional derivative of 
the Hartree functional 
\begin{align}
    \raisebox{-6pt}{\includegraphics[width=0.23\textwidth]{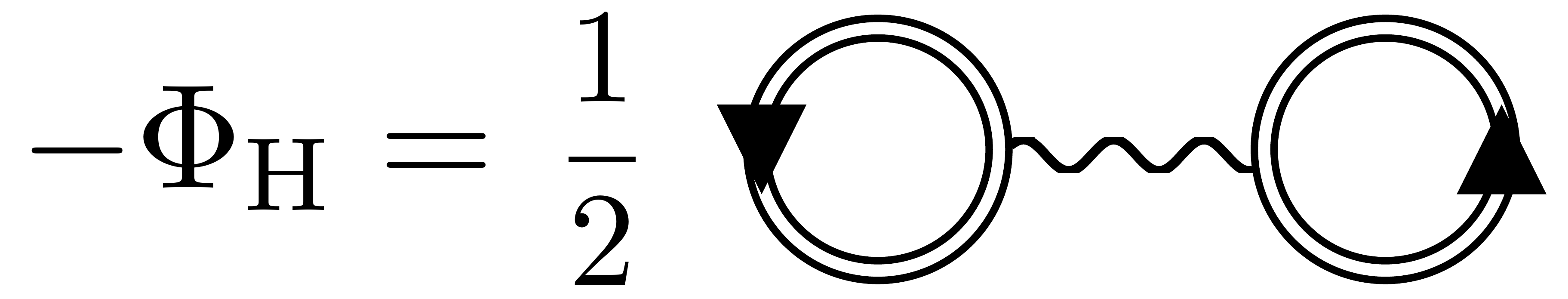}} \;\;,
    \label{PhiHartree}
\end{align}
whereas the Ehrenfest 
self-energy is obtained from the functional derivative of 
the Ehrenfest functional
\begin{align}
    \raisebox{-6pt}{\includegraphics[width=0.32\textwidth]{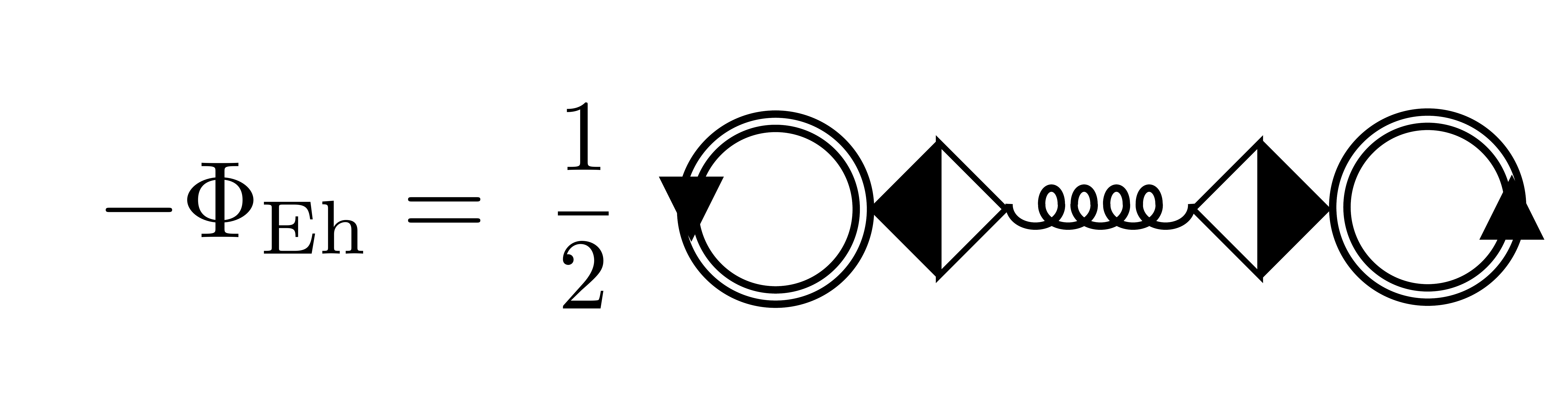}} \;\;.
    \label{Phitilde}
\end{align}
Therefore the full self-energy $\S$ is the functional 
derivative of the $\F$-functional defined as
\begin{align}
\F\equiv\F_{\rm H}[G,v]+\F_{\rm Eh}[G,D_{0},g]+\F_{\rm xc}[G,D,v,g].
\end{align}

In most cases we can only  deal with approximate functionals.
These are obtained by selecting 
an appropriate subset of $\F$-diagrams.  We say that the 
self-energies 
are $\F$-derivable whenever there exists an approximate functional $\F$ 
such that $\S_{\rm xc}$ and $\P$ can be 
written as in Eqs.~(\ref{fderpeb}).
The $\F$-functional is invariant under gauge transformations and 
contour-time deformations, i.e., $z\to w(z)$ with 
$w(t_{0})=t_{0}$ and $w(t_{0}-i\b)=t_{0}-i\b$,
implying the fulfillment of 
the continuity equation and energy conservation for the GF 
that satisfy the equations of motion Eqs.~(\ref{contourkbeGephon2}) 
and (\ref{contourkbeDephon}) with $\F$-derivable 
self-energies~\cite{baym_self-consistent_1962}. 
We mention here that the use of $\F$-derivable 
self-energies evaluated at different input GF 
still guarantee the satisfaction of all conservation laws
provided that they are convoluted with the same input GF~\cite{karlsson_fast_2021}. 
In other words, self-consistency is {\em not} required for having a 
conserving theory. 

\begin{figure}[tbp]
    \centering
\fbox{\includegraphics[width=0.46\textwidth]{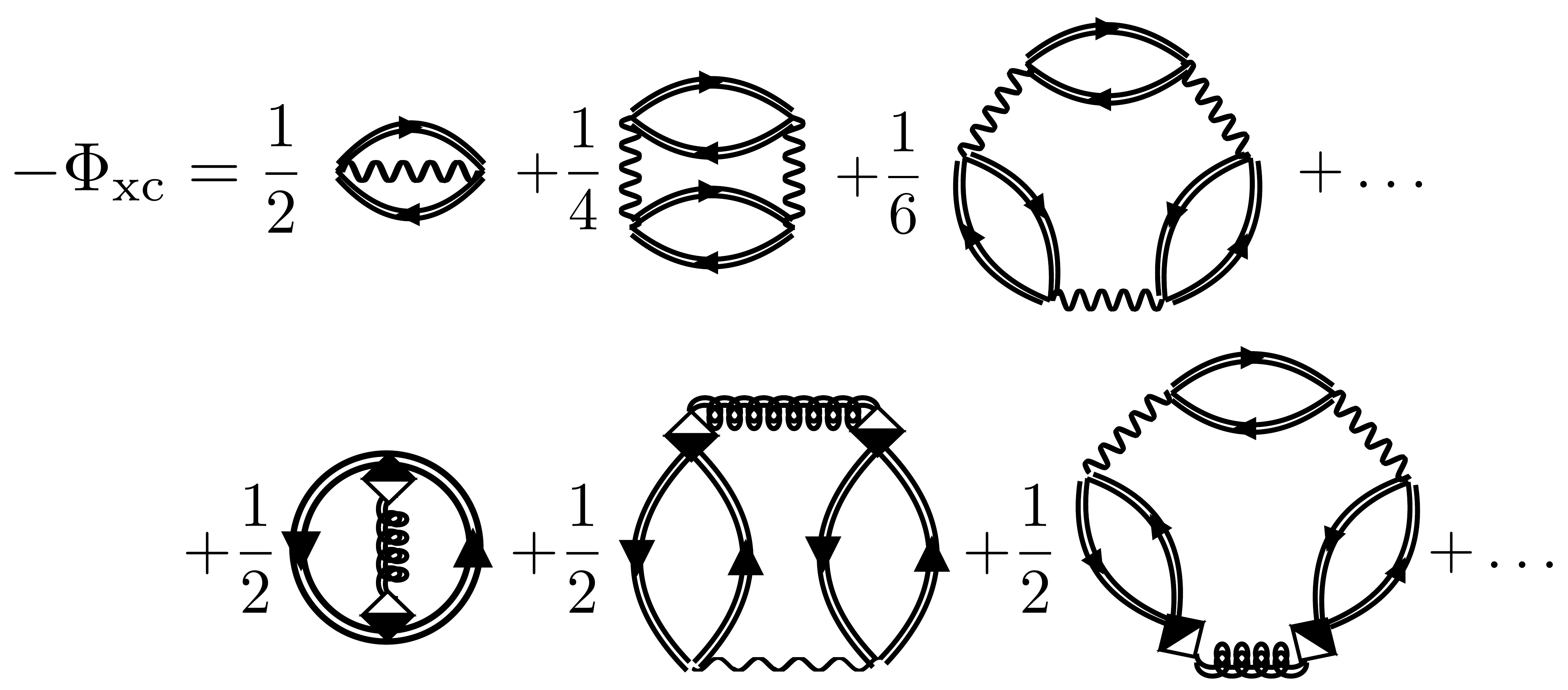}}
\caption{Diagrams of the $\F_{\rm xc}$ functional leading to the 
self-energies Eqs.~(\ref{GW+FM}) and (\ref{aGGad}) through the functional 
derivatives in Eqs.~(\ref{fderpeb}).}
\label{PhidiagGWFM}
\end{figure}

We conclude by observing that the approximation to the 
self-energies discussed at the end of Section~\ref{hedinsec} (derived 
by setting $\tilde{K}_{r}=0$ in the Hedin-Baym equations) is 
$\F$-derivable. 
The diagrams for $\F_{\rm xc}$  are those of the GW 
approximation plus a second infinite sum 
of ring diagrams in which one $v$-line is replaced by 
$(gDg)$, see Fig.~\ref{PhidiagGWFM}.

\section{Kadanoff-Baym equations}
\label{kbesec}

Placing the arguments on different branches of the contour and 
using the Langreth rules~\cite{svl-book,l.1976} we 
can convert the equations of motion Eqs.~(\ref{contourkbeGephon2}) 
and (\ref{contourkbeDephon})
into a coupled system of equations 
for the Keldysh components of $G$ and $D$.
These are the {\em Kadanoff-Baym equations} (KBE)
for systems of electrons and 
phonons to be solved with KMS boundary conditions. 
As for the case of only electrons the equations for the Matsubara 
components decouple. The Matsubara GF 
$G^{\rm M}$ and $D^{\rm M}$ allow for 
calculating the initial thermal average of any one-body operator 
for electrons and of any quadratic operator in the displacements and 
momenta
for the nuclei. The KBE for the GF with 
times on the horizontal branches (hence the left/right and 
lesser/greater components) allow for monitoring the 
system evolution as well as for calculating electronic and phononic 
spectral functions of the system in any stationary state. 
As already pointed out below Eqs.~(\ref{tdhampar}) 
we can study how the system responds  to 
different kind of external perturbations, e.g., interaction quenches, 
laser fields, phonon drivings, etc.. 

\subsection{Self-consistent Matsubara equations}
\label{ephmatzprobsec}

The preliminary step to solve the 
equations of motion for the GF consists 
in solving the Matsubara problem. The Matsubara self-energies do 
indeed depend only on the Matsubara GF~\cite{svl-book} and 
therefore the equations for the Matsubara components are closed.
The Matsubara component of any correlator $X(z_{1},z_{2})$
with arguments on $\g$ is 
defined as $X^{\rm M}(\t_{1},\t_{2})
\equiv X(z_{1}=t_{0}-i\t_{1},z_{2}=t_{0}-i\t_{2})$.
The KMS boundary conditions allow for 
expanding the Matsubara GF and  
self-energies according 
\begin{align}
X^{\rm M}(\t_{1},\t_{2})=\frac{1}{-i\b}
\sum_{m=-\iif}^{\iif}e^{-\w_{m}(\t_{1}-\t_{2})}
X^{\rm M}(\w_{m})
\label{matsexp}
\end{align}
where the Matsubara frequencies 
$\w_{m}=2mi\p/\b$ for periodic functions like $D$ and $\P$ 
and $\w_{m}=(2m+1)i\p/\b$
for antiperiodic functions	like $G$ and $\S_{\rm c}$.
Taking into account that along the vertical track 
$U_{\blq\n}=P_{\blq\n}=0$ since $\D n=0$, see Eq.~(\ref{uknpknexpl}),  
the equations of motion Eqs.~(\ref{contourkbeGephon2}) and 
(\ref{contourkbeDephon}) yield
\begin{subequations}
\begin{align}
G^{\rm M}_{\blk}(\w_{m})&=\frac{1}{\w_{m}-
h_{\rm HF}(\blk)-\m-\S_{\rm c,\blk}^{\rm M}(\w_{m})},
\\
D^{\rm M}_{\blq}(\w_{m})&=\frac{1}{\w_{m}\a-Q(\blq)-
\P^{\rm M}_{\blq}(\w_{m})},
\label{intDcM}
\end{align}
\end{subequations}
where $Q(\blq)=\left(\begin{array}{cc}
K(\blq) & 0 \\ 0 & 1 \end{array}\right)$, see Eq.~(\ref{Qiipqnunup}).
For any approximation to $\S_{\rm c}[G,D,v,g]$ 
and $\P[G,D,v,g]$ these equations 
can be solved self-consistently. We recall that $K=K[n^{0}]$ depends 
on the equilibrium density 
\begin{align}
n^{0}(\blx)
=-i \sum_{\blk\m\m'}
\Q_{\blk\m}(\blx)G^{\rm 
M}_{\blk\m\m'}(\t,\t^{+})\Q^{\ast}_{\blk\m'}(\blx),
\label{n0fromNEGF}
\end{align}
see Eq.~(\ref{elasticmat}). Thus, the 
Matsubara equations are coupled even if we set $\S_{\rm c}=\P=0$.
It is only 
in the partial self-consistent scheme discussed at the 
end of Section~\ref{qselnusec} that the  
Matsubara equations with $\S_{\rm c}=\P=0$ 
decouple (the elastic tensor $K$ is not 
updated in this case).

{\em Phononic self-energy in the clamped+static approximation.--}
In all physical situations of relevance setting $\S_{\rm c}=\P=0$ is 
a very poor approximation.
For the electronic self-energy the $GW$ approximation is a ``gold 
standard'' for obtaining accurate or at least reasonable 
results~\cite{hybertsen_electron_1986,ag.1998,shishkin_self-consistent_2007,nabok_accurate_2016,golze_the-gw_2019,rasmussen_toward_2021}.
What about the phononic self-energy? Let us explore the physics 
of Eq.~(\ref{elphonexPi}) when $\chi$ is calculated by summing all diagrams 
{\em without} $e$-$ph$ coupling, i.e., $\chi\simeq 
\chi_{\rm clamp}$. This approximation for $\chi$  corresponds to the 
response function of a system of only electrons interacting through 
the Coulomb repulsion and feeling the 
potential $V$ generated by clamped nuclei in 
$\blR^{0}$~\cite{feliciano_electron-phonon_2017}. Using the definition in Eq.~(\ref{Pidiagq})
the Matsubara phononic self-energy in the {\em clamped 
approximation}  reads  
\begin{align}
\P_{\blq\n_{1}\n_{2}}^{i_{1}i_{2},\rm M}(\w_{m})\simeq\d_{i_{1},1}\d_{i_{2},1}
&\int \!d\bar{\blx}_{1}d\bar{\blx}_{2}\,g_{\blq \n_{1}}(\bar{\blr}_{1})
\nn\\
\times& \chi_{\rm clamp}^{\rm M}(\bar{\blx}_{1},\bar{\blx}_{2};\w_{m})
g_{-\blq\n_{2}}(\bar{\blr}_{2}),
\end{align}
where we use that along the vertical track $g_{\blq 
\n}(\bar{\blr}_{1},t_{0}-i\t)=g_{\blq \n}(\bar{\blr}_{1})$ is 
independent of $\t$. Let us further approximate 
the r.h.s. with its value at $\w_{m}=0$~\cite{hl.1970}. This is the {\em 
static approximation} and it is similar in spirit to the statically 
screened approximation of $W$. The response function
$\chi^{\rm M}(\bar{\blx}_{1},\bar{\blx}_{2};0)$ calculated 
at a zero frequency is identical to the retarded or advanced response 
function
$\chi^{\rm R/A}(\bar{\blx}_{1},\bar{\blx}_{2};0)$ also calculated at 
zero frequency~\cite{svl-book}. Therefore the phononic self-energy in
the clamped+static approximation can be written as
\begin{align}
\P_{\blq\n_{1}\n_{2}}^{i_{1}i_{2},\rm M}(\w_{m})\simeq
\d_{i_{1},1}\d_{i_{2},1}\P^{\rm clamp+stat}_{\blq\n_{1}\n_{2}}
\end{align}
with
\begin{align}
\P^{\rm clamp+stat}_{\blq\n_{1}\n_{2}}\equiv
\int \!d\bar{\blx}_{1}d\bar{\blx}_{2}\,g_{\blq \n_{1}}(\bar{\blr}_{1})
\chi_{\rm clamp}^{\rm R}(\bar{\blx}_{1},\bar{\blx}_{2};0)
g_{-\blq\n_{2}}(\bar{\blr}_{2}).
\label{clamp+statPi}
\end{align}
Inserting this approximation into Eq.~(\ref{intDcM}) we see that the  
effect of $\P$ is to renormalize the $(1,1)$ block of $Q$. In other 
words the interacting $D$ in the clamped+static approximation 
has the same form as the noninteracting $D_{0}$ with a renormalized 
elastic tensor
\begin{align}
K_{\n_{1}\n_{2}}(\blq)\to K_{\n_{1}\n_{2}}^{\rm renorm}(\blq)=K_{\n_{1}\n_{2}}(\blq)+
\P^{\rm clamp+stat}_{\blq\n_{1}\n_{2}}.
\label{Kren}
\end{align}

{\em Connection with the Born-Oppenheimer approximation.--}
If we use the Born-Oppenheimer (BO) approximation
to evaluate the equilibrium nuclear positions and electronic density 
then $\blR^{0}\simeq \blR^{0,\rm BO}$ and $n^{0}\simeq n^{0,\rm BO}$. 
Correspondingly we have an approximation to the $e$-$ph$ coupling 
$g\simeq g^{\rm BO}$ and to the elastic tensor $K\simeq K^{\rm BO}$.
In the BO approximation the 
renormalized elastic tensor is exactly the Hessian $\callH$ of the BO 
energy calculated in $\blR^{0,\rm BO}$, i.e.,
$K_{\n_{1}\n_{2}}^{\rm 
renorm}(\blq)=\callH_{\n_{1}\n_{2}}(\blq)$, see 
Ref.~\cite{feliciano_electron-phonon_2017} or Appendix~\ref{BOapp}. 
The Hessian has positive 
eigenvalues $\w^{2}_{\blq\n}$ since $\blR^{0,\rm BO}$ is 
the global minimum of the BO energy. The frequencies 
$\w_{\blq\n}\equiv \sqrt{\w^{2}_{\blq\n}}\geq 0$ are called the {\em phonon frequencies} and they 
 provide an excellent starting point already for a clamped 
response function $\chi_{\rm clamp}$ evaluated at the RPA level. 
Nonetheless, 
the importance of going beyond the static approximation (especially 
for metallic systems) has been 
reported in the 
literature~\cite{maksimov_nonadiabatic_1996,lazzeri_nonadiabatic_2006,pisana_breakdown_2007}.  
These considerations make it clear that in order to extract 
physical phonons from the {\em ab initio} $e$-$ph$
Hamiltonian, it is necessary to treat the $e$-$ph$ coupling in a 
nonperturbative manner.   

In the basis of the normal modes of the Hessian 
we have 
$\callH_{\n_{1}\n_{2}}(\blq)=\d_{\n_{1}\n_{2}}\w^{2}_{\blq\n_{1}}$. Thus the 
$(1,1)$ block of the interacting phononic GF 
in the clamped+static approximation for $\P$ is 
simply, see Eq.~(\ref{11blockDM}),
\begin{align}
D_{\blq\n\n'}^{11,\rm M}(\w_{m})=
\frac{\d_{\n\n'}}{\w^{2}_{m}-\w^{2}_{\blq\n}}.
\label{Dcsapp}
\end{align}
As the phonon frequencies are all positive we can use them to 
construct the phononic annihilation operators 
\begin{align}
\hat{b}_{\blq\n}=\sqrt{\frac{\w_{\blq\n}}{2}}
\hat{U}_{\blq\n}+i\frac{\hat{P}_{\blq\n}}{\sqrt{2\w_{\blq\n}}}
\end{align}
and creation operators 
$\hat{b}^{\dag}_{\blq\n}$ with commutation relations 
$[\hat{b}_{\blq\n},\hat{b}^{\dag}_{\blq'\n'}]=\d_{\blq,\blq'}\d_{\n\n'}$.
These results has led 
several authors to partition the low-energy 
Hamiltonian in Eq.~(\ref{el-phonham}) in a slightly different way, see for instance 
Refs.~\cite{feliciano_electron-phonon_2017,baroni_phonons_2001,keating_dielectric_1968,marini_many-body_2015,marini_equilibrium_2023}.
The main difference consists in using the 
Hessian to define the phononic Hamiltonian
\begin{align}
\hat{H}_{ph}&=
\frac{1}{2}\sum_{\blq \n}\hat{P}^{\dag}_{\blq \n}
\hat{P}_{\blq \n}
+\frac{1}{2}\sum_{\blq\n\n'}
\hat{U}^{\dag}_{\blq \n}\callH_{\n\n'}(\blq)\,
\hat{U}_{\blq \n'}
\nn\\&=
\sum_{\blq \n}\w_{\blq\n}
(\hat{b}^{\dag}_{\blq\n}\hat{b}_{\blq\n}+\frac{1}{2}).
\end{align}
In such alternative partitioning the remainder (a quadratic form in 
the nuclear displacements)
\begin{align}
\D\hat{H}_{ph}\equiv \hat{H}_{0,ph}-\hat{H}_{ph}=
\frac{1}{2}\sum_{\blq\n\n'}
\hat{U}^{\dag}_{\blq \n}\big(K_{\n\n'}(\blq)-\callH_{\n\n'}(\blq)\big)
\hat{U}_{\blq \n'}
\label{phonhamremain}
\end{align}
must be treated somehow.
As we have shown the NEGF formalism and related diagrammatic 
expansions are most easily 
formulated with the partitioning of Eq.~(\ref{el-phonham3}).
In fact, there is no particular convenience in rewriting the full 
Hamiltonian in terms of the operators 
$\hat{b}_{\blq\n}$ and $\hat{b}^{\dag}_{\blq\n}$ since
$\hat{H}_{0,ph}$ is not diagonal. 
There is instead a convenience in using 
the Hessian eigenbasis to define 
$\hat{U}_{\blq\n}$ and $\hat{P}_{\blq\n}$
since the interacting 
phononic GF in the clamped+static approximation is 
diagonal, see Eq.~(\ref{Dcsapp}). 
In this approximation $\hat{b}_{\blq\n}$  
($\hat{b}^{\dag}_{\blq\n}$) annihilates (creates) 
a quantum of vibration (the phonon) characterized by 
a well defined energy (the frequency $\w_{\blq\n}$)
and hence an infinitely long life-time.

{\em Conserving approximations.--}
\index{phonon!dressed}
The clamped+static approximation is {\em not} a conserving approximation.
Going beyond it the concept of 
phonons as infinitely long-lived lattice excitations is no longer 
justified. Phonons become {\em quasi-phonons} or {\em dressed phonons} by 
acquiring a finite life-time~\cite{feliciano_electron-phonon_2017}. 
Still, the clamped+static approximation remains an excellent starting 
point, often providing quantitative interpretations of Raman spectra.
Accordingly, 
the minimal phononic self-energy which is at the same time conserving 
and physically sensible
is $\P=g\chi g$ with the 
RPA $\chi=\chi^{0}+\chi^{0}v\chi$, or equivalently $\P=g\chi^{0} 
g^{d}$ with $g^{d}=g+v\chi g=g+W\chi^{0}g$ and $W=v+v\chi^{0}W$. 
We emphasize that $\chi$ is not the RPA response function at clamped 
nuclei since the GF in $\chi^{0}=-iGG$ are evaluated with the 
Fan-Migdal self-energy.
As pointed out in Section~\ref{conservingsec} this phononic self-energy is 
$\F$-derivable, the xc functional being the sum of all diagrams in the 
second row of Fig.~\ref{PhidiagGWFM}. For the theory to be conserving the 
electronic self-energy must be consistently derived from the 
same functional $\F$. Therefore any calculation with $\P=g\chi g$ should 
be done with an electronic Fan-Migdal self-energy $\S_{\rm FM}=i g^{d}GD 
g^{d}$. We can add to $\S_{\rm FM}$ the $GW$ self-energy and still be 
conserving since the diagrams in the first row of 
Fig.~\ref{PhidiagGWFM} do not contribute to $\P$.
\index{self-energy!Fan-Migdal}\index{approximation!Fan-Migdal}
\index{self-energy!GW}\index{approximation!GW}

\subsection{Time-dependent evolution and steady-state solutions}
\label{tdephkbesec}

With the Matsubara GF at our disposal  we can proceed 
with the calculation of all other Keldysh components by 
time-propagation. The {\em right} component of any correlator 
$X(z_{1},z_{2})$ with 
arguments on $\g$ is defined as 
$X^{\rceil}(t,\t)\equiv X(t_{\pm},t_{0}-i \t)$. The equations for the right 
components of $G$ and $D$ follow from Eqs.~(\ref{contourkbeGephon2}) and 
(\ref{contourkbeDephon}) when setting $z_{1}=t_{+}$ or $z_{1}=t_{-}$ and 
$z_{2}=t_{0}-i\t$. Using the Langreth rules~\cite{svl-book} we find
\begin{subequations}
\begin{align}
\Big[i\frac{d}{dt}-h_{\rm HF}(\blk,t)
&+\sum_{\n}\tilde{g}_{\n}(\blk,t)U_{{\mathbf 0}\n}(t)
\Big]G^{\rceil}_{\blk}(t,\t),
\nn\\&=
\Big[\S_{\rm c,\blk}^{\rm R}\cdot G_{\blk}^{\rceil}+
\S_{\rm c,\blk}^{\rceil}\star G_{\blk}^{\rm M}\Big](t,\t)
\label{eomGright}
\end{align}
\begin{align}
\Big[i
\frac{d}{dt}\,\a -Q(\blq,t)
\Big]
D^{\rceil}_{\blq}(t,\t)=\left[
\P^{\rm R}_{\blq}\cdot D^{\rceil}_{\blq}
+\P^{\rceil}_{\blq}\star D^{\rm M}_{\blq}
\right](t,\t).
\label{eomDright}
\end{align}
\label{eomXright}
\end{subequations}
Henceforth we use the short-hand notation `` $\cdot$ '' for 
time convolutions between $t_{0}$ and $\iif$ and `` $\star$ '' for 
convolutions between $0$ and $\b$~\cite{svl-book,sa-1.2004}.
Similarly, the equations for the left component, defined for any 
correlator as $X^{\lceil}(\t,t)\equiv X(t_{0}-i\t,t_{\pm})$,
follow when setting
$z_{1}=t_{0}-i\t$ and $z_{2}=t_{-}$ or $t_{+}$ in the counterparts 
of the equations of motion Eqs.~(\ref{contourkbeGephon2}) and 
(\ref{contourkbeDephon}) with derivative with respect to $z_{2}$. We find
\begin{subequations}
\begin{align}
G_{\blk}^{\lceil}(\t,t)
\Big[-i\frac{\overleftarrow{d}}{dt}-&h_{\rm HF}(\blk,t)
+\sum_{\n}\tilde{g}_{\n}(\blk,t)U_{{\mathbf 0}\n}(t)\Big]
\nn\\
&=\Big[ 
G_{\blk}^{\lceil}\cdot\S_{\rm c,\blk}^{\rm A}+
G_{\blk}^{\rm M}\star\S_{\rm c,\blk}^{\lceil}\Big]
(\t,t),
\label{eomGleft}
\end{align}
\begin{align}
D^{\lceil}_{\blq}(\t,t)\Big[-i
\frac{\overleftarrow{d}}{dt}\,\a -Q(\blq,t)
\Big]=\left[
D^{\lceil}_{\blq}\cdot \P^{\rm A}_{\blq}
+D^{\rm M}_{\blq}\star\P^{\lceil}_{\blq}
\right](\t,t),
\label{eomDleft}
\end{align}
\label{eomXleft}
\end{subequations}
where the left arrow over the derivative indicates that the derivative 
acts on the quantity to its left.
At fixed $\t$ these equations are first order 
integro-differential 
equations in $t$ which must be solved with initial conditions
\begin{subequations}
\begin{align}
G_{\blk}^{\rceil}(t_{0},\t)&=G_{\blk}^{\rm M}(0,\t),
\quad\;\;\,
G_{\blk}^{\lceil}(\t,t_{0})=G_{\blk}^{\rm M}(\t,0),
 \\
D_{\blq}^{\rceil}(t_{0},\t)&=D_{\blq}^{\rm M}(0,\t),
\quad\;\;\,
D_{\blq}^{\lceil}(\t,t_{0})=D_{\blq}^{\rm M}(\t,0).
\end{align}
\end{subequations}
The dependence on time in $h_{\rm HF}(\blk,t)$ and $Q(\blq,t)$ 
may be due to some external laser field and/or phonon driving.

The retarded/advanced as well as the left/right components of the 
self-energies depend not only on the left and right GF
but also on the lesser and greater GF. 
For any correlator the lesser component is defined as 
$X^{<}(t,t')\equiv X(t_{-},t'_{+})$ whereas the greater component is 
defined as $X^{>}(t,t')\equiv X(t_{+},t'_{-})$. The retarded and 
advanced components are not independent quantities since 
$X^{\rm R/A}(t,t')=\d(t-t')X^{\d}(t)\pm\th(\pm t\mp 
t')[X^{>}(t,t')-X^{<}(t,t')]$, where $X^{\d}$ is the weight of a 
possible singular part of the correlator $X(z_{1},z_{2})$
[for, e.g., $G$ and $D$ we have $G^{\d}=D^{\d}=0$ while for the electronic 
self-energy $\S$ we have that $\S^{\d}$
is the sum of the HF and Ehrenfest diagrams, see 
Eq.~(\ref{S=Sd+Sc})].
To close the set of equations we 
need the equations of motion for $G_{\blk}^{\lessgtr}$ and 
$D_{\blq}^{\lessgtr}$. These are
obtained by setting $z_{1}=t_{1\pm}$ and $z_{2}=t_{2\mp}$ 
in Eqs.~(\ref{contourkbeGephon2}) and 
(\ref{contourkbeDephon}) and in their counterparts with derivative 
with respect to $z_{2}$. We find
\begin{subequations}
\begin{align}
\Big[i\frac{d}{dt_{1}}-&h_{\rm HF}(\blk,t_{1})
+\sum_{\n}\tilde{g}_{\n}(\blk,t_{1})U_{{\mathbf 0}\n}(t_{1})\Big]
G_{\blk}^{\lessgtr}(t_{1},t_{2})
\nn \\
&=\left[
\S_{\rm c,\blk}^{\rm R}\cdot G_{\blk}^{\lessgtr}+
\S_{\rm c,\blk}^{\lessgtr}\cdot G_{\blk}^{\rm A}+
\S_{\rm c,\blk}^{\rceil}\star G_{\blk}^{\lceil}
\right](t_{1},t_{2}),
\label{eomG<>1}
\end{align}
\begin{align}
\Big[i
\frac{d}{dt_{1}}\,\a -&Q(\blq,t_{1})
\Big]
D^{\lessgtr}_{\blq}(t_{1},t_{2})
\nn\\&=\left[
\P^{\rm R}_{\blq}\cdot D^{\lessgtr}_{\blq}+
\P^{\lessgtr}_{\blq}\cdot D^{\rm A}_{\blq}
+\P^{\rceil}_{\blq}\star D^{\lceil}_{\blq}
\right](t_{1},t_{2}),
\label{eomD<>1}
\end{align}
\begin{align}
G_{\blk}^{\lessgtr}(t_{1},t_{2})
\Big[\!&-i\frac{\overleftarrow{d}}{dt_{2}}-h_{\rm HF}(\blk,t_{2})
+\sum_{\n}\tilde{g}_{\n}(\blk,t_{2})U_{{\mathbf 0}\n}(t_{2})\Big]
\nn \\
&=\left[G_{\blk}^{\lessgtr}\cdot\S_{\rm c,\blk}^{\rm A}+
G_{\blk}^{\rm R}\cdot\S_{\rm c,\blk}^{\lessgtr}+
G_{\blk}^{\rceil}\star\S_{\rm c,\blk}^{\lceil}
\right](t_{1},t_{2}),
\label{eomG<>2}
\end{align}
\begin{align}
D^{\lessgtr}_{\blq}(t_{1},t_{2})\Big[\!&-i
\frac{\overleftarrow{d}}{dt_{2}}\,\a -Q(\blq,t_{2})
\Big]\nn\\
&=\left[D^{\lessgtr}_{\blq}\cdot \P^{\rm A}_{\blq}
+D^{\rm R}_{\blq}\cdot \P^{\lessgtr}_{\blq}+
D^{\rceil}_{\blq}\star\P^{\lceil}_{\blq}
\right](t_{1},t_{2}),
\label{eomD<>2}
\end{align}
\label{eomXlessgtr}
\end{subequations}
which must be solved with initial conditions
\begin{subequations}
\begin{align}
G_{\blk}^{<}(t_{0},t_{0})&=G_{\blk}^{\rm M}(0,0^{+}),
\quad
G_{\blk}^{>}(t_{0},t_{0})=G_{\blk}^{\rm M}(0^{+},0),
\\
D_{\blq}^{<}(t_{0},t_{0})&=D_{\blq}^{\rm M}(0,0^{+}),
\quad
D_{\blq}^{>}(t_{0},t_{0})=D_{\blq}^{\rm M}(0^{+},0).
\end{align}
\end{subequations}

Finally, we need the equation of motion for the displacement and 
momentum at $\blq=0$ since $U_{{\mathbf 0}\n}(t)$ appears explicitly in the 
equations for the electronic GF. These  are given by
Eqs.~(\ref{eomupelphon}) when setting $z=t_{\pm}$ and read
\begin{subequations}
\begin{align}
\frac{d P_{{\mathbf 0}\n}(t)}{dt}&=	
\int d\blx\,g_{{\mathbf 0}\n}(\blr,t)\D n(\blx,t)
-\sum_{\n'}K_{\n\n'}({\mathbf 0},t)U_{{\mathbf 0}\n'}(t)
\\
\frac{d U_{{\mathbf 0}\n}(t)}{dt}&=P_{{\mathbf 0}\n}(t)  
\label{eomPU}
\end{align}
\end{subequations}
where [compare with Eq.~(\ref{n0fromNEGF})]
\begin{align}
\D n(\blx,t)=-i \sum_{\blk\m\m'}\Q_{\blk\m}(\blx)
G^{<}_{\blk\m\m'}(t,t)
\Q^{\ast}_{\blk\m'}(\blx)-n^{0}(\blx).
\end{align}
From these equations we see that the equilibrium density $n^{0}$
appearing in the full Hamiltonian $\hat{H}$ {\em must be} the same as the 
self-consistent density in Eq.~(\ref{n0fromNEGF}) for otherwise the r.h.s. 
of the equation of motion for $P_{{\mathbf 0}\n}$ does not vanish at 
$U_{{\mathbf 0}\n}=0$. 

The set of equations Eqs.~(\ref{eomXright}), (\ref{eomXleft}) and 
(\ref{eomXlessgtr})
are the KBE for systems of 
electrons and phonons.
The KBE, together with the initial conditions  
for the Keldysh components of $G_{\blk}$ and 
$D_{\blq}$, completely determine the electronic and phononic GF
with one and two real times once a
choice for the self-energies is made. These equations have been so 
far solved only in relatively simple model 
systems~\cite{schuler_time-dependent_2016,sakkinen_thesis,sakkinen_many-body_2015-1,sakkinen_many-body_2015-2}.
Under certain approximations (Generalized Kadanoff-Baym 
Ansatz~\cite{lipavsky_generalized_1986,karlsson_fast_2021}, diagonal 
density matrices and Markov approximation) the KBE can be shown to 
reduce to the well 
known Boltzmann equations.

In the clamped+static approximation for $\P_{\blq}$ the 
KBE for $D_{\blq}$ simplify considerably since 
$\P_{\blq}^{\lessgtr}=\P_{\blq}^{\rceil}=\P_{\blq}^{\lceil}=0$ and 
$\P^{i_{1}i_{2},\rm R/A}_{\blq\n_{1}\n_{2}}(t_{1},t_{2})=
\d_{i_{1},1}\d_{i_{2},1}\d(t_{1}-t_{2})\P^{\rm 
clamp+stat}_{\blq\n_{1}\n_{2}}$, see Eq.~(\ref{clamp+statPi}). 
We recall that 
in this approximation the resulting GF are not 
conserving since $\P^{\rm clamp+stat}$ is not the functional 
derivative of any $\F$ functional. 
In particular the total energy of the unperturbed system
is not constant in time.

{\em Steady-state solution.--}
The KBE 
for the lesser/greater GF can be formally solved. 
This is done in 
Refs.~\cite{sa-1.2004,svl-book} for $G$ where it is also shown that 
$\lim_{t,t'\to\iif}G_{\blk}^{\lessgtr}(t,t')=
\Big[G_{\blk}^{\rm R}\cdot\S_{\blk}^{\lessgtr}\cdot G_{\blk}^{\rm 
A}\Big](t,t')$ provided that 
$\S_{\rm c,\blk}$ vanishes when the 
separation between its time arguments go to infinity.  
Following  the same mathematical steps one can show a similar 
relation for the phononic GF
\begin{align}
\lim_{t,t'\to\iif}D_{\blq}^{\lessgtr}(t,t')=
\Big[D_{\blq}^{\rm R}\cdot\P_{\blq}^{\lessgtr}\cdot D_{\blq}^{\rm 
A}\Big](t,t'),
\end{align}
which is valid provided that $\P_{\blq}$ 
vanishes when the separation between its time arguments go to 
infinity.
Further assuming that in the long-time limit the system attains a 
steady-state we can Fourier transform with respect to the 
time-difference and find
\begin{align}
D_{\blq}^{\lessgtr}(\w)=
D_{\blq}^{\rm R}(\w)\P_{\blq}^{\lessgtr}(\w) D_{\blq}^{\rm 
A}(\w),
\nn
\end{align}
where 
\begin{align}
D_{\blq}^{\rm R/A}(\w)=\frac{1}{(\w\pm i\eta)\a-Q(\blq)-\P^{\rm 
R/A}_{\blq}(\w)},
\label{ssradispg}
\end{align}
and $Q(\blq)=\lim_{t\to\iif}Q(\blq,t)$. In most physical 
situations $Q(\blq,t)=Q(\blq)$ is independent of time. By 
construction, see Eq.~(\ref{connectedDcdef}) and Appendix~\ref{symmapp}, 
$D^{\rm R}_{\blq}(\w)=\big[D^{\rm 
A}_{\blq}(\w)\big]^{\dag}$ and hence $\P^{\rm 
R}_{\blq}(\w)=\big[\P^{\rm A}_{\blq}(\w)\big]^{\dag}$. 
We can then uniquely write the retarded/advanced
phononic self-energy 
as
\begin{align}
\P^{\rm R/A}_{\blq}(\w)	= \L_{ph,\blq}(\w)\mp\frac{i}{2}
\G_{ph,\blq}(\w),
\label{Pi=L+iG}
\end{align}
where $\L_{ph,\blq}$ and $\G_{ph,\blq}$ are self-adjoint matrices. 
Accordingly, the phononic spectral function of the system at the 
steady state reads
\begin{align}
A_{ph,\blq}(\w)&\equiv i\Big[D_{\blq}^{\rm R}(\w)-
D_{\blq}^{\rm A}(\w)\Big]=
i\Big[D_{\blq}^{>}(\w)-D_{\blq}^{<}(\w)\Big]
\nn\\
&=D_{\blq}^{\rm R}(\w)\Big[
\G_{ph,\blq}(\w)+2\eta\a\Big]
D_{\blq}^{\rm A}(\w).
\label{phonspectr}
\end{align}

The phononic self-energy has only one nonvanishing block, which is 
the block $(1,1)$. If $\G^{11}_{ph,\blq}(\w)\neq 0$ we can discard 
the positive infinitesimal $\eta$ in Eq.~(\ref{phonspectr}) and find for the 
block $(1,1)$  of the spectral function
\begin{align}
A^{11}_{ph,\blq}(\w)
=D^{11,\rm R}_{\blq}(\w)\G^{11}_{ph,\blq}(\w)
D^{11,\rm A}_{\blq}(\w),
\label{spectrphon11}
\end{align}
with
\begin{align}
D^{11,\rm 
R/A}_{\blq}(\w)=\frac{1}{(\w\pm i\eta)^{2}-K(\blq)-\P^{11,\rm 
R/A}_{\blq}(\w)}.
\label{ssD11}
\end{align}
We observe that the retarded/advanced phononic GF 
differs from its more conventional form in that
it is not multiplied by the BO frequency $\w_{\blq\n}$, see 
e.g., Refs.~\cite{feliciano_electron-phonon_2017,marini_equilibrium_2023}. This is 
simply due to a different definition of the 
correlator where $\sqrt{\w_{\blq\n}}\,\hat{U}_{\blq\n}$ is used in 
place of $\hat{U}_{\blq\n}$. We prefer to stick to our original definition as 
it does not contain any notion of the BO approximation. 
The phononic GF in
Eq.~(\ref{ssD11}) is written in an arbitrary basis of normal modes; 
it reduces to the BO diagonal form in the basis of the BO normal modes 
if $\P^{11,\rm R/A}_{\blq}(\w)\to 
\P^{\rm clam+stat}_{\blq}$. 

Among the possible steady states we have thermal equilibrium. In this 
case we can obtain all Keldysh components from the spectral function. 
Indeed the fluctuation-dissipation theorem implies that
\begin{subequations}
\begin{align}
D^{<}_{\blq}(\w)&=-i f(\w)A_{ph,\blq}(\w),
\\
D^{>}_{\blq}(\w)&=-i \bar{f}(\w)A_{ph,\blq}(\w),
\end{align}
\end{subequations}
where $f(\w)=1/(e^{\b\w}-1)$ is the Bose function and 
$\bar{f}(\w)=e^{\b\w}f(\w)$.
Taking into account that $\bar{f}(\w)-f(\w)=1$ we also have
\begin{align}
D^{\rm R/A}_{\blq}(\w)&=i \int\frac{d\w'}{2\p}
\frac{D^{>}_{\blq}(\w')-D^{<}_{\blq}(\w')}{\w-\w'\pm i\eta}
\nn\\
&=\int\frac{d\w'}{2\p}
\frac{A_{ph,\blq}(\w')}{\w-\w'\pm i\eta}.
\end{align}

{\em Quasi-phonons and life-times.--}
From Eq.~(\ref{ssD11}) we also see that 
any frequency-dependent phononic self-energy gives rise to a finite 
life-time ($\L_{ph}$ and $\G_{ph}$ are related by a Hilbert 
transformation) in the same way as any approximation beyond HF for 
the electronic self-energy gives rise to a finite life-time for the 
electrons.

We can estimate the frequency renormalization  and the phononic life-time  
assuming that the correction to the clamped+static Born-Oppenheimer 
approximation is small. Working in the basis of the normal modes 
of the BO Hessian we approximate~\cite{feliciano_electron-phonon_2017} 
\begin{align}
\P^{11,\rm R/A}_{\blq\n\n'}(\w)\simeq 
\P^{\rm clam+stat}_{\blq\n\n'}+
\d_{\n\n'}\Big(\L^{\rm dyn}_{\blq\n}(\w)\mp
\frac{i}{2}\,\G^{\rm dyn}_{\blq\n}(\w)\Big),
\end{align}
where $\L^{\rm dyn}_{\blq\n}(\w)$ and 
$\G^{\rm dyn}_{\blq\n}(\w)$ are real functions. 
Bearing in mind that $[K(\blq)+\P^{\rm 
clam+stat}_{\blq}]_{\n\n'}=\d_{\n\n'}\w^{2}_{\blq\n}$, see 
Section~\ref{ephmatzprobsec}, we see that the block $(1,1)$ of the 
retarded/advanced 
phononic GF is diagonal in the chosen basis. To lowest 
order in $\L^{\rm dyn}_{\blq\n}$ and $\G^{\rm dyn}_{\blq\n}$ 
Eq.~(\ref{ssD11}) yields 
\begin{widetext}
\begin{align}
D^{11,\rm R/A}_{\blq\n\n}(\w)&=\frac{1}
{(\w\pm i\frac{\G^{\rm 
dyn}_{\blq\n}(\w)}{4\w})^{2}-(\w_{\blq\n}+\frac{\L^{\rm 
dyn}_{\blq\n}(\w)}{2\w_{\blq\n}})^{2}}
\nn\\
&=
\frac{1}{2\big(\w_{\blq\n}+
\frac{\L^{\rm dyn}_{\blq\n}(\w)}{2\w_{\blq\n}}\big)}
\Big[\frac{1}{\w-\w_{\blq\n}-
\frac{\L^{\rm dyn}_{\blq\n}(\w)}{2\w_{\blq\n}}\pm i\frac{\G^{\rm 
dyn}_{\blq\n}(\w)}{4\w}}
-\frac{1}{\w+\w_{\blq\n}+
\frac{\L^{\rm dyn}_{\blq\n}(\w)}{2\w_{\blq\n}}\pm i\frac{\G^{\rm 
dyn}_{\blq\n}(\w)}{4\w}}
\Big].
\label{approxD11RA2}
\end{align}
\end{widetext}
Let us define the {\em quasi-phonon frequencies} $\W^{\pm}_{\blq\n}$
as the solution of
$\W^{\pm}_{\blq\n}=\pm\big(\w_{\blq\n}+
\frac{\L^{\rm dyn}_{\blq\n}(\W^{\pm}_{\blq\n})}{2\w_{\blq\n}}\big)$.
To first 
order in $\w- \W^{\pm}_{\blq\n}$ we can then write
\begin{align}
\w\mp \big(\w_{\blq\n}+
\frac{\L^{\rm dyn}_{\blq\n}(\w)}{2\w_{\blq\n}}\big)\simeq 
\frac{\w- \W^{\pm}_{\blq\n}}{Z^{\pm}_{ph,\blq\n}},
\end{align}
where 
\begin{align}
Z^{\pm}_{ph,\blq\n}\equiv
\frac{1}{\left(1\mp\frac{1}{2\w_{\blq\n}}
\frac{\de \L^{\rm dyn}_{\blq\n}}{\de \w}
\right)_{\w=\W^{\pm}_{\blq\n}}}.
\end{align}
The {\em quasi-phonon weight} $Z^{\pm}_{ph,\blq\n}$ 
gives the probability that a lattice vibration with 
momentum $\blq$ along the normal mode $\n$ excites a phonon with 
quantum numbers $\blq\n$. The remaining spectral weight 
$(1-Z^{\pm}_{ph,\blq\n})$ is absorbed by collective phononic 
excitations arising from correlation effects.

If the function $\G^{\rm dyn}_{\blq\n}(\w)/\w$
is small for $\w\simeq 
\W^{\pm}_{\blq\n}$ and slowly varying in $\w$ then we 
can approximate it with its value in $\W^{\pm}_{\blq\n}$ 
for $\w\simeq 
\W^{\pm}_{\blq\n}$ and rewrite Eq.~(\ref{approxD11RA2}) as
\begin{align}
D^{11,\rm R/A}_{\blq\n\n}(\w)&=
\frac{1}{2\l^{+}_{\blq\n}}
\frac{Z^{+}_{ph,\blq\n}}{\w-\W^{+}_{\blq\n}\pm i/(2\t^{+,\rm phon}_{\blq\n})}
\nn\\
&-
\frac{1}{2\l^{-}_{\blq\n}}
\frac{Z^{-}_{ph,\blq\n}}{\w-\W^{-}_{\blq\n}\pm i/(2\t^{-,\rm phon}_{\blq\n})},
\end{align}
where $\l^{\pm}_{\blq\n}=\w_{\blq\n}+\frac{\L^{\rm 
dyn}_{\blq\n}(\W^{\pm}_{\blq\n})}{2\w_{\blq\n}}$ and 
\begin{align}
\t^{\pm,\rm phon}_{\blq\n}&\equiv Z^{\pm}_{ph,\blq\n}\,
\frac{\G^{\rm dyn}_{\blq\n}(\W^{\pm}_{\blq\n})}{2\W^{\pm}_{\blq\n}}.
\end{align}
is the {\em phononic life-time}. This formula agrees with the 
expression in 
Refs.~\cite{allen_theory_1976,grimvall_the-electron-phonon_1981,feliciano_electron-phonon_2017} for $Z^{\pm}_{ph,\blq\n}=1$.
We observe that, in general, $\W^{+}_{\blq\n}\neq -\W^{-}_{\blq\n}$, and therefore the 
energy lost from the emission of a phonon $\blq\n$  is not the same 
as the energy gained by the absorption of the same phonon. Energy is 
however conserved since the same analysis carried for 
$[D^{11,\rm R/A}_{-\blq}(\w)]_{\n\n}$ leads to
\begin{align}
[D^{11,\rm R/A}_{-\blq}(\w)]_{\n\n}&=
\frac{1}{2\l^{-}_{\blq\n}}
\frac{Z^{-}_{ph,\blq\n}}{\w+\W^{-}_{\blq\n}\pm i/(2\t^{-,\rm phon}_{\blq\n})}
\nn\\
&-
\frac{1}{2\l^{+}_{\blq\n}}
\frac{Z^{+}_{ph,\blq\n}}{\w+\W^{+}_{\blq\n}\pm i/(2\t^{+,\rm 
phon}_{\blq\n})},
\end{align}
where we use the property $\w_{\blq\n}=\w_{-\blq\n}$
and the properties $\L^{\rm dyn}_{\blq\n}(\w)=
\L^{\rm dyn}_{-\blq\n}(-\w)$ and $\G^{\rm dyn}_{\blq\n}(\w)=
-\G^{\rm dyn}_{-\blq\n}(-\w)$, see Appendix~\ref{symmapp} for the 
derivation of the symmetry properties of the phononic GF and 
self-energy.
In crystals with time-reversal symmetry we also have 
$\L^{\rm dyn}_{\blq\n}(\w)=\L^{\rm dyn}_{-\blq\n}(\w)$, 
and therefore 
$\L^{\rm dyn}_{\blq\n}(\w)=\L^{\rm dyn}_{\blq\n}(-\w)$. 
In this case $\W^{+}_{\blq\n}=- 
\W^{-}_{\blq\n}$ and no correlation-induced splitting occurs.

\section{Summary and outlook}
\label{outlooksec}

In this work we have developed the many-body theory of 
the electron-phonon problem. The first step was to clarify the 
dependence of the {\em ab initio} $e$-$ph$ Hamiltonian on the equilibrium 
electronic density. Only through a self-consistent procedure it is 
possible to determine the $e$-$ph$ Hamiltonian and 
hence to make fair comparisons between different many-body methods as well as 
between different approximations within the same method.
The analysis has also highlighted an issue affecting those semi-empirical 
approaches that include the $e$-$ph$ effects through the addition of 
a term of the form 
$\sum_{\l}\w_{\l}\hat{b}^{\dag}_{\l}\hat{b}_{\l}+\sum_{\l,k}
[g_{\l,k}
\hat{O}_{e}^{k}\,\hat{b}^{\dag}_{\l}+{\rm h.c.}]$, with 
$\hat{O}_{e}^{k}$ purely electronic operators, to the electronic 
Hamiltonian at clamped nuclei. 
Beside being devoid of a theoretical foundation, the
phonon frequencies $\w_{\l}$ may get significantly renormalized by the 
phononic self-energy, possibly becoming complex.

We identify in Eq.~(\ref{el-phonham3})
the most suitable partitioning of the {\em ab initio} $e$-$ph$ 
Hamiltonian and in Eq.~(\ref{ndispGF}) the most suitable phononic GF 
for the NEGF formulation. After mapping the cumbersome 
many-body expansion onto a diagrammatic theory 
we have delved into the diagrammatic content of the self-energies until 
reaching a closed form in terms of skeleton diagrams. We have in this 
way provided the 
diagrammatic derivation of the Hedin-Baym equations for in- and 
out-of-equilibrium systems at any temperature. The main differences 
with respect to case of  equilibrium at 
zero-temperature~\cite{feliciano_electron-phonon_2017} is the domain 
of the time-integration for the internal vertices and the appearance 
of the Ehrenfest self-energy.
The merits of the  diagrammatic approach are (i) it provides a 
systematic way for improving many-body approximations through 
a proper selection of physically insightful Feynman diagrams (ii) it 
naturally combines with the $\F$-derivable theory to generate fully 
conserving GF and (iii) it is versatile in the resummation of 
different diagrammatic series. The last point is especially relevant 
for the skeletonic expansion in terms of only $G$ and $D$, 
which is crucial for closing the KBE. In fact, 
the Hedin-Baym equations are of scarce 
practical use to study the time-dependent evolution of the $e$-$ph$ system.
On the contrary the KBE, being them 
integro-differential equations, are better suited for real-time simulations.  
Considering the fast pace of progress in time-resolved experiments we  
foresee a growing interest in this direction.  
The interest is also fueled by the possibility of solving the KBE 
using a time-linear scheme for a 
large number of self-energy 
approximations~\cite{schlunzen_achieving_2020,joost_g1-g2_2020,karlsson_fast_2021,pavlyukh_photoinduced_2021,pavlyukh_time-linear_2022,pavlyukh_interacting_2022,perfetto_real_2022},
thus making NEGF competitive with the fastest quantum methods 
currently available.  

The NEGF formalism presented in this work is applicable to 
Hamiltonians that are far more general than the $e$-$ph$ Hamiltonian. 
One straightforward extension is to make the one-electron Hamiltonian 
nonlocal in space and nondiagonal in spin, i.e., $H_{0,e}^{s}(z)=\int 
d\blx d\blx' 
\hat{\q}^{\dag}(\blx)h(\blx,\blx',z)\hat{\q}^{\dag}(\blx')$, and to 
consider a spin-dependent $e$-$e$ interaction $v(\blx,\blx')$. This 
allows for including the coupling with vector potentials as well as 
relativistic corrections like the spin-orbit interaction, and hence to 
deal with magnetic systems and noncollinear spins~\cite{aryasetiawan_generalized_2008}.
It also makes possible to use pseudo-potentials, thus eliminating the core degrees of 
freedom. Another extension consists in considering an 
$e$-$ph$ interaction Hamiltonian like $\hat{H}_{e-ph}
=\sum_{\blq\n,i}\int d\blx d\blx' 
g_{-\blq\n}^{i}(\blx,\blx')\hat{\q}^{\dag}(\blx)\hat{\q}(\blx')\hat{\f}^{i}_{\blq\n}$.
In fact, the only property required by our derivation is 
$g_{-\blq\n}^{i}(\blx,\blx')=g_{\blq\n}^{i\ast}(\blx',\blx)$, which is 
clearly satisfied by the $e$-$ph$ coupling, see Eq.~(\ref{proelphoncoup}).
The spatial nonlocality of $g$ allows for coupling bosonic particles 
like photons
to nonlocal electronic operators like the current, thereby providing a 
quantum treatment of the light-matter interaction. We can also study 
exotic couplings between electrons and the bosonic momentum since 
we do not need to assume that $g_{-\blq\n}^{i}=0$ for $i=2$.
A further extension concerns $H_{0,ph}^{s}$ as the 
only properties required by
our derivation are $Q(\blq)=Q^{\dag}(\blq)=Q^{\ast}(-\blq)$.
Thus the formalism can deal
with Hamiltonians containing terms proportional to
$\hat{P}_{\blq\n}^{\dag}\hat{U}_{\blq\n}$ and 
$\hat{U}_{\blq\n}^{\dag}\hat{P}_{\blq\n}$. 
If the bosonic particles are photons these 
terms arise in the effective Hamiltonian for squeezed 
light~\cite{andersen_30-years_2016}. We finally observe that all these 
extensions widen the class of external drivings to those that
break the crystal periodicity.  

Coming back to the $e$-$ph$ Hamiltonian, the theory presented in this 
work indicates the route for the treatment of anharmonic 
effects~\cite{monserrat_anharmonic_2013,antonius_dynamical_2015}.
To third order in the fluctuation operators we must add to the 
Hamiltonian in Eq.~(\ref{el-phonham3}) the Debye-Waller interaction,
which is proportional to $\hat{U}^{2}\D\hat{n}$, and the 
next-to-quadratic term in the expansion of $E_{n-n}$, which is 
proportional to $\hat{U}^{3}$. These two extra pieces in the 
Hamiltonian can be treated in precisely the same manner as we have 
treated $\hat{H}_{e-e}$ and $\hat{H}_{e-ph}$ in 
Section~\ref{exactexpsec}. The diagrammatic theory will be enriched 
by two more couplings, the Debye-Waller one 
connecting two $G$-lines to two $D$-lines, and the one stemming from the 
cubic displacement connecting three $D$-lines. 
The systematic analysis of the resulting diagrammatic expansion is 
a topic for future research.  Another topic for future research is the 
generalization of the NEGF formalism for superconductors. Starting as always 
from the Hamiltonian Eq.~(\ref{el-phonham3}) we here outline the 
steps for this achievement (i) add to $\hat{H}$ an infinitesimally 
small term breaking the $U(1)$ gauge symmetry, e.g., 
$\int d\blx d\blx' 
\D(\blx,\blx')\hat{\q}^{\dag}(\blx)\hat{\q}^{\dag}(\blx')+h.c.$; (ii) 
derive the noninteracting Martin-Schwinger hierarchy for 
the many-particle GF containing $n$ annihilation operators and $m$ 
creation operators with $n\neq m$; (iii) establish the Wick's theorem 
as the solution of the generalized noninteracting Martin-Schwinger 
hierarchy and (iv) expand the GF with $n+m=1$ (this is the so called 
Nambu GF) as outlined in Section~\ref{exactexpsec}.

We conclude with our perspective on the combination of the NEGF formalism with 
DFT, which we intentionally left aside in this work as it was not
necessary in developing the theory.
Zero-temperature~\cite{hohenberg_inhomogeneous_1964,kohn_self-consistent_1965} or 
finite-temperature~\cite{m.1965} DFT as well as Density Functional 
Perturbation 
Theory (DFTP)~\cite{zein_1984,baroni_green_1987,blat_calculations_1991,giannozzi_ab-initio_1991,zein_ab-initio_1992} 
provide an invaluable capacity to obtain the {\em ab initio} 
parameters of the 
Hamiltonian~\cite{epwcode,zhou_perturbo_2021,cepellotti_phoebe_2022,epiqcode,nomura_ab-initio_2015,petretto_high-throughput_2018,mounet_two-dimensional_2018}. Through DFT and DFPT one can minimize the total 
energy with respect to the nuclear positions and find an approximation 
for $\blR^{0}$, and hence for the electronic 
potential $V$ and coupling $g$ through Eqs.~(\ref{V(r)}) and 
(\ref{gia(r)}). In the partial self-consistent scheme for the 
determination of the $e$-$ph$ Hamiltonian, see end of 
Section~\ref{qselnusec},  the equilibrium 
DFT density can also be used to calculate the elastic tensor $K$. 
However, the DFT density cannot be used in the term containing 
$\D\hat{n}$, see Eq.~(\ref{el-phonintham2}) and following discussion,
for otherwise the nuclei would start drifting away from $\blR^{0}$. 
DFPT can also be used to approximate the dressed $e$-$ph$ coupling, 
necessary for the calculation of the phononic self-energy and 
phonon-induced $e$-$e$ interaction, see Table~\ref{Hedinebtab}. In 
fact, $g^{d}=(\d+WP)g=(\d+v\chi)g$. However, DFPT provides $g^{d}$ in the 
clamped+static approximation which we have already observed to be 
nonconserving. One may think to restore 
the conserving properties by resorting to Time-Dependent (TD) 
DFT~\cite{RungeGross:84,onida_electronic_2002,Ullrich:12,maitra_perspective_2016}, which gives us access to  
the full frequency dependence of the clamped response 
function. This is unfortunately not so. The problem lies in the fact 
that from the TDDFT expression $\chi=\chi_{0}+\chi_{0}(v+f_{\rm xc})\chi$, with 
$f_{\rm xc}$ the xc kernel, it is not obvious to trace the 
$\F$-functional fulfilling $\chi=-\frac{1}{2}\d\F_{\rm xc}/\d 
v$~\cite{svl-book}. One possibility would be to construct the kernel from the 
conserving linearized Sham-Schl\"uter 
equation~\cite{sham_density-functional_1983} as discussed in 
Ref.~\cite{vonbarth_conserving_2005}. However, 
the linearized Sham-Schl\"uter equation is itself an approximate 
equation, hence the resulting theory will never be exact. 
We finally remark that if one is only interested in spectral 
properties then the use of non-conserving 
approximations is much less 
critical. In this case the fundamental requirement is to use 
positive-semidefinite (PSD) self-energies~\cite{stefanucci_diagrammatic_2014,uimonen_diagrammatic_2015}. 
PSD approximations can be evaluated 
with the dressed DFTP $e$-$ph$ coupling by simply paying 
attention to the double-counting problem.

We hope that our contribution to the $e$-$ph$ problem will stimulate 
further research both at a fundamental and computational level 
and that it will be of help in the development and in the 
implementation of accurate approximation 
schemes to face the novel challenges posed by new materials and time-resolved 
experiments.

\begin{acknowledgments}
G.S. and E.P. acknowledge the financial support from MIUR PRIN (Grant No. 
20173B72NB) and from INFN
through the TIME2QUEST project.
R.v.L. would like to thank the Finnish Academy for support under 
Project No. 317139.
\end{acknowledgments}

\appendix

\section{Noninteracting Green's functions}
\label{nonintDsec}

The GF $G_{0}^{s}$ and $D_{0}$ are the 
building blocks of the diagrammatic expansions. In 
Refs.~\cite{sa-1.2004,svl-book} we derive the explicit form of $G_{0}^{s}(z,z')$ 
with arguments $z,z'\in\g$  and then extract all its Keldysh 
components. In this appendix we proceed along the lines outlined in 
Ref.~\cite{sakkinen_many-body_2015-2} and derive 
the explicit form of $D_{0}(z,z')$ 
with arguments on the contour $\g$. We further extract all Keldysh 
components of $D_{0}(z,z')$ by choosing $z$ and $z'$ on 
the different branches of $\g$. 

The phononic GF $D_{0}$ satisfies 
Eq.~(\ref{eomford01}).
 Mutatis mutandis we can derive the 
equation of motion with derivative 
with respect to $z'$ (in matrix form)
\begin{align}
D_{0,\blq}(z,z')\Big[-i
\frac{\overleftarrow{d}}{dz'}\a-
Q(\blq,z')\Big]=\mathbb{1}\d(z,z'),
\label{eomd0ephon2}
\end{align}
where $\mathbb{1}_{\n,\n'}^{ii'}=\d_{\n\n'}\d_{ii'}$.
According to the definition in Eq.~(\ref{phicomp}) and the result in 
Eq.~(\ref{connectedDcdef})  the 
matrix elements of $D_{0}$ are correlators between 
fluctuation operators of displacements 
and momenta. Taking into account  
Eq.~(\ref{connctD0D0}) we have for instance
\begin{align}
D_{0,\blq\n\n'}^{11}(z,z')=\frac{1}{i}\frac{1}{\callZ^{s}_{0,ph}}
\bra\callT\Big\{\,\D\hat{U}_{\blq\n}(z)\D\hat{U}_{-\blq\n'}(z')\Big\}
\ket_{0,ph}^{s}\;,
\label{dispdispcorr}
\end{align}
where the independent average $\bra\ldots\ket_{0,ph}^{s}$
is defined below Eq.~(\ref{facteb}).
We define the  non-unitary matrices $W_{L\blq}$ and $W_{R\blq}$ as 
the solution of
\begin{subequations}
\begin{align}
i\frac{d}{dz}W_{L\blq}(z)&=
Q(\blq,z)\a
W_{L\blq}(z),
\\
-i\frac{d}{dz}W_{R\blq}(z)
&=
W_{R\blq}(z)
Q(\blq,z)\a,
\end{align}	
\end{subequations}
with boundary conditions 
$W_{L\blq}(t_{0-})=W_{R\blq}(t_{0-})=\mathbbm{1}$. The explicit 
expression for these matrices is given in terms of the 
contour-ordering and anti-contour-ordering operators
\begin{subequations}
\begin{align}
W_{L\blq}(z)&=\callT\left\{e^{-i\int_{t_{0-}}^{z}d\bar{z}\;
Q(\blq,\bar{z})\a}\right\},
\\
W_{R\blq}(z)&=\bar{\callT}\left\{e^{i\int_{t_{0-}}^{z}d\bar{z}\;
Q(\blq,\bar{z})\a}\right\},
\end{align}
\end{subequations}
from which it follows that $W_{L\blq}(z)W_{R\blq}(z)=
W_{R\blq}(z)W_{L\blq}(z)=\mathbbm{1}$.
We look for solutions of the form
\begin{align}
D_{0,\blq}(z,z')=-i\a W_{L\blq}(z)
F_{\blq}(z,z')W_{R\blq}(z').
\end{align}
Inserting this expression into Eqs.~(\ref{eomford01}) and 
(\ref{eomd0ephon2}) we obtain a couple of equations for the unknown 
matrix function $F_{\blq}$
\begin{align}
\frac{d}{dz}F_{\blq}(z,z')=-\frac{d}{dz'}F_{\blq}(z,z')=
\mathbbm{1}\d(z,z'),
\end{align}
which is solved by 
$F_{\blq}(z,z')=\th(z,z')F^{>}_{\blq}+\th(z',z)F^{<}_{\blq}$,
with 
\begin{align}
F^{>}_{\blq}-F^{<}_{\blq}=\mathbbm{1}.
\label{F>F<=1}
\end{align}
Imposing the KMS relations
$D_{0,\blq}(t_{0-},z')=D_{0,\blq}(t_{0}-i\b,z')$ 
we find
\begin{align}
F^{<}_{\blq}=W_{L\blq}(t_{0}-i\b)
F^{>}_{\blq}=e^{-\b Q(\blq)\a }F^{>}_{\blq},
\label{kmsF>F<}
\end{align}
where $Q(\blq)=Q(\blq,t_{0}-i\t)$ is independent of $\t$.
Equations (\ref{F>F<=1}) and (\ref{kmsF>F<}) can be solved for 
$F^{<}_{\blq}$ and $F^{>}_{\blq}$, and the final expression for the 
phononic GF reads
\begin{align}
D_{0,\blq}(z,z')=-i\a W_{L\blq}(z)
\Big[&\th(z,z')\bar{f}(Q(\blq)\a)
\nn\\+&
\th(z',z)f(Q(\blq)\a)\Big]W_{R\blq}(z'),
\label{D0contour}
\end{align}
where $f(\w)=1/(e^{\b\w}-1)$ is the Bose function and 
$\bar{f}(\w)=e^{\b\w}f(\w)$. Having the GF on the 
contour we can now extract all its Keldysh components.

{\em Matsubara component.--} The Matsubara component 
$D_{0,\blq}^{\rm M}(\t,\t')$ is obtained by 
setting $z=t_{0}-i\t$ and $z'=t_{0}-i\t'$ in Eq.~(\ref{D0contour}). 
Alternatively, we can set $z=t_{0}-i\t$ and $z'=t_{0}-i\t'$ in 
one of the equations of motion, and then solve for $D_{0,\blq}^{\rm M}$. 
Choosing the equation of motion
Eq.~(\ref{eomford01}) we have
\begin{align}
\Big[-
\frac{d}{d\t}\,\a -Q(\blq)
\Big]
D_{0,\blq}^{\rm M}(\t,\t')
=\mathbbm{1}i\d(\t,\t')
\label{eomd0ephon1M}
\end{align}
Expanding the Matsubara GF and the Dirac-delta function 
in bosonic Matsubara frequencies, see Eq.~(\ref{matsexp}),
we find an algebraic equation 
for the coefficients of the expansion 
\begin{align}
\Big[\w_{m}\a-Q(\blq)\Big]D_{0,\blq}^{\rm M}(\w_{m})=\mathbbm{1}.
\label{Matzcoeffephon}
\end{align}
Taking into account the explicit form of the matrices $\a$ and $Q(\blq)$ 
Eq.~(\ref{Matzcoeffephon}) is converted into a set of 
algebraic equations for the four blocks of the matrix $D_{0,\blq}^{\rm M}$
\begin{align}
\left(\begin{array}{cc}
-K(\blq) & i\w_{m}\mathbbm{1} \\
-i\w_{m}\mathbbm{1} & -\mathbbm{1}
\end{array}\right)
\left(\begin{array}{cc}
D_{0,\blq}^{11,\rm M}(\w_{m}) & D_{0,\blq}^{12,\rm M}(\w_{m}) \\
& \\
D_{0,\blq}^{21,\rm M}(\w_{m}) & D_{0,\blq}^{22,\rm M}(\w_{m})
\end{array}\right)=\left(\begin{array}{cc}
\mathbbm{1} & 0 \\
0 & \mathbbm{1}
\end{array}\right).
\end{align}
In this formula each of the entries of the $2\times 2$ matrices is 
itself a matrix with indices $\n,\n'$. The 
$(1,1)$ block corresponding to the displacement-displacement 
correlator in Eq.~(\ref{dispdispcorr}) reads
\begin{align}
D_{0,\blq}^{11,\rm M}(\w_{m})=\frac{1}{\w^{2}_{m}-K(\blq)}.
\label{11blockDM}
\end{align}
If we choose to work in the basis of the normal modes of $K$
then $K_{\n\n'}(\blq)=\d_{\n\n'}\w^{2}_{0\blq\n}$ and 
the displacement-displacement correlator simplifies to
\begin{align}
D_{0,\blq\n\n'}^{11,\rm M}(\w_{m})=
\frac{\d_{\n\n'}}{\w^{2}_{m}-\w^{2}_{0\blq\n}}.
\end{align}
As already emphasized in Section~\ref{qselnusec}  the 
eigenvalues $\w^{2}_{0\blq\n}$ are not physical and can even be 
negative. Therefore, there is no particular convenience to work in 
the basis of the normal modes of $K$.

{\em Lesser and greater components.--}
The lesser (greater) component is 
obtained by setting $z=t_{-}$ ($z=t_{+}$) 
and $z'=t'_{+}$ ($z'=t'_{-}$) in Eq.~(\ref{D0contour}). 
For times on the horizontal branches of the contour the matrices 
$W_{L\blq}$ and $W_{R\blq}$ are independent of the branch, i.e.,
$W_{L\blq}(t_{\pm})=W_{L\blq}(t)$ and 
$W_{R\blq}(t_{\pm})=W_{R\blq}(t)$ with
\begin{subequations}
\begin{align}
W_{L\blq}(t)&=T\left\{e^{-i\int_{t_{0}}^{t}d\bar{t}\;
Q(\blq,\bar{t})\a}\right\},
\\
W_{R\blq}(t)&=
\bar{T}\left\{e^{i\int_{t_{0}}^{t}d\bar{t}\;
Q(\blq,\bar{t})\a}\right\},
\end{align}
\end{subequations}
and $T$ and $\bar{T}$ the time-ordering and anti-time-ordering 
operators. We then have
\begin{subequations}
\begin{align}
D_{0,\blq}^{<}(t,t')&=-i\a W_{L\blq}(t)
f\big(Q(\blq)\a\big)W_{R\blq}(t'),
\\
D_{0,\blq}^{>}(t,t')&=-i\a W_{L\blq}(t)
\bar{f}\big(Q(\blq)\a\big)W_{R\blq}(t').
\end{align}
\label{D0><ttp}
\end{subequations}
The greater component is obtained from the lesser component by 
replacing $f\to\bar{f}$. In the following we then consider only 
$D_{0,\blq}^{<}$.

Let us discuss the special, yet most common, case of 
$Q(\blq,t)=Q(\blq)$ 
independent of time (no phonon driving). Then 
$W_{L\blq}(t)=\exp[-i Q(\blq)\a (t-t_{0})]$ and 
$W_{R\blq}(t)=\exp[i Q(\blq)\a (t-t_{0})]$ commute with the 
Bose function, and the lesser component simplifies to 
\begin{align}
D_{0,\blq}^{<}(t,t')&=-i\a 
f\big(Q(\blq)\a\big)e^{-i Q(\blq)\a (t-t')}.
\end{align}
As this function depends only on the time difference it can be 
Fourier transformed with respect to $t-t'$. 
In frequency space the lesser GF
reads
\begin{align}
D_{0,\blq}^{<}(\w)&=-2\pi \a
f\big(Q(\blq)\a\big)\,\d\big(\w- Q(\blq)\a\big)
\nn\\ &=
-2\pi\a
f(\w)\,\d\big(\w- Q(\blq)\a\big).
\label{D0<w}
\end{align}
We can easily work out the four blocks of $D_{0,\blq}^{<}(\w)$ 
using the Cauchy relation
\begin{align}
-2\pi\,\d(\w&- Q(\blq)\a)=\left[
\frac{1}{\w- Q(\blq)\a+i\eta}-
\frac{1}{\w- Q(\blq)\a-i\eta}\right]
\nn\\
&=\a\left[\frac{1}{(\w+i\eta)\a- Q(\blq)}-
\frac{1}{(\w-i\eta)\a- Q(\blq)}\right].
\end{align}
In the last equality we have used that
$\a^{-1}=\a$ and hence $\a A^{-1}=\a^{-1}A^{-1}=(A\a)^{-1}$ for any 
matrix $A$.
The fraction with $(\w\pm i\eta)$ is the same as 
$D_{0,\blq}^{\rm M}(\w_{m})$ calculated in $\w_{m}=\w\pm i\eta$, 
compare with Eq.~(\ref{Matzcoeffephon}). Therefore the $(1,1)$ block of 
$D_{0,\blq}^{<}(\w)$ can be read off from Eq.~(\ref{11blockDM})
\begin{align}
D_{0,\blq}^{11,<}(\w)&=
f(\w)\left[\frac{1}{(\w+i\eta)^{2}-K(\blq)}-
\frac{1}{(\w-i\eta)^{2}-K(\blq)}\right].
\end{align}
Using the same strategy one can work out the explicit form of 
the remaining three blocks. The greater component has the same form 
as the lesser component with $f(\w)\to\bar{f}(\w)$.

{\em Retarded and advanced components.--} The retarded and 
advanced components can be calculated from the lesser and greater 
components. We have 
\begin{align}
D_{0,\blq}^{\rm R}(t,t')&=\th(t-t')[D_{0,\blq}^{>}(t,t')-D_{0,\blq}^{<}(t,t')]
\nn\\
&=-i\th(t-t') \a W_{L\blq}(t)W_{R\blq}(t'),
\end{align}
and similarly
\begin{align}
D_{0,\blq}^{\rm A}(t,t')
=i\th(t'-t) \a W_{L\blq}(t)W_{R\blq}(t'),
\end{align}
from which it follows that we can write the lesser/greater components 
in Eq.~(\ref{D0><ttp}) as 	
\begin{align}	
D^{\lessgtr}_{0,\blq}(t,t')=D^{\rm R}_{0,\blq}(t,t_{0})\,\a\,
D^{\lessgtr}_{0,\blq}(t_{0},t_{0})\,\a\,D^{\rm A}_{0,\blq}(t_{0},t').
\label{D<DRD<DA}
\end{align}
This result is at the basis of the recently proposed Generalized 
Kadanoff-Baym Ansatz for bosons~\cite{karlsson_fast_2021}. In fact, 
Eq.~(\ref{D<DRD<DA}) can equivalently be written as 
\begin{align}
D^{\lessgtr}_{0,\blq}(t,t')=i D^{\rm R}_{0,\blq}(t,t')\,\a\,
D^{\lessgtr}_{0,\blq}(t',t')-iD^{\lessgtr}_{0,\blq}(t,t)\,\a\,
D^{\rm A}_{0,\blq}(t,t').
\end{align}	

In the special case $Q(\blq,t)=Q(\blq)$ the retarded/advanced 
components depend only on the time difference and they can be Fourier 
transformed. In Fourier space we have 
\begin{align}
D^{\rm R/A}_{0,\blq}(\w)
=\frac{1}{(\w\pm i\eta)\a-Q(\blq)}=D^{\rm 
M}_{0,\blq}(\w\pm i\eta).
\label{D0RAw}
\end{align}
The $(1,1)$ block can be read off from Eq.~(\ref{11blockDM})
\begin{align}
D^{11,\rm R/A}_{0,\blq}(\w)=	
\frac{1}{(\w\pm i\eta)^{2}-K(\blq)}.
\nn
\end{align}

We close the appendix by observing 
that from Eqs.~(\ref{D0<w}) 
and (\ref{D0RAw}) it follows the fluctuation-dissipation 
theorem
\begin{align}
D^{<}_{0,\blq}(\w)=f(\w)\,
\Big[D^{\rm R}_{0,\blq}(\w)-D^{\rm A}_{0,\blq}(\w)\Big],
\nn
\end{align}
and the like for the greater component.

\section{Born-Oppenheimer approximation}
\label{BOapp}

The Born-Oppenheimer (BO) Hamiltonian is given by Eq.~(\ref{elnucham}) in 
the limit of infinite nuclear masses, and it is therefore independent 
of the nuclear momenta
\begin{align}
\hat{H}^{\rm BO}(\hat{\blR})\equiv 
\lim_{\{M_{i}\}\to\iif}
\hat{H}.
\end{align}
The BO Hamiltonian  commutes with all 
the nuclear position operators $\hat{\blR}_{i}$. Therefore the 
eigenkets of $\hat{H}^{\rm BO}(\hat{\blR})$ have the form 
$|\Q(\blR)\ket|\blR_{1}\ldots\blR_{N_{n}}\ket$ where the electronic 
ket $|\Q(\blR)\ket\in\mathbb{F}$ is an eigenket of 
$\hat{H}^{\rm BO}(\blR)$, and 
$|\blR_{1}\ldots\blR_{N_{n}}\ket\in\mathbb{D}_{N_{n}}$
is a ket describing $N_{n}$ distinguishable nuclei in positions 
$\blR=(\blR_{1},\ldots,\blR_{N})$. 

Let $|\Q^{\rm BO}(\blR)\ket$ 
be the ground state of $\hat{H}^{\rm BO}(\blR)$ with ground-state 
energy  
$E^{\rm BO}(\blR)$. 
An approximation for the 
equilibrium positions of the nuclei and for the equilibrium 
electronic density is found by minimizing $E^{\rm 
BO}(\blR)$ over all $\blR$. Using the Hellmann-Feynman 
theorem 
\begin{align}
\frac{\de E^{\rm BO}(\blR)}{\de R_{i,\a}}
&=\bra\Q^{\rm BO}(\blR)|\frac{\de \hat{H}^{\rm BO}(\blR)}{\de 
R_{i,\a}}|\Q^{\rm BO}(\blR)\ket
\nn\\
&=-\int \!d\blx\,n^{\rm BO}(\blx,\blR)\frac{\de 
V(\blr,\blR)}{\de R_{i,\a}}+\frac{\de E_{n-n}(\blR)}{\de R_{i,\a}},
\label{nuclforce}
\end{align}
where we define the BO density
\begin{align}
n^{\rm BO}(\blx,\blR)\equiv\bra\Q^{\rm BO}(\blR)|
\hat{n}(\blx)|\bra\Q^{\rm BO}(\blR)\ket.
\end{align}
Let $\blR^{0,\rm BO}$ be the solution of $\de E^{\rm BO}(\blR)/\de R_{i,\a}=0$ for all
$i,\a$.
The BO approximation consists in approximating
$\blR^{0}\simeq \blR^{0,\rm BO}$ and 
$n^{0}(\blx)\simeq n^{0,\rm 
BO}(\blx)\equiv n^{\rm 
BO}(\blx,\blR^{0,\rm BO})$. The approximated nuclear positions and density can 
then be used to obtain an approximation for $V\simeq V^{\rm BO}$, 
$g\simeq g^{\rm BO}$ and $K\simeq K^{\rm BO}$. 

{\em On the stability of the BO approximation.--}
As already pointed out in Section~\ref{qselnusec}
the 
equilibrium density $n^{0}$ enters the low-energy Hamiltonian through $K$ 
{\em as well as} 
$\D\hat{n}$. The replacement 
$n^{0}\simeq n^{0,\rm BO}$ in $\D\hat{n}$ 
leads to an 
inconsistency. In fact, any exact or even approximate  
treatment of the 
low-energy Hamiltonian Eq.~(\ref{el-phonham})
gives a ground state density $n^{0}$ which  
is, in general, different from  $n^{0,\rm BO}$. 
In equilibrium we then  
have $\D n=n^{0}-n^{0,\rm BO}\neq 0$, and the nuclei move away
from $\{\blR\}=\{\blR^{0,\rm BO}\}$
in accordance with the 
equation of motion Eq.~(\ref{eomfornuclp2}). 
The minimal effort to get rid of the inconsistency
consists in treating $n^{0}$ in $\D\hat{n}$ self-consistently 
while keeping  $V$, $g$ and $K$ fixed at the BO values.
Whether this partial self-consistent scheme leads to a stable dynamics when the 
system is perturbed by weak external fields remains to be checked 
case by case. 

{\em Elastic tensor and Hessian of the BO energy.--}
The elastic tensor $K^{\rm BO}$ can alternatively be obtained from 
the Hessian $\callH$ of 
$E^{\rm BO}(\blR)$ and the density-density response function {\em 
at clamped nuclei}.
Taking into account the definitions in Eq.~(\ref{nucldenscoeff}), 
a second differentiation of 
Eq.~(\ref{nuclforce}) yields~\cite{baroni_phonons_2001}
\begin{align}
\callH_{i,\a;j,\b}&\equiv
\left.\frac{\de^{2} E^{\rm BO}(\blR)}{\de R_{i,\a}\de R_{j,\b}}
\right|_{\blR=\blR^{0,\rm BO}}	
\nn\\&=
-\int \!d\blx\left.\frac{\de n^{\rm BO}(\blx,\blR)}{\de R_{j,\b}}
\right|_{\blR=\blR^{0,\rm BO}}\!
g^{\rm BO}_{i,\a}(\blr)
\nn\\&-\int \!d\blx\,n^{0,\rm BO}(\blx)g^{\rm DW,BO}_{i,\a;j,\b}(\blr)+
\left.\frac{\de^{2} E_{n-n}(\blR)}{\de R_{i,\a}\de 
R_{j,\b}}\right|_{\blR=\blR^{0,\rm BO}}.
\label{2derboener}
\end{align}
The first term on the r.h.s. can be rewritten in a more symmetric 
form using linear response theory. Let us imagine to manually move the 
infinitely heavy nuclei from $\blR^{0,\rm BO}$ to $\blR^{0,\rm BO}+\d\blR$. 
We indicate with $\d \blR(t)$  the 
extremely slow 
time-dependent function with the property 
that $\d \blR(t)=0$ for $t=-\iif$ and $\d \blR(t)=\d \blR$ 
for $t=\iif$. This nuclear rearrangement induces a change in the electronic density.
For the extremely slow (adiabatic) change considered here
the time-dependent electronic density $n(\blx,t)$ is identical to
the ground-state electronic density $n^{\rm BO}(\blx,\blR(t))$ 
corresponding to a nuclear geometry $\blR(t)=\blR^{0,\rm BO}+\d\blR(t)$
(instantaneous relaxation). We then have
\begin{align}
\d n^{\rm BO}(\blx,t)&=
-\!
\int \!d\blx'dt'\chi^{\rm R}_{\rm clamp}(\blx,t;\blx',t')
\d V(\blr',\blR(t'))
\nn\\
&=-\!
\int \!d\blx'dt'\chi^{\rm R}_{\rm clamp}(\blx,t;\blx',t')
\!\sum_{j\b}g^{\rm BO}_{j,\b}(\blr')\d R_{j,\b}(t'),
\end{align}
where $\chi^{\rm R}_{\rm clamp}$ is the  equilibrium density-density 
response function of the electronic system with clamped nuclei in 
$\blR^{0,\rm BO}$.
For a response function with the property that 
$\chi^{\rm R}_{\rm clamp}(\blx,t;\blx',t')\to 0$ for $|t-t'|\to \iif$
we can replace 
$\d R_{j,\b}(t')\to \d R_{j,\b}(t)$ (adiabatic change). 
Performing the integral over 
$t'$ and taking the limit $t\to\iif$ we get 
\begin{align}
\d n^{\rm BO}(\blx)
=-\int \!d\blx'\chi^{\rm R}_{\rm clamp}(\blx,\blx';0)
\sum_{j\b}g^{\rm BO}_{j,\b}(\blr')\d R_{j,\b},
\label{nboadiab}
\end{align}
where $\chi^{\rm R}_{\rm clamp}(\blx,\blx';0)$ is the Fourier 
transform of the response function calculated at zero frequency. 
Using Eq.~(\ref{nboadiab}) to evaluate the derivative of the ground-state 
density in Eq.~(\ref{2derboener}) and comparing with 
Eq.~(\ref{elasticmat}) we conclude that
\begin{align}
\callH_{i,\a;j,\b}=K^{\rm BO}_{i,\a;j,\b}
+\int d\blx d\blx' g^{\rm BO}_{i,\a}(\blr)
\chi^{\rm R}_{\rm clamp}(\blx,\blx';0)g^{\rm BO}_{j,\b}(\blr').
\label{elasticmat2}
\end{align}
The elastic tensor in the BO approximation
coincides with the Hessian of the BO 
energy only if the electron-nuclear coupling $g^{\rm BO}$ vanishes.

In crystals the relation in Eq.~(\ref{elasticmat2})  reads
\begin{align}
\callH_{s,\a;s',\b}(\bln-\bln')&=K^{\rm BO}_{s,\a;s',\b}(\bln-\bln')
\nn\\
&+\int d\blx d\blx' g^{\rm BO}_{\bln s,\a}(\blr)
\chi^{\rm R}_{\rm clamp}(\blx,\blx';0)g^{\rm BO}_{\bln's',\b}(\blr').
\label{elhesrel}
\end{align}
The Hessian tensor satisfies the same properties as the elastic 
tensor. Let then be $\w^{2}_{\blq\n}$ and $\ble^{\n}(\blq)$ the 
eigenvalues and normal modes of $\callH(\blq)$, which is defined in 
terms of 
$\callH(\bln)$ like in Eq.~(\ref{Kprop}). 
The eigenvalues $\w^{2}_{\blq\n}$ are
by construction non-negative 
since the Hessian of a function calculated in the 
global minimum is positive semidefinite.
Multiplying Eq.~(\ref{elhesrel}) by $e^{\n\ast}_{s,\a}(\blq)
e^{-\iu\blq\cdot(\bln-\bln')}e^{\n'}_{s'\b}(\blq)/\sqrt{M_{s}M_{s'}}$ 
and summing over $\bln s,\a$ and $\bln's',\b$ we obtain
\begin{align}
\d_{\n\n'}\w^{2}_{\blq\n}=K^{\rm BO}_{\n\n'}(\blq)	+\!
\int\! d\blx d\blx' g^{\rm BO}_{\blq\n}(\blr)\,\chi^{\rm 
R}_{\rm clamp}(\blx,\blx';0)\,g^{\rm BO}_{-\blq\n'}(\blr'),
\label{KHesrel}
\end{align}
to be compared with Eqs.~(\ref{clamp+statPi}) and (\ref{Kren}).

\section{Symmetry properties of the phononic Green's function and self-energy}
\label{symmapp}

From the definition in Eq.~(\ref{connectedDcdef}) and the property in 
Eq.~(\ref{Ddiagq}) we have
\begin{align}
D_{\blq\n\n'}^{ii'}(z,z')
=\frac{1}{i}\Tr\Big[\r\callT\Big\{\D\hat{\f}^{i}_{\blq\n,H}(z)
\D\hat{\f}^{i'}_{-\blq\n',H}(z')\Big\}\Big].
\label{explD}
\end{align}
It is then straightforward to derive 
\begin{subequations}
\begin{align}
D^{ii',<}_{\blq\n\n'}(t,t')&=D^{i'i,>}_{-\blq\n'\n}(t',t),
\\
D^{ii',\gtrless}_{\blq\n\n'}(t,t')&=-
[D^{i'i,\gtrless}_{\blq\n'\n}(t',t)]^{\ast},
\end{align}
\label{Dcprop}
\end{subequations}
The phononic Green's function satisfies also another important 
property if the Hamiltonian is invariant under time-reversal. 
Let $\hat{\Th}$ be the anti-unitary time-reversal operator.
Time reversal symmetry implies that we can choose the many-body eigenstates 
$|\Q_{A}\ket$ of 
$\hat{H}$ such that $|\Q_{A}\ket=\hat{\Th}|\Q_{A}\ket$, and 
therefore we have the property~\cite{sakurai_napolitano_2017} 
\begin{align}
\bra\Q_{A}|\hat{O}|\Q_{B}\ket=\bra\Q_{B}|\hat{\Th}\,\hat{O}^{\dag}\hat{\Th}^{-1}|\Q_{A}\ket,
\label{trprop}
\end{align}
for any operator $\hat{O}$. Under a time-reversal transformation the
displacement operators are even whereas the momentum operators are 
odd, i.e., $\hat{\Th}\,\hat{U}_{\bln s,a}\hat{\Th}^{-1}=\hat{U}_{\bln 
s,a}$ and $\hat{\Th}\,\hat{P}_{\bln s,a}\hat{\Th}^{-1}=-\hat{P}_{\bln 
s,a}$. Expanding these operators like in Eqs.~(\ref{momdisplexp}) we 
immediately find 
\begin{subequations}
\begin{align}
\hat{\Th}\,\hat{U}_{\blq \n}\hat{\Th}^{-1}&=\hat{U}_{-\blq 
\n}=\hat{U}^{\dag}_{\blq \n},
\\
\hat{\Th}\,\hat{P}_{\blq \n}\hat{\Th}^{-1}&=-\hat{P}_{-\blq 
\n}=-\hat{P}^{\dag}_{\blq \n},
\end{align}
\end{subequations}
where we take into account that $\hat{\Th} c = c^{\ast}\hat{\Th}$ for 
any complex number $c$ and we use the property 
Eq.~(\ref{quasfinconst}) of the normal modes. From Eq.~(\ref{trprop}) 
we then infer that
\begin{subequations}
\begin{align}
\bra\Q_{A}|\hat{U}_{\blq\n}|\Q_{B}\ket&=\bra\Q_{B}|\hat{U}_{\blq\n}|\Q_{A}\ket,
\\
\bra\Q_{A}|\hat{P}_{\blq\n}|\Q_{B}\ket&=-\bra\Q_{B}|\hat{P}_{\blq\n}|\Q_{A}\ket.
\end{align}
\label{trpropU&P}
\end{subequations}
Let $E_{A}$ be the eigenenergy of $|\Q_{A}\ket$. If the system is 
in equilibrium then $|\Q_{A}\ket$ is also an eigenket of the density 
matrix $\hat{\r}$, and we denote by $\r_{A}$ its eigenvalue. Consider 
the $(1,1)$ block of the greater GF in Eq.~(\ref{explD}). We have
\begin{align}
D_{\blq\n\n'}^{11,>}(t,t')
&=\frac{1}{i}\sum_{A,B}\r_{A}e^{i(E_{A}-E_{B})(t-t')}
\nn\\&\times
\bra\Q_{A}|\hat{U}_{\blq\n}|\Q_{B}\ket\bra\Q_{B}|\hat{U}_{-\blq\n'}|\Q_{A}\ket
=D_{-\blq\n'\n}^{11,>}(t,t'),
\label{explDlehm}
\end{align}
where in the second equality we use Eq.~(\ref{trpropU&P}). We can 
analogously derive the relations for all other blocks and for the 
lesser GF. The final result is
\begin{align}
D_{\blq\n\n'}^{ii',\gtrless}(t,t')=(-)^{i+i'}D_{-\blq\n'\n}^{i'i,\gtrless}(t,t').
\label{trpropD><}
\end{align}

To highlight the mathematical structure of the derived relations we 
denote by $D^{T}$ the transpose matrix of $D$, i.e., 
$[D^{T}]^{ii'}_{\n\n'}\equiv D^{i'i}_{\n'\n}$, and 
$D^{\dag}=[D^{T}]^{\ast}$. Then Eqs.~(\ref{Dcprop}) and 
(\ref{trpropD><}) take the following compact form
\begin{subequations}
\begin{align}
D^{<}_{\blq}(t,t')&=[D^{>}_{-\blq}(t',t)]^{T},
\\
D^{\gtrless}_{\blq}(t,t')&=-
[D^{\gtrless}_{\blq}(t',t)]^{\dag},
\\
D^{\gtrless}_{\blq}(t,t')&=\s_{z}[D^{\gtrless}_{-\blq}(t,t')]^{T}\s_{z}
\quad {\rm[ \hat{\Th}\;invariance]},
\label{trsms}
\end{align}
\label{Dcpropcompact}
\end{subequations}
where $[\s_{z}]^{ii'}_{\n\n'}=\d_{\n\n'}\left(\begin{array}{cc} 1 & 0 
\\ 0 & -1\end{array}\right)_{ii'}$.
In frequency space the first two relations in Eqs.~(\ref{Dcpropcompact}) read
\begin{align}
D^{<}_{\blq}(\w)=[D^{>}_{-\blq}(-\w)]^{T},
\quad\quad
D^{\lessgtr}_{\blq}(\w)=-[D^{\lessgtr}_{\blq}(\w)]^{\dag},
\end{align}
which imply
\begin{align}
D^{\rm R}_{\blq}(\w)=[D^{\rm A}_{-\blq}(-\w)]^{T}
,\quad\quad
D^{\rm R}_{\blq}(\w)=[D^{\rm A}_{\blq}(\w)]^{\dag},
\end{align}
which in turn lead to the following properties of the phononic 
self-energy through Eq.~(\ref{ssradispg}) [take into account that 
$\a^{T}=-\a$ and $Q^{T}(-\blq)=Q(\blq)$]
\begin{align}
\P^{\rm R}_{\blq}(\w)=[\P^{\rm A}_{-\blq}(-\w)]^{T}
,\quad\quad	
\P^{\rm R}_{\blq}(\w)=[\P^{\rm A}_{\blq}(\w)]^{\dag}.
\end{align}
Writing the retarded/advanced phononic self-energy like in 
Eq.~(\ref{Pi=L+iG}) we can deduce the following properties of 
$\L_{ph}$ and $\G_{ph}$
\begin{subequations}
\begin{align}
\L_{ph,\blq}(\w)&=[\L_{ph,-\blq}(-\w)]^{T},
\quad\quad
\L_{ph,\blq}(\w)=[\L_{ph,\blq}(\w)]^{\dag},
\\
\G_{ph,\blq}(\w)&=-[\G_{ph,-\blq}(-\w)]^{T},
\quad\quad
\G_{ph,\blq}(\w)=[\G_{ph,\blq}(\w)]^{\dag}.
\end{align}
\label{LphonGphonprop}
\end{subequations}

For systems with time-reversal symmetry 
we have also the properties, see Eq.~(\ref{trsms}),
\begin{align}
D^{\lessgtr}_{\blq}(\w)=\s_{z}[D^{\lessgtr}_{-\blq}(\w)]^{T}\s_{z},
\end{align}
which implies
\begin{align}
D^{\rm R/A}_{\blq}(\w)=\s_{z}[D^{\rm R/A}_{-\blq}(\w)]^{T}\s_{z},
\end{align}
and therefore 
\begin{align}
\P^{\rm R/A}_{\blq}(\w)=\s_{z}[\P^{\rm R/A}_{-\blq}(\w)]^{T}\s_{z},
\end{align}
or equivalently
\begin{subequations}
\begin{align}
\L_{ph,\blq}(\w)&=\s_{z}[\L_{ph,-\blq}(\w)]^{T}\s_{z},
\\
\G_{ph,\blq}(\w)&=\s_{z}[\G_{ph,-\blq}(\w)]^{T}\s_{z}.
\end{align}
\label{LphonGphonproptr}
\end{subequations}


%

\end{document}